\documentstyle[preprint,tighten,eqsecnum,aps,floats,psfig]{revtex}

\def\mpalla{\vbox{\hbox{\null\kern0.25em$\scriptstyle\circ$}%
\vskip-1.65ex\hbox{$m$}}}

\begin{document}

\draft

\preprint{IFUP-TH 22/99\ \ \ \ \ \ ROM 99/1253}

\title{
Improved high-temperature expansion and critical equation of state 
of three-dimensional Ising-like systems}
\author{Massimo Campostrini$\,^1$, Andrea Pelissetto$\,^2$, 
        Paolo Rossi$\,^1$, Ettore Vicari$\,^1$ }
\address{
$^1$ Dipartimento di Fisica dell'Universit\`a di Pisa and I.N.F.N., 
I-56126 Pisa, Italy
}
\address{$^2$ Dipartimento di Fisica dell'Universit\`a di Roma I
and I.N.F.N., I-00185 Roma, Italy \\
{\bf e-mail: \rm
{\tt campo@mailbox.difi.unipi.it}, 
{\tt rossi@mailbox.difi.unipi.it},
{\tt Andrea.Pelissetto@roma1.infn.it},
{\tt vicari@mailbox.difi.unipi.it}
}}
\date{June 18th, 1999}

\maketitle

\begin{abstract}
High-temperature series are computed for a generalized $3d$ Ising
model with arbitrary potential.  Two specific ``improved'' potentials
(suppressing leading scaling corrections) are selected by Monte Carlo
computation.  Critical exponents are extracted from high-temperature
series specialized to improved potentials, achieving high accuracy;
our best estimates are: $\gamma=1.2371(4)$, $\nu=0.63002(23)$,
$\alpha=0.1099(7)$, $\eta=0.0364(4)$, $\beta=0.32648(18)$.  By the
same technique, the coefficients of the small-field expansion for the
effective potential (Helmholtz free energy) are computed.  These
results are applied to the construction of parametric representations
of the critical equation of state.  A systematic approximation scheme,
based on a global stationarity condition, is introduced (the
lowest-order approximation reproduces the linear parametric model).
This scheme is used for an accurate determination of universal ratios
of amplitudes.  A comparison with other theoretical and experimental
determinations of universal quantities is presented.
\end{abstract}

\vskip1.5pc
\bgroup\small
\leftskip=0.10753\textwidth \rightskip\leftskip
\par\egroup
\vskip1.5pc



\section{Introduction}
\label{introduction}

According to the universality hypothesis, critical phenomena can be
described by quantities that do not depend on the microscopic details
of a system, but only on global properties such as the dimensionality
and the symmetry of the order parameter. Many three-dimensional
systems characterized by short-range interactions and a scalar order
parameter (such as density or uniaxial magnetization) belong to
the Ising universality class. This implies that the critical
exponents, as well as other universal quantities, are the same for all
these models.  Their precise determination is therefore important in
order to test the universality hypothesis.

The high-temperature (HT) expansion is one of the most effective
approaches to the study of critical phenomena.  Much work (even
recently) has been devoted to the computation of HT series,
especially for $N$-vector models and in particular the Ising
model.  An important issue in the analysis of the HT series is related
to the presence of non-analytic corrections to the leading power-law
behavior.  For instance, according to the renormalization group theory
(see e.g.\ Ref.\ \cite{ZJ-book}), the magnetic susceptibility should
behave as
\begin{equation}
\chi = C t^{-\gamma} \left( 1 + a_{0,1} t + a_{0,2}t^2 + ... + a_{1,1} 
t^\Delta + a_{1,2} t^{2\Delta} + ... + a_{2,1} t^{\Delta_2} + ... \right),
\label{chiwexp}
\end{equation}
where $t\equiv (T-T_c)/T_c$ is the reduced temperature.  The leading
critical exponent $\gamma$, and the correction exponents
$\Delta,\Delta_2,...$, are universal, while the amplitudes $C$ and
$a_{i,j}$ are non-universal and should depend smoothly on any
subsidiary parameter that may change $T_c$, but does not affect the
nature of the transition.  Non-analytic correction terms to the
leading power-law behavior, represented by noninteger powers of $t$,
are related to the presence of irrelevant operators.  For
three-dimensional Ising-like systems the existence of leading
corrections described by $\Delta\simeq 0.5$ is well established.  In
order to obtain precise estimates of the critical parameters, the
approximants of the HT series should properly allow for the confluent
non-analytic corrections
\cite{Nickel-82,Gaunt-82,ZJ-79,C-F-N-82,Adler-83,G-R-84,F-C-85}.  The
so-called integral approximants \cite{int-appr-ref} can, in principle,
allow for them (see e.g.\ Ref.\ \cite{Guttrev} for a review).
However, they require long series to detect non-leading effects, and
in practice they need to be biased to work well.  Analyses meant to
effectively allow for confluent corrections are generally based on
biased approximants where the value of $\beta_c$ and the first
non-analytic exponent $\Delta$ is given (see e.g.\ Refs.\ 
\cite{Roskies-81,A-M-P-82,B-C-97,P-V-gr-98,B-C-g-98}).  It is indeed
expected that the leading non-analytic correction is the dominant
source of systematic error.

An alternative approach to this problem is the construction of a HT
expansion where the dominant confluent correction is suppressed.  If
the leading non-analytic terms are not anymore present in the
expansion, the analysis technique based on integral approximants
should become much more effective, since the main source of systematic
error has been eliminated.  In order to obtain an improved
high-temperature (IHT) expansion, we may consider improved
Hamiltonians characterized by a vanishing coupling with the irrelevant
operator responsible for the leading scaling corrections.  This idea
has been pursued by Chen, Fisher, and Nickel \cite{C-F-N-82}, who
studied classes of two-parameter models (such as the bcc scalar
double-Gaussian and Klauder models).  Such models interpolate between
the spin-$1/2$ Ising model and the Gaussian model, and they are all
expected to belong to the Ising universality class.  The authors of
Ref.\ \cite{C-F-N-82} showed that improved models with suppressed
leading corrections to scaling can be obtained by tuning the
parameters (see also Refs.\ \cite{F-C-85,N-R-90}).  This approach has
been recently considered in the context of Monte Carlo simulations
\cite{H-P-V-98,B-F-M-M-98,B-F-M-M-P-R-99,Hasenbusch-99}, using lattice
$\phi^4$ models.  It is also worth mentioning the recent work
\cite{B-N-97} of Belohorec and Nickel in the context of dilute
polymers, where a substantial improvement in the determination of the
critical exponents $\nu$ and $\omega$ was achieved by simulating the
two-parameter Domb-Joyce model.

We consider the class of scalar models defined on the simple cubic
lattice by the Hamiltonian
\begin{equation}
{\cal H} = -\beta \sum_{<i,j>} \phi_i \phi_j + 
\sum_i V(\phi_i^2),\label{hamiltonian}
\end{equation}
where $\beta\equiv 1/T$, $<i,j>$ indicates nearest-neighbor sites,
$\phi_i$ are real variables, and $V(\phi^2)$ is a generic potential
(satisfying appropriate stability constraints).  The critical limit of
these models is expected to belong to the Ising universality class
(apart from special cases corresponding to multicritical points).
Using the linked cluster expansion technique, we calculated the
high-temperature expansion to 20th order for an arbitrary potential,
generalizing the existing expansions for the standard Ising model (see
e.g.\ Ref.\ \cite{B-C-97} for a review of the existing HT
calculations).  In this work we will essentially consider and present
results for a potential of the form
\begin{equation}
V(\phi^2) = \phi^2 + \lambda_4 (\phi^2-1)^2 + \lambda_6(\phi^2-1)^3 ;
\label{potential}
\end{equation}
such a potential will be assumed in the following, unless otherwise
stated.  Within this family of potentials, improved Hamiltonians can
be obtained by looking for values of the parameters $\lambda_4$ and
$\lambda_6$ for which leading scaling corrections are suppressed.  In
particular we may keep $\lambda_6$ fixed and look for the
corresponding value $\lambda_4^*$ of $\lambda_4$ that produces an
improved Hamiltonian.  Notice that, for generic choices of the
Hamiltonian, $\lambda_4^*$ may not exist. This is the case of the
$O(N)$ $\phi^4$ theory with nearest-neighbor couplings on a cubic
lattice in the large-$N$ limit, where it is impossible to find a
positive value of $\lambda_4$ achieving the suppression of the
dominant scaling corrections. Using the estimates of the leading
scaling correction amplitudes reported in Ref.\ \cite{B-C-g-98}, one
can argue that the same is true for finite $N>3$.  As shown
numerically by Monte Carlo simulations
\cite{H-P-V-98,B-F-M-M-98,B-F-M-M-P-R-99,Hasenbusch-99}, $\lambda_4^*$
exists in the case $N=1$, which is the single-component $\phi^4$ model
(i.e.\ the model presented above with $\lambda_6=0$).  By using
finite-size techniques, Hasenbusch obtained a precise estimate of
$\lambda_4^*$: $\lambda_4^*=1.10(2)$ \cite{Hasenbusch-99}.  In our
work we will also consider the spin-1 (or Blume-Capel) Hamiltonian
\begin{equation}
{\cal H} = -\beta \sum_{<i,j>}s_i s_j + D\sum_i s^2_i,
\label{spin1action}
\end{equation}
where the variables $s_i$ take the values $0,\pm 1$.
In this case the value of $D$ for which the leading scaling corrections 
are suppressed 
is $D^*=0.641(8)$~\cite{Hasenbusch-99-2}.

From the point of view of the HT expansion technique, the main problem
is the determination of the improved Hamiltonian.  Once the improved
Hamiltonian is available, the analysis of its HT series leads, as we
shall see, to much cleaner and therefore reliable results.  A precise
estimate of the parameters associated with an improved Hamiltonian is
crucial in order to obtain a substantial improvement of the IHT
results.  As shown in Refs.\ \cite{H-P-V-98,Hasenbusch-99}, Monte
Carlo simulations using finite-size scaling techniques seem to provide
the most efficient tool for this purpose.  For comparison, in the case
of the pure $\phi^4$ theory ($\lambda_6=0$), our best estimate of
$\lambda_4^*$ from the HT expansion is consistent with the
above-mentioned Monte Carlo result $\lambda_4^*=1.10(2)$, but it is
affected by an uncertainty of about 10\% (see Sec.\ \ref{lambdadet}).
So we decided to follow the strategy of determining the improved
Hamiltonian by Monte Carlo simulations employing finite-size scaling
techniques.  For the $\phi^4$ model, this work has been satisfactorily
done by Hasenbusch \cite{Hasenbusch-99}.  For the $\phi^6$ model with
$\lambda_6=1$, we performed Monte Carlo simulations in order to
calculate $\lambda_4^*$, obtaining $\lambda_4^*=1.90(4)$.  

The comparison of the results obtained from the 
three improved Hamiltonians considered
strongly supports our working
hypothesis of the reduction of systematic errors in the IHT estimates,
and provides an estimate of the residual errors due to the subleading
confluent corrections to scaling.

The analysis of our 20th order IHT series allows us to obtain very
precise estimates of the critical exponents $\gamma$, $\nu$ and
$\eta$.  Our estimates substantially improve previous determinations
by HT or other methods.

We extended our study to the small-field expansion of the effective
potential, which is the Helmholtz free energy of the model.  This
expansion can be parametrized in terms of the zero-momentum $n$-point
couplings $g_n$ in the symmetric phase.  The analysis of the IHT
series provides new results for the couplings $g_n$, and leads to interesting
comparisons with the estimates from other approaches based on field
theory and lattice techniques.  Moreover we improved the knowledge of
the universal critical low-momentum behavior of the two-point function
of the order parameter, which is relevant for critical scattering
phenomena.

By exploiting the known analytic properties of the critical equation
of state, one may reconstruct the full critical equation of state from
the small-field expansion of the effective potential, which is related
to the behavior of the equation of state for small magnetization in
the symmetric phase. This can be achieved by using parametric
representations implementing in a rather simple way the known
analytic properties of the equation of state.  Effective parametric
representations can be obtained by parametrizing the magnetization $M$
and the reduced temperature $t$ in terms of two variables $R$ and
$\theta$, setting $M\propto R^\beta\theta$, $t=R(1-\theta^2)$, and
$H\propto R^{\beta\delta}h(\theta)$.  In this framework, following
Guida and Zinn-Justin \cite{G-Z-97}, one may
develop an approximation scheme based on truncations of the Taylor
expansion of the function $h(\theta)$ around $\theta=0$.
Knowing a given number of terms in the small-field expansion of
the effective potential, one can derive the same number of terms in
the small-$\theta$ expansion of $h(\theta)$, with a dependence on an
arbitrary normalization parameter $\rho$. One can try to fix
$\rho$ so that this small-$\theta$ expansion has the fastest
possible convergence.  We propose a prescription based
on the global stationarity of the truncated equation of state with
respect to the arbitrary parameter $\rho$.  This extends the
stationarity condition of the linear model (i.e.\ the lowest-order
non-trivial approximation) discussed in Refs.\  
\cite{Schofield-69,S-L-H-69,Josephson-69,W-Z-74}.  Using the IHT
results for $\gamma$, $\nu$ and the first few coefficients of the
small-field expansion of the effective potential, we constructed
approximate representations of the full critical equation of state.
From them we obtained accurate estimates of many ratios of universal
amplitudes.  Varying the truncation order of $h(\theta)$, we observed
a fast convergence, supporting our arguments.

For our readers' convenience, we collected in Table \ref{eqst} a
summary of all the results obtained in this paper.  There they can
find new estimates of most of the universal quantities (exponents and
ratios of amplitudes) introduced in the literature to describe
critical phenomena in $3d$ Ising-like systems.  The only important
quantity for which we have not been able to give a good estimate is
the exponent $\omega$, which is related to the leading scaling
corrections.  We mention that a precise estimate of $\omega$ has been
reported recently in Ref.\ \cite{Hasenbusch-99}: $\omega=0.845(10)$.
It has been obtained by a Monte Carlo study using a finite-size
scaling method.

The paper is organized as follows.  

In Sec.\ \ref{sec2} we discuss the main features of improved
Hamiltonians from the point of view of the renormalization group.  

In Sec.\ \ref{lambdadet} we describe our Monte Carlo simulations and
present the estimates of $\lambda_4^*$ for the potential
(\ref{potential}) with $\lambda_6=1$.

Sec.\ \ref{ce} is dedicated to the determination of the critical
exponents from the IHT expansion.  We present estimates of all
relevant critical exponents (except for $\omega$), and compare our
results with other theoretical approaches and experiments.

In Sec.\ \ref{effpot} we study the small-field expansion of the
effective potential.  We present estimates of the
first few coefficients of the expansion.  We discuss the relevance of
the determination of the zero-momentum four-point renormalized
coupling for field-theoretical approaches (Sec.\ \ref{subef2}).

Sec.\ \ref{twopointf} presents a study of the low-momentum behavior of
the two-point function in the critical region.  Estimates of the first
few coefficients of its universal low-momentum expansion are given.

In Sec.\ \ref{eqofstate} we study the critical equation of state,
which gives a description of the whole critical region, including the
low-temperature phase.  Using the estimates of the critical exponents
and of the first few coefficients of the small-field expansion of the
effective potential, the critical equation of state is reconstructed
employing approximate parametric representations (Sec.\ 
\ref{parrepth}).  In Sec.\ \ref{stationarity} we present our method,
based on the global stationarity of the approximate equation of state.
Relevance to the $\epsilon$-expansion is discussed in Sec.\ 
\ref{epsilon}.  In Sec.\ \ref{eqstres} we apply the results of Sec.\ 
\ref{stationarity} to the computation of universal ratios of
amplitudes, using as inputs the results of the IHT expansion.  The
results are then compared with other theoretical estimates and with
experimental determinations. For sake of comparison, we also present
results for the two-dimensional Ising model.

Many details of our calculations are reported in the Appendices.
App.\ \ref{seriesanalysis} contains a detailed description of our HT
calculations, i.e.\ the list of the quantities we have considered and
the description of the method we used to generate and analyze the HT
series.  We report many details and intermediate results so that the
reader can judge the quality of the results we will present.  In App.\ 
\ref{univra} we present the notations for the critical amplitudes, and
report the expressions of the universal ratios of amplitudes in terms
of the parametric representation of the critical equation of state.
In App.\ \ref{rhotheta} we discuss in more detail the approximation
scheme for the parametric representation of the equation of state
based on stationarity.

\section{Improved Hamiltonians}
\label{sec2}

As discussed in the introduction, we will work with ``improved''
Hamiltonians, i.e.\ with models in which the leading correction to
scaling has a vanishing (in practice very small) amplitude.

To clarify the basic idea, let us consider a model with two relevant
operators (the thermal and the magnetic ones) and one irrelevant
operator.  If $\tau$, $\kappa$, and $\mu$ are the associated
non-linear scaling fields, the singular part of the free energy
$F_{\rm sing}$ has the scaling form \cite{Wegner-76}
\begin{equation}
F_{\rm sing}(\tau,\kappa,\mu) = 
  |\tau|^{d\nu} f_\pm\!\left(\kappa |\tau|^{-(d+2-\eta)\nu/2},
                           \mu |\tau|^{\Delta}\right).
\end{equation}
where the function $f_\pm$ depends on the phase of the model.  Since
the operator associated with $\mu$ is irrelevant, $\Delta$ is positive
and $\mu |\tau|^{\Delta}\to 0$ at the critical point. Therefore we can
expand the free energy obtaining
\begin{equation}
F_{\rm sing}(\tau,\kappa,\mu) =
  |\tau|^{d\nu} \sum_{n=0}^\infty 
   f_{n,\pm}\!\left(\kappa |\tau|^{-(d+2-\eta)\nu/2}\right) 
        \mu^n |\tau|^{n \Delta}.
\label{F-expansion}
\end{equation}
The presence of the irrelevant operator induces non-analytic
corrections proportional to $|\tau|^{n \Delta}$. Now, let us suppose
that the Hamiltonian of our model depends on three parameters $r$,
$h$, and $\lambda$, where $r$ is associated to the temperature, $h$ is
the magnetic field and $\lambda$ is an irrelevant parameter.  For each
value of $\lambda$ and for $h=0$, the theory has a critical point for
$r=r_c(\lambda)$.  The non-linear scaling fields $\tau$, $\kappa$, and
$\mu$ are analytic functions of the parameters appearing in the
Hamiltonian, and therefore we can write
\begin{eqnarray}
\tau &=& t + t^2 g_{1\tau}(\lambda) + h^2 g_{2\tau}(\lambda) + 
         O(t^3,t h^2,h^4), 
\\
\kappa &=& h\left[1 + t g_{1\kappa}(\lambda) + h^2 g_{2\kappa}(\lambda) + 
          O(t^2, t h^2, h^4)\right],
\\
\mu  &=& g_{1\mu} (\lambda) + t g_{2\mu} (\lambda) + 
         h^2 g_{3\mu} (\lambda) + O(t^2, t h^2, h^4),
\label{mu-expansion}
\end{eqnarray}
where $t\equiv r - r_c(\lambda)$.  Substituting these expressions into
Eq.\ (\ref{F-expansion}), we see that, if 
$g_{1\mu} (\lambda)\not=0$, the free energy has corrections of order
$t^{n\Delta}$. For the susceptibility in zero magnetic field we obtain
the explicit formula \cite{A-F-83}
\begin{equation}
\chi =\, t^{-\gamma} \sum_{m,n=0}^\infty 
    \chi_{1,mn}(\lambda) t^{m\Delta + n} \,
     + t^{1 -\alpha} \sum_{m,n=0}^\infty
    \chi_{2,mn}(\lambda) t^{m\Delta + n} + 
     \sum_{n=0}^\infty \chi_{3,n}(\lambda) t^{n},
\end{equation}
where the contribution proportional to $t^{1 -\alpha}$ stems from the
terms of order $h^2$ appearing in the expansion of $\tau$ and $\mu$,
and the last term is the contribution of the regular part of the free
energy.  Notice that
it is often assumed that the regular part of the free energy does not
depend on $h$. If this were the case, we would have
$\chi_{3,n}(\lambda) = 0$. However, for the two-dimensional Ising
model, one can prove rigorously that $\chi_{3,0}\not=0$
\cite{Kong-etal_86,Gartenhaus-McCullough_88}, showing the
incorrectness of this conjecture. For a discussion, see Ref.\ 
\cite{Salas-Sokal_99}.

In many interesting instances, it is possible to cancel the leading
correction due to the irrelevant operator by choosing $\lambda =
\lambda^*$ such that $g_{1\mu}(\lambda^*)=0$.  In this case $\mu
\tau^\Delta \sim t^{1 + \Delta}$, so that no term of the form
$t^{m\Delta + n}$, with $n < m$, will be present.  In particular the
leading term proportional to $t^\Delta$ will not appear in the
expansion.

In general other irrelevant operators will be present in the theory,
and therefore we expect corrections proportional to $t^\rho$ with
$\rho = n_1 + n_2 \Delta + \sum_i m_i \Delta_i$, where $\Delta_i$ are
the exponents associated to the additional irrelevant operators.  For
$\lambda=\lambda^*$ the expansion will contain only terms with 
$n_1 \ge n_2$.

It is important to notice that, by working with $\lambda = \lambda^*$,
we use a Hamiltonian such that the non-linear scaling field $\mu$
vanishes at the critical point. This property is independent of the
observable we are considering. Therefore all quantities will be
improved, in the sense that the leading correction to scaling,
proportional to $t^\Delta$, will vanish.  We will call the
Hamiltonians with $\lambda = \lambda^*$ ``improved Hamiltonians''.

\section{Determination of the improved parameters}
\label{lambdadet}

The Hamiltonian defined by Eqs.\ (\ref{hamiltonian}) and
(\ref{potential}) with $\lambda_6=0$ was considered in Ref.\ 
\cite{Hasenbusch-99}, where it was shown that the leading correction
to scaling cancels for $\lambda_4^* = 1.10(2)$. Here we will also
consider the case $\lambda_6=1$, and determine the corresponding
$\lambda^*_4$ using a method similar to the one discussed in Ref.\ 
\cite{H-P-V-98}.

The idea is the following: consider a renormalization-group invariant
observable $\cal O$ on a finite lattice $L$ and let ${\cal O}^*$ be
its value at the critical point, i.e.\ 
\begin{equation}
{\cal O}^*\, =\, 
  \lim_{L\to\infty} \lim_{\beta \to\beta_c(\lambda_4)} 
  {\cal O} (\beta,\lambda_4,L).
\end{equation}
The quantity ${\cal O}^*$ is a universal number and therefore it will
be independent of $\lambda_4$.  The standard scaling arguments predict
\begin{equation}
{\cal O} (\beta_c(\lambda_4),\lambda_4,L) \approx {\cal O}^* + 
   a_1 (\lambda_4) L^{-\omega} + 
   a_2 (\lambda_4) L^{-2 \omega} + \ldots +
   b_1 (\lambda_4) L^{-\omega_2} \ldots
\end{equation}
where $\omega = \Delta/\nu$, $\omega_2 = \Delta_2/\nu$, $\Delta_2$
being the next-to-leading correction-to-scaling exponent. 
Since for $\lambda_4 = \lambda_4^*$,
$a_1 (\lambda_4^*) = a_2 (\lambda_4^*)= \ldots =0$, for
$\lambda_4 \approx \lambda_4^*$ we can rewrite the previous equation
as
\begin{equation}
{\cal O} (\beta_c(\lambda_4),\lambda_4,L) \approx {\cal O}^* +
   (\lambda_4 - \lambda_4^*) 
   \left( a_{11} L^{-\omega} + a_{21} L^{-2 \omega} + \ldots \right) + 
    b_1(\lambda_4^*) L^{-\omega_2} \ldots
\label{calO-expansion}
\end{equation}
Now, suppose we know the exact value ${\cal O}^*$, and let us define 
$\lambda_4^{\rm eff}(L)$ as the solution of the equation
\begin{equation}
{\cal O} (\beta_c(\lambda_4^{\rm eff} (L)),
          \lambda_4^{\rm eff} (L),L) = {\cal O}^*.
\label{equation-lambdastar}
\end{equation}
From Eq.\ (\ref{calO-expansion}) we obtain immediately 
\begin{equation}
\lambda_4^{\rm eff} (L) = \lambda_4^* - 
   {b_1(\lambda_4^*)\over a_{11}} L^{\omega - \omega_2} + \ldots
\end{equation}
Since $\omega_2 > \omega$, $\lambda_4^{\rm eff} (L)$ converges to
$\lambda_4^*$ as $L\to\infty$.  For the $3d$ Ising universality class,
$\omega_2 \simeq 2 \omega$ \cite{G-R-76,N-R-84} and 
$\omega \simeq 0.85$ \cite{Hasenbusch-99}.

In order to apply this method in practice we need two ingredients: 
a precise determination of $\beta_c(\lambda_4)$ and an estimate 
of ${\cal O}^*$.

Very precise estimates of $\beta_c(\lambda_4)$ can be obtained from
the analysis of the HT series of the susceptibility
$\chi$, that we have calculated to $O(\beta^{20})$.  For $\lambda_6=0$
and $1.0\le \lambda_4\le 1.2$ the values of $\beta_c(\lambda_4)$ can
be interpolated by the polynomial
\begin{equation}
\beta_c(\lambda_4) = 
0.40562043 + 
0.00819000\, \lambda_4 -
0.04626355\, \lambda_4^2 +
0.01235674\, \lambda_4^3 \pm 0.0000014.
\label{betacr-HT-lambda6=0}
\end{equation}
In particular, for $\lambda_4 = 1.10$, we have 
$\beta_c(1.10) = 0.3750973(14)$, to be compared with
$\beta_c(1.10) = 0.3750966(4)$ of Ref.\ 
\cite{Hasenbusch-99}.  For $\lambda_6 = 1$ and 
$1.8\le \lambda_4\le 2.0$ --- as we shall see this is the relevant
interval --- we have
\begin{equation}
\beta_c(\lambda_4) =
0.68612192 - 
0.18274273\, \lambda_4 +
0.02634688\, \lambda_4^2 -
0.00102710\, \lambda_4^3 \pm 0.0000018.
\end{equation}

The second quantity we need is an observable ${\cal O}$ such 
that ${\cal O}^*$ can be computed with high precision.
We have chosen the Binder parameter
\begin{equation}
Q = \, {\langle m^4 \rangle\over \langle m^2\rangle^2},
\end{equation}
where $m$ is the magnetization. A precise estimate of $Q$ was obtained
in Ref.\ \cite{H-P-V-98} by means of a large-scale simulation of the
spin-1 model. They report
\begin{equation}
Q^* = 0.62393 (13{+}35{+}5),
\label{Qstar-HPV98}
\end{equation}
where the error is given as the sum of three contributions: the first
is the statistical error; the second and the third account for
corrections to scaling.  We have tried to improve this estimate by
performing a high-precision Monte Carlo simulation\footnote{We
used the Brower-Tamayo algorithm \cite{Brower-Tamayo_89}, 
each iteration consisting of a Swendsen-Wang update of the sign
of $\phi$ and of a Metropolis sweep.} of the Hamiltonian
(\ref{hamiltonian}) for $\lambda_6 = 0$ and by computing $Q$ for
$\lambda_4=1.10$, which is the best estimate of $\lambda_4^*$. Since
this Hamiltonian is improved (i.e.\ the leading correction to scaling
vanishes), we expect to be able to obtain a reliable estimate from
simulations on small lattices for which it is possible to accumulate a
large statistics.  The results\footnote{The simulations were performed
before the appearance of Ref.\ \cite{Hasenbusch-99} and the generation
of the HT series, when only a very approximate expression for
$\lambda_4^*$ existed. Therefore the runs were not made at the correct
values of $\lambda_4$ and $\beta_c$. The values reported in Table
\ref{table_QBinder-MC} have been obtained from the Monte Carlo data by
means of a standard reweighting technique.}  are reported in Table
\ref{table_QBinder-MC}.
\begin{table}[tbp]
\caption{
For several values of lattice size $L$ and for $\lambda_6 = 0$, we
report: the values of the parameters used in the Monte Carlo
simulation $\lambda_{4,\rm run},\beta_{\rm run}$, the number of Monte
Carlo iterations $N_{\rm iter}$, each iteration consisting of a
standard Swendsen-Wang update and of a Metropolis sweep, and the the
estimate of the Binder parameter $Q$ at $\lambda=1.10$, 
$\beta = 0.3750973$. The reported error on $Q$ is the sum of three
terms: the statistical error, and the errors due to the uncertainty
of $\lambda^*$ and $\beta_c(\lambda)$. 
}
\label{table_QBinder-MC}
\begin{tabular}{cccrc}
\multicolumn{1}{c}{$L$}&
\multicolumn{1}{c}{$\lambda_{4,{\rm run}}$}&
\multicolumn{1}{c}{$\beta_{\rm run}$}&
\multicolumn{1}{c}{$N_{\rm iter}$}&
\multicolumn{1}{c}{$Q$}\\
\tableline \hline
6  &  1.100  & 0.375 & $  91 \times 10^6$ & 0.62370(15+37+2) \\
7  &  1.100  & 0.375 & $  78 \times 10^6$ & 0.62386(17+33+2) \\
9  &  1.080  & 0.376 & $ 178 \times 10^6$ & 0.62389(12+26+3) \\
12 &  1.105  & 0.375 & $ 305 \times 10^6$ & 0.62387(10+25+5) \\
\end{tabular}
\end{table}
There are three different sources of error: we
report the statistical error, the variation of the estimate of the
Binder parameter when $\lambda_4$ varies within the interval 
$1.08 \hbox{--} 1.12$ (due to corrections to scaling of order
$L^{-\omega}$, which are not completely suppressed, since the value
used for $\lambda_4$ is not exactly equal to $\lambda_4^*$), and the
variation of $Q$ when $\beta_c$ varies within one error bar.  The
values of $L$ we use are relatively small ($L\le 12$) and one could be
afraid that next-to-leading corrections still give a non-negligible
systematic deviation. Our data do not show any evidence of such an
effect, and the estimates for different values of $L$ are consistent.
Using the estimate obtained for $L=12$, we get the final result
\begin{equation}
Q^* = 0.62388(32)
\end{equation}
(the uncertainty is obtained assuming independence of systematic and
statistical errors), which is in agreement with the estimate
(\ref{Qstar-HPV98}) with a slightly smaller error bar.

We have next determined $\lambda_4^*$ for the model with Hamiltonian
(\ref{hamiltonian}) and $\lambda_6 = 1$ using the method presented
above.  Estimates of $\lambda^{\rm eff}_4(L)$ are reported in Table
\ref{table_lambdastar-lambda6=1}, from which we conclude
\begin{equation}
\lambda_4^* = 1.90(4).
\end{equation}
\begin{table}[tbp]
\caption{
For several values of lattice size $L$ and for $\lambda_6 = 1$, we
report: the values of the parameters used in the Monte Carlo
simulation $\lambda_{4,\rm run}, \beta_{\rm run}$, the number of Monte
Carlo iterations $N_{\rm iter}$,  each iteration consisting of a
standard Swendsen-Wang update and of a Metropolis sweep, and the
estimate of $\lambda^{\rm eff}(L)$, the solution of Eq.\
(\ref{equation-lambdastar}). The error is reported as the sum of
three terms: the statistical error on $Q$, the error on $Q^*$, and
the uncertainty of the critical value $\beta_c(\lambda)$.
 }
\label{table_lambdastar-lambda6=1}
\begin{tabular}{cccrc}
\multicolumn{1}{c}{$L$}&
\multicolumn{1}{c}{$\lambda_{4,{\rm run}}$}&
\multicolumn{1}{c}{$\beta_{\rm run}$}&
\multicolumn{1}{c}{$N_{\rm iter}$}&
\multicolumn{1}{c}{$\lambda_4^{\rm eff}(L)$}\\
\tableline \hline
6  &  1.900  & 0.427 & $ 100 \times 10^6$ & 1.915(9+19+1) \\  
7  &  1.900  & 0.427 & $ 252 \times 10^6$ & 1.894(6+21+2) \\
9  &  1.920  & 0.425 & $ 214 \times 10^6$ & 1.897(9+25+3) \\
12 &  1.920  & 0.425 & $ 294 \times 10^6$ & 1.904(9+32+6) \\
\end{tabular}
\end{table}
Notice that the last three points show a small upward trend which, although
consistent with a statistical effect, could be a systematic
increase\footnote{This effect could also be due to the fact that $Q^*$
is only approximately known.  To check if this is the case, we have
computed $\lambda_{4,\pm}^{\rm eff}(L)$, the solutions of Eq.\ 
(\ref{equation-lambdastar}) with the r.h.s.\ replaced by 
$Q^*\pm \sigma_Q$, $\sigma_Q$ being the error on $Q^*$.  If the
increase is associated with the uncertainty of $Q^*$, we should
observe that $\lambda_{4,\pm}^{\rm eff}(L)$ have opposite trends, one
increasing, the other decreasing. In the present case
$\lambda_{4,\pm}^{\rm eff}(L)$ are both increasing, thereby excluding
that this effect is due to the uncertainty of $Q^*$.}  due to the
corrections of order $L^{-\omega_2 + \omega}$.  With the present
statistical errors we cannot distinguish between these two
possibilities: we have considered as our final estimate the average of
the results obtained for $L=7,9,12$, but we cannot exclude that the
correct $\lambda_4^*$ is slightly larger than our estimate. However
the quoted error should be large enough to include this systematic
increase.

As explained in the introduction, the analysis of HT series for the
determination of universal quantities is sensitive to non-analytic
scaling corrections.  As we will discuss below, one can use this fact
to obtain a rough estimate of the optimal value of $\lambda$.

Consider for example the 
zero-momentum four-point coupling constant $g_4$ defined by
\begin{equation}
g_4 = - {\chi_4\over \chi^2 \xi^3},
\label{grdef}
\end{equation}
where $\chi$, $\xi^2$ and $\chi_4$ are, respectively, the magnetic
susceptibility, the second-moment correlation length and the
zero-momentum four-point connected correlation function (definitions
can be found in App.\ \ref{series}).  We have chosen this observable
because it appears to be affected by large corrections to scaling, but
the method can be applied to any universal quantity.  From the
discussion of Sec.\ \ref{sec2}, we have for $\beta\to\beta_c$
\begin{equation}
g_4(\beta) = g_4^* + c_\Delta (\beta_c - \beta)^\Delta + \ldots 
\label{conf}
\end{equation}
where $g_4^*$ is a universal constant, and $c_\Delta$ is a
non-universal amplitude depending on the Hamiltonian. For improved
models, as discussed before, $c_\Delta = 0$.  The traditional methods
of analysis, e.g.\ those based on Pad\`e (PA) and Dlog-Pad\`e (DPA)
approximants, are unable to handle an asymptotic behavior like
(\ref{conf}) unless $\Delta$ is an integer number, thus leading to a
systematic error. Integral approximants allow for non-analytic scaling
corrections, but, as already said, with the series of moderate length
available today, they need to be biased to give correct results:
without any bias they give estimates that are similar
to those of PA's and DPA's \cite{P-V-gr-98}.  At present the only
analyses that are able to effectively take into account the confluent
corrections use biased approximants, fixing the value of $\beta_c$ and
of the first non-analytic exponent $\Delta$ (see e.g.\ Refs.\ 
\cite{Guttrev,Roskies-81,A-M-P-82,P-V-gr-98,B-C-g-98,B-C-g-97} for a
discussion of this issue and for a presentation of the different
methods used in the literature).  The method we use has been proposed
in Ref.\ \cite{Roskies-81} and generalized in Ref.\ \cite{A-M-P-82}.
The idea is to perform a Roskies transform (RT), i.e.\ the change of
variables
\begin{equation}
z= 1 - (1 - \beta/\beta_c)^\Delta,
\label{RTr}
\end{equation}
so that the non-analytic terms in $\beta_c - \beta$ become analytic in
$1-z$.  Therefore the analysis of the resulting series by means of
standard approximants should give correct results.  For the models we
are considering the exponent $\Delta$ is approximately $1/2$ 
(e.g., Ref.\ \cite{Hasenbusch-99} reports $\Delta=0.532(6)$);
for simplicity we have used the transformation (\ref{RTr}) with
$\Delta = 1/2$.

\begin{figure}[tb]
\centerline{\psfig{width=15truecm,angle=-90,file=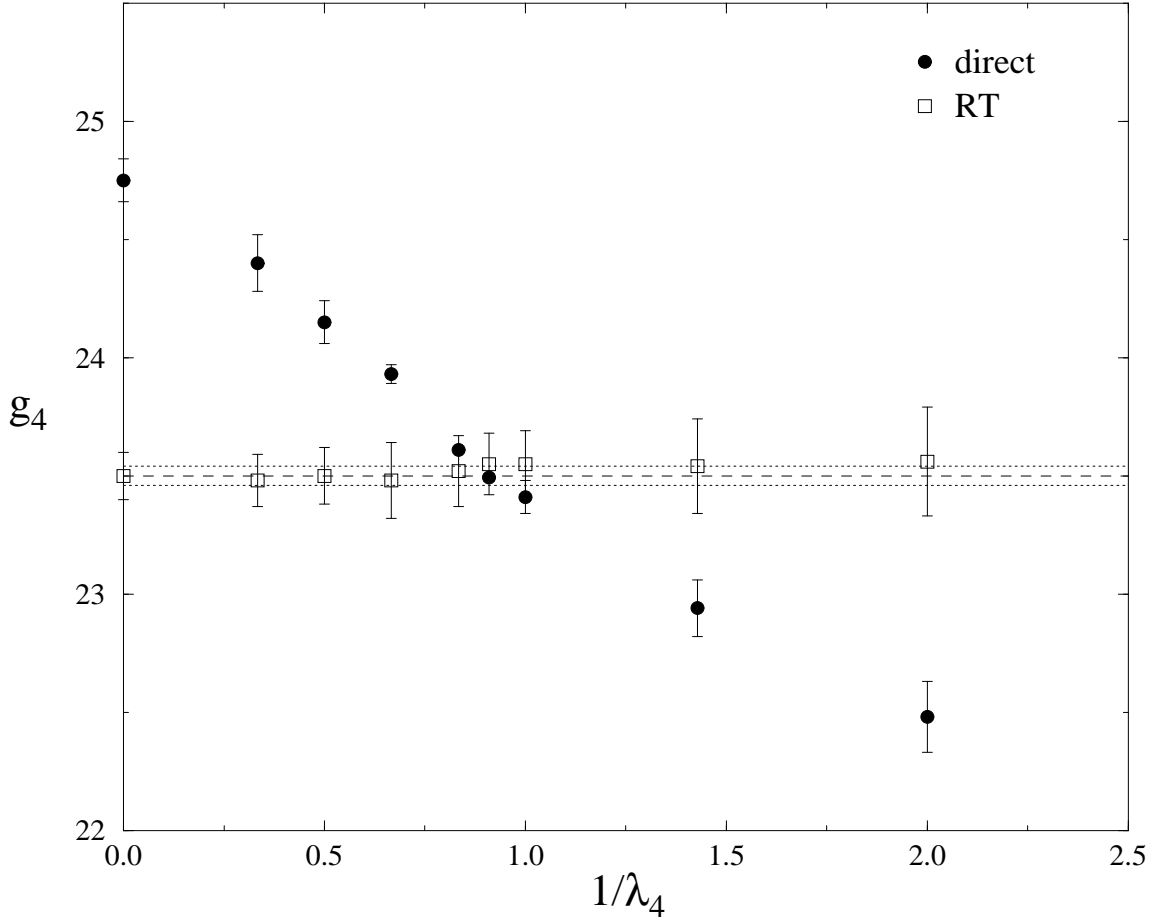}}
\caption{
Comparison of the determination of $g_4$ (plotted vs.\ $1/\lambda_4$)
from HT series without (direct) and with the Roskies transform (RT)
for the pure $\phi^4$ lattice  model. The dashed line marks the 
more precise estimate (with its error) we derived from the analysis of
the IHT expansion.
}
\label{figg}
\end{figure}

We have analyzed the HT expansion of $g_4$ for the model with
Hamiltonian (\ref{hamiltonian}) and $\lambda_6=0$ for several values
of $\lambda_4$.  We computed PA's, DPA's, and first-order integral
(IA1) approximants of the series in $\beta$ and of its Roskies
transform (RT) in $z$.  In Fig.\ \ref{figg} we plot the results as a
function of $\lambda_4^{-1}$.  The reported errors are related to the
spread of the results obtained from the different approximants, see
App.\ \ref{ratioofamp} for details.  The estimates obtained from the
RT'ed series are independent of $\lambda_4$ within error bars, giving
the estimate $g_4^*\simeq 23.5$, in agreement with previous analyses
of the HT expansion of the Ising model on various lattices using the
RT or other types of biased approximants
\cite{P-V-gr-98,B-C-g-98,B-C-g-97} (in Sec.\ \ref{subef2} we will
improve this estimate by analyzing the IHT expansion).  The
independence of the result from the value of $\lambda_4$ clearly
indicates that the RT is effectively able to take into account the
non-analytic behavior (\ref{conf}). On the other hand the analysis of
the series in $\beta$ gives results which vary with $\lambda_4$ more
than the spread of the approximants: for instance, the analysis of the
series of the standard Ising model, corresponding to
$\lambda_4=\infty$, gives results that differ by more than 5\% from
the estimate quoted above, while the spread of the approximants is
much smaller.  Clearly there is a large systematic error. It is
important to notice that the direct analysis and the RT one coincide
when $1.0\lesssim \lambda_4\lesssim 1.2$, i.e.\ in the region in which
the leading non-analytic corrections are small. This fact confirms our
claim that the observed discrepancies are an effect of the confluent
corrections.

The results presented above can be used to obtain an estimate of
$\lambda_4^*$ from the HT series alone: $\lambda_4^*$ should fall in
the interval in which the direct analysis gives results compatible
with those obtained from the RT'ed series.  As we already mentioned,
for $\lambda_6 = 0$ we obtain $\lambda_4^* = 1.1(1)$, while for
$\lambda_6 = 1$ we get $\lambda_4^* = 1.9(1)$. The latter
estimate was indeed the starting point of our Monte Carlo simulation.
We have also tried to estimate $\lambda_4^*$ in more direct ways, but
all methods we tried were even less precise.

A similar method for the determination of the improved Hamiltonian
from the HT series was presented in Ref.\ \cite{F-C-85}. The optimal
value of the parameter (called $y$ in Ref.\ \cite{F-C-85}) was
determined comparing the results for the critical point $\beta_c(y)$
obtained using IA1's and DPA's: $y^*$ is estimated from the value at
which DPA and IA1 estimates of $\beta_c(y)$ agree between each other.
It should be noticed that, for the double-Gaussian model, partial
differential approximants and a later analysis of Nickel and
Rehr~\cite{N-R-90} using a different method gave significantly
different estimates of $y^*$.

\section{Critical exponents}
\label{ce}

The analysis of Sec.\ \ref{lambdadet} is encouraging and supports our
basic assumption that the systematic error due to confluent
singularities is largely reduced when analyzing IHT expansions.  To
further check this hypothesis we will compare results obtained from
different improved Hamiltonians. This will provide an estimate of the
remaining systematic error which is not covered by the spread of the
results from different approximants.

The definition of the quantities we have considered and a detailed
description of the method we used to generate and analyze the HT
series is presented in App.\ \ref{seriesanalysis}.

We computed $\beta_c$ and $\gamma$ from the analysis of the HT
expansion of the magnetic susceptibility. We cosidered integral
approximants of first, second and third order. After a careful
analysis we preferred the second-order integral approximants (IA2's),
which turned out to be the most stable: most of the results we present
in this section and in the related App.\ \ref{exponents} have been
obtained by using IA2's.  As a further check of the effectiveness of the
approximants employed, we made use of the fact that $\chi$ (and
$\xi^2$) must present an antiferromagnetic singularity at
$\beta_c^{\rm af}=-\beta_c$ of the form \cite{Fisher-62}
\begin{equation}
\chi = c_0 + c_1 \left( \beta - \beta_c^{\rm af}\right)^{1-\alpha} + ... \,,
\label{fisheraf}
\end{equation}
where $\alpha$ is the specific heat exponent, $c_i$ are constants and
the ellipses represent higher-order singular or analytic
corrections.  We verified the existence of a singularity at
$\beta\simeq -\beta_c$ in the approximants, and calculated the
associated exponent.  
We also considered approximants that were biased requiring the presence of two
symmetric singularities at $\beta=\pm\beta_c$ \cite{F-C-85}; the
results obtained are consistent with the predicted behavior
(\ref{fisheraf}) (see App.\ \ref{exponents} and related Tables).

The exponent $\nu$ was obtained from the series of the second-moment
correlation length
\begin{equation}
\xi^2 = {m_2\over 6\chi}\sim (\beta_c-\beta)^{2\nu},
\label{xidef}
\end{equation}
where $m_i$ are the moments of the two-point function.  We followed
the procedure suggested in Ref.\ \cite{Guttmann-87}, i.e.\ we used the
estimate of $\beta_c$ obtained from $\chi$ to bias the analysis of
$\xi^2$.  For this purpose we used IA's biased to $\beta_c$.  We also
considered approximants biased to have a pair of singularities at
$\pm\beta_c$.

In Table \ref{summaryexp} we report the results obtained for the
Hamiltonians (\ref{hamiltonian}) with $\lambda_6=0$, $\lambda_4=1.10$
and with $\lambda_6=1$, $\lambda_4=1.90$, and for the Hamitonian
(\ref{spin1action}) with $D=0.641$.
\begin{table}[tbp]
\caption{
Our final estimates of $\gamma$, $\nu$, $\eta$ and $\sigma$.
The error is reported as a sum of two terms: the first one is related
to the spread of the approximants; the second one is
related to the uncertainty of the value of $\lambda_4^*$.
}
\label{summaryexp}
\begin{tabular}{cr@{}lr@{}lr@{}lr@{}l}
\multicolumn{1}{c}{}&
\multicolumn{2}{c}{$\gamma$}&
\multicolumn{2}{c}{$\nu$}&
\multicolumn{2}{c}{$\eta$}&
\multicolumn{2}{c}{$\sigma$}\\
\tableline \hline
$\lambda_6=0$ &   1&.23732(24+16) &  0&.63015(13+12) & 0&.0364(3+1) &  0&.0213(13+1)\\
$\lambda_6=1$ &   1&.23712(26+31) &  0&.63003(13+23) & 0&.0363(3+2) &  0&.0213(14+2)\\
spin-1       &    1&.23680(30+12) &  0&.62990(15+8)  & 0&.0366(3+1) &  0&.0202(10+1)\\
\end{tabular}
\end{table}
The errors are given as a sum of two terms: the first one is computed
from the spread of the approximants; the
second one is related to the uncertainty of the value of
$\lambda_4^*$ and $D^*$, and it is evaluated by changing $\lambda_4$ in the
range $1.08 \hbox{--} 1.12$ for $\lambda_6=0$ and $1.86 \hbox{--}
1.94$ for $\lambda_6=1$, and $D$ in the range $0.633\hbox{--}0.649$
for the spin-1 model.  There is a good agreement among the
estimates of $\gamma$ and $\nu$ obtained from the three improved
Hamiltonians considered. This is
an important check of our working hypothesis, i.e.\ that systematic
errors due to confluent corrections are largely reduced. This will be
also confirmed by the results for the universal ratios of amplitudes.
We determine our final estimates by 
combining the results of the three improved Hamiltonians: 
as estimate we take the weighted average of the three results, 
and as estimate of the uncertainty the smallest of
the three errors.
We obtain for $\gamma$ and $\nu$
\begin{eqnarray}
\gamma &=& 1.2371(4),\label{resga}\\
\nu &=& 0.63002(23), \label{resnu}
\end{eqnarray}
and by the hyperscaling relation $\alpha=2-3\nu$
\begin{equation}
\alpha=0.1099(7).
\label{resalpha}
\end{equation}
In App.\ \ref{exponents} we also report some further checks using the
Monte Carlo estimate of $\beta_c$ reported in Ref.\ 
\cite{Hasenbusch-99} to bias the analysis of the series. The results
are perfectly consistent.
We mention that from the analysis of the antiferromagnetic singularity,
cf.\ Eq.\ (\ref{fisheraf}), we obtain the estimate $\alpha=0.105(10)$,
which is consistent with result (\ref{resalpha}) obtained
assuming  hyperscaling.

From the results for $\gamma$ and $\nu$, we can obtain $\eta$ by the
scaling relation $\gamma=(2-\eta)\nu$. This gives
$\eta=0.0364(10)$, where the error is estimated by considering the
errors on $\gamma$ and $\nu$ as independent, which is of course not
true.  We can obtain an estimate of $\eta$ with a smaller, yet reliable, 
error using the so-called critical point renormalization method (CPRM)
(see Ref.\ \cite{int-appr-ref} and references therein).  We obtain
the results reported in Table \ref{summaryexp}, with considerably
smaller errors.  Our final estimate is
\begin{equation}
\eta=0.0364(4).
\label{reseta}
\end{equation}
Moreover using the scaling relations we obtain
\begin{eqnarray}
\delta &=& {5- \eta\over 1 +\eta}= 4.7893(22), \label{deltaex}\\
\beta &=& {\nu\over 2} \left( 1 + \eta\right) = 0.32648(18) \label{betaex}
\end{eqnarray}
(the error on $\beta$ has been estimated by considering the errors of
$\nu$ and $\eta$ as independent).

Finally we consider the universal critical exponent describing how the
spatial anisotropy, which is present in physical systems with cubic
symmetry (e.g.\ uniaxial magnets), vanishes when approaching the
rotationally-invariant fixed point \cite{C-P-R-V-98}.  For this class
of systems the two-point function $G(x)$ is not rotationally
invariant.  Therefore non-spherical moments are in general
non-vanishing, but near the critical point they are depressed with
respect to spherical moments carrying the same naive physical
dimensions by a factor $\xi^{-\rho}$, where $\rho$ is a universal
critical exponent.  From a field-theoretical point of view, space
anisotropy is due to non rotationally invariant irrelevant operators
in the effective Hamiltonian, whose presence depends essentially on
the symmetries of the physical system, or of the lattice formulation.
In Table \ref{summaryexp} we report the results for 
$\sigma\equiv 2-\rho$ as obtained by analyses of the first
non-spherical moments (cf.\ Eq.\ (\ref{q4})) using the CPRM.  The
exponent $\sigma$ turns out to be very small:
\begin{equation}
\sigma= 0.0208(12),\label{ressigma}
\end{equation}
and $\rho=1.9792(12)$.

In Table \ref{avaexp} we compare our results  with some of the most recent
estimates of the critical exponents $\gamma$, $\nu$, $\eta$, $\alpha$,
and $\beta$.  The table should give an overview of the state of the
art for the various approaches.  
\begin{table}[tbp]
\caption{
Theoretical estimates of critical exponents.  See text for explanation 
of symbols in the first column.  For values marked with an asterisk,
the error is not quoted explicitly in the reference.
}
\label{avaexp}
\begin{tabular}{lr@{}lr@{}lr@{}lr@{}lr@{}l}
\multicolumn{1}{c}{}&
\multicolumn{2}{c}{$\gamma$}&
\multicolumn{2}{c}{$\nu$}&
\multicolumn{2}{c}{$\eta$}&
\multicolumn{2}{c}{$\alpha$}&
\multicolumn{2}{c}{$\beta$}\\
\tableline \hline
IHT                    &    1&.2371(4)  &  0&.63002(23) & 0&.0364(4) & 0&.1099(7) & 0&.32648(18)\\\hline
HT (sc)\cite{B-C-97}   &    1&.2388(10)   &  0&.6315(8)  &&    &  & &&  \\
HT (bcc)\cite{B-C-97}  &    1&.2384(6)    &  0&.6308(5)  &&    &  & &&  \\
HT \cite{N-R-90}  &    1&.237(2)     &  0&.6300(15) & 0&.0359(7) & 0&.11(2)  && \\
HT\cite{Guttmann-87}   &    1&.239(3)    &  0&.632$^{+0.002}_{-0.003}$ &  &  & & &&  \\
HT\cite{F-C-85}     &    1&.2395(4)    &  0&.632(1)   &  &  & 0&.105(7) &&  \\
HT\cite{G-R-84}     &    1&.2378(6)    &  0&.63115(30)   &  &  & & &&  \\
HT\cite{C-F-N-82}     &    1&.2385(15)    &  &   &  &  & & &&  \\
HT\cite{ZJ-79}               &    1&.2385(25)   &  0&.6305(15) &&    &  & && \\\hline
MC \cite{Hasenbusch-99} & 1&.2367(11) &  0&.6296(7) & 0&.0358(9) && && \\
MC \cite{H-P-V-98} & & & 0&.6298(5) & 0&.0366(8) && && \\
MC \cite{B-F-M-M-P-R-99} & & &  0&.6294(10) & 0&.0374(12) && && \\
MC \cite{H-P-97} & & &  0&.6308(10) &&  & & && \\
MC \cite{T-B-96} & & &  & & & & && 0&.3269(6)  \\
MC \cite{G-T-96} & & &  0&.625(1) & 0&.025(6) && && \\
MC \cite{B-L-H-95} & 1&.237(2) &  0&.6301(8) & 0&.037(3) & 0&.110(2) & 0&.3267(10) \\\hline
$\epsilon$-exp.$_{\rm free}$ \cite{G-Z-98}  &    1&.2355(50)  &  0&.6290(25) & 0&.0360(50) && & 0&.3257(25) \\
$\epsilon$-exp.$_{\rm bc}$ \cite{G-Z-98}    &    1&.2380(50)  &  0&.6305(25) & 0&.0365(50) && & 0&.3265(15) \\
$\epsilon$-exp.$_{\rm bc}$ \cite{P-V-gr-98} &    1&.240(5)   & 0&.631(3)    & &  && && \\\hline
$d$=3 $g$-exp.\ \cite{G-Z-98}     &    1&.2396(13)  &  0&.6304(13)  & 0&.0335(25)  & 0&.109(4) & 0&.3258(14) \\
$d$=3 $g$-exp.\ \cite{Kleinert-98}&    1&.241$^*$   &  0&.6305$^*$   & 0&.0347(10)  && && \\
$d$=3 $g$-exp.\ \cite{M-N-91}     &    1&.2378(6+18)&  0&.6301(5+11)& 0&.0355(9+6) && &&  \\
$d$=3 $g$-exp.\ \cite{L-Z-77}     &    1&.2405(15)  &  0&.6300(15)  & 0&.032(3)    && && \\\hline
ERG \cite{Morris-97}              &     &           &  0&.618(14)   & 0&.054$^*$ && && \\
ERG \cite{T-W-94}                 &    1&.247$^*$   &  0&.638$^*$   & 0&.045$^*$ && && \\
\end{tabular}
\end{table}
Let us first note the good agreement of our IHT estimates with the
very precise results of the recent Monte Carlo simulations (MC) of
Refs.\ \cite{H-P-V-98,B-F-M-M-P-R-99,Hasenbusch-99}.  The small
difference with the HT estimates of Ref.\ \cite{B-C-97} (obtained from
the standard Ising model) may be explained by the difficulty of
controlling the effects of the confluent singularities, and by a
systematic error induced by the uncertainty on the external input
parameters ($\beta_c$ and $\Delta$) that are used in their biased
analysis.  The estimates of Refs.\ 
\cite{C-F-N-82,G-R-84,F-C-85,N-R-90} have been obtained from a HT
analysis of two families of models, the Klauder and the
double-Gaussian models on the bcc lattice. The results of these
analyses are in good agreement with our IHT estimates,
especially those by Nickel and Rehr~\cite{N-R-90}.
The HT series for the double-Gaussian model were 
analyzed also in Ref.~\cite{F-C-85} where
a higher estimate of $\gamma$ was obtained.
As pointed out in Ref.~\cite{N-R-90}, the discrepancy is 
essentially due to the use of a higher estimate of the improvement
parameter $y^*$ with respect to that used in Ref.~\cite{N-R-90}
(see the discussion at the end of Sec.
\ref{lambdadet}).
Refs.\ \cite{F-C-85,N-R-90} report also estimates of $\alpha$ obtained
analyzing the singularity of the susceptibility at the
antiferromagnetic critical point. The result agrees with our estimate.

The agreement with the field-theoretical calculations is overall good.
The slightly larger result for $\gamma$ obtained in the analyses of
Refs.\ \cite{G-Z-98,L-Z-77} (using $O(g^7)$ series
\cite{M-N-91,B-N-G-M-77}) may be due to an underestimate of the
systematic error due to the non-analyticity of the Callan-Symanzik
$\beta$-function. Similar results have been obtained by Kleinert, who
resummed the $O(g^7)$ expansion by a variational method
\cite{Kleinert-98}, still neglecting confluent singularities at the
infrared-stable fixed point.  We shall return on this point later.  A
better agreement is found with the analysis of the $d$=3 $g$-expansion
performed by Murray and Nickel, who allow for a more general
non-analytic behavior of the $\beta$-function \cite{M-N-91}. In Table
\ref{avaexp}, we quote two errors on the results of Ref.\ 
\cite{M-N-91}: the first one is the resummation error, and the second
one takes into account the uncertainty of $g^*$, which is estimated to
be $\sim 0.01$.  The results of the $\epsilon$-expansion were obtained
from the $O(\epsilon^5)$ series calculated in Refs.\ 
\cite{C-G-L-T-83,K-N-S-C-L-93}. We report estimates obtained by
performing standard analyses (denoted as ``free'') and constrained
analyses \cite{L-Z-89} (denoted by ``bc'') that incorporate the
knowledge of the exact two-dimensional values.  Both are essentially
consistent with our IHT estimates, 
but present a significantly larger uncertainty.
In Table \ref{avaexp} we also report the results obtained by
approximately solving the exact renormalization-group equation (ERG);
they seem to be much less precise.  A more complete list of references
pertaining to the theoretical determination of the critical exponents
can be found in Ref.\ \cite{G-Z-98}.  Concerning the exponent $\sigma$
related to the rotational symmetry, the IHT results represent a
substantial improvement of the estimates obtained by various
approaches (HT and field theory) presented in Ref.\ \cite{C-P-R-V-98}.

Experimental results have been obtained studying the 
liquid-vapor transition in simple fluids, and the different 
critical transitions in multicomponent fluid mixtures, 
uniaxial antiferromagnetic materials and 
micellar systems.  Many recent estimates can be found in
Refs.\ \cite{B-L-H-95,Anisimov,P-H-A-91}.  In Table \ref{expexp} we
report some experimental results, most of them published after 1990. It
is not a complete list of the published results, but it
may be useful to get an overview of the experimental state of the art.
\begin{table}[tbp]
\caption{
Experimental estimates of critical exponents. lv\ denotes the liquid-vapor 
transition in simple fluids, bm\ refers to a binary fluid mixture, 
ms\ to a uniaxial magnetic system, and  mi\  to a micellar system.
}
\label{expexp}
\begin{tabular}{llr@{}lr@{}lr@{}lr@{}lr@{}l}
\multicolumn{1}{c}{}&
\multicolumn{1}{c}{Ref.}&
\multicolumn{2}{c}{$\gamma$}&
\multicolumn{2}{c}{$\nu$}&
\multicolumn{2}{c}{$\eta$}&
\multicolumn{2}{c}{$\alpha$}&
\multicolumn{2}{c}{$\beta$}\\
\tableline \hline
lv & \cite{H-S-99}   &  && && && 0&.1105$^{+0.0250}_{-0.0270}$ &&  \\
 &\cite{S-N-93}      &  && && && 0&.1075(54)  && \\
 &\cite{Edwards-84}  &  && && && 0&.1084(23) && \\
 &\cite{A-P-B-94}    &  && && && 0&.111(1) & 0&.324(2)  \\
 &\cite{K-A-S-W-95}  &  && && && && 0&.341(2)  \\
 &\cite{Damay-etal_98}& && && 0&.042(6) && && \\ 
\hline
bm& \cite{R-J-98}    &  && && && 0&.104(11) &&  \\
  & \cite{P-C-84}    &  1&.233(10) && && && &0&.327(2) \\
  & \cite{K-M-T-O-94}&  1&.09(3) && && && && \\
  & \cite{Jacobs-86} &  1&.26(5) & 0&.64(2) && && && \\
&\cite{H-K-K-K-85}   &  1&.24(1) & 0&.606(18) & && 0&.077(44) & 0&.319(14) \\
&\cite{W-G-W-92}     &  && && &&  0&.105(8) &&  \\
&\cite{A-S-94}       &  && && && && 0&.324(5)$,\;\;$0.329(2)  \\
&\cite{A-S-W-Z-94}   &  && && && && 0&.329(4)$,\;\;$0.333(2)  \\
&\cite{S-W-W-93}     &  && 0&.610(6) && && && \\
&\cite{B-W-W-92}     &  && & & &&  & & 0&.336(30) \\
&\cite{D-L-C-89}     &  1&.228(39) & 0&.628(8) & 0&.0300(15) && && \\\hline
ms& \cite{B-Y-87}       &   1&.25(2) & 0&.64(1) & && & && \\
 &\cite{B-N-K-J-L-B-83}&&& && && 0&.115(4) & 0&.331(6)  \\
&\cite{M-M-B-95}     &  && && && 0&.11(3) &&  \\
&\cite{M-D-N-94}     &  && && && && 0&.325(2) \\
&\cite{S-A-A-S-C-E-93}& && && && 0&.11(3) &&  \\
&\cite{S-P-K-T-93}   &   1&.25(2) && && && & 0&.315(15)  \\\hline
mi& \cite{A-B-93}    &   && && && && 0&.34(8) \\
  & \cite{A-B-93b}   &   1&.18(3) & 0&.60(2) && && && \\
&\cite{S-B-W-94}     &   1&.216(13) & 0&.623(13) & 0&.039(4) && && \\
&\cite{Z-B-W-94}     &   1&.237(7) & 0&.630(12) && && && \\
&\cite{H-K-K-K-91}   &   1&.25(2) & 0&.63(1)    && && && \\
&\cite{S-W-92}       &   1&.17(11) & 0&.65(4)   && && && \\
\end{tabular}
\end{table}
Even if the systems studied are quite different, the results
substantially agree, although, looking in more detail, as already
observed in Ref.\ \cite{B-L-H-95}, one can find small discrepancies.
Moreover they substantially agree with the theoretical predictions 
discussed above, confirming the fact that all these transitions 
are in the Ising universality class. It should also be noticed that the 
experimental results are less accurate than the theoretical estimates.

\section{The effective potential}
\label{effpot}

\subsection{Small-field expansion of the effective potential in the
high-temperature phase}
\label{subef1}

The effective potential (Helmholtz free energy) is related to the
(Gibbs) free energy of the model.  Indeed, if $M\equiv\langle
\phi\rangle$ is the magnetization and $H$ the magnetic field, one
defines
\begin{equation}
{\cal F} (M) = M H - {1\over V} \log Z(H),
\end{equation}
where $Z(H)$ is the partition function and the dependence on the
temperature is always understood in the notation.

The global minimum of the effective potential determines the value of
the order parameter which characterizes the phase of the model.  In
the high-temperature or symmetric phase the minimum is unique with
$M=0$. According to the Ginzburg-Landau theory, as the temperature
decreases below the critical value, the effective potential takes a
double-well shape. The order parameter does not vanish anymore and the
system is in the low-temperature or broken phase.  Actually in the
broken phase the double-well shape is not correct because the
effective potential must be convex \cite{Griffiths-66}.  In this phase
it should present a flat region around the origin.

In the high-temperature phase the effective potential admits 
an expansion around $M=0$:
\begin{equation}
\Delta {\cal F} \equiv {\cal F} (M) - {\cal F} (0) = 
\sum_{j=1}^\infty {1\over (2j)!} a_{2j} M^{2j}.
\end{equation}
The coefficients $a_{2j}$ can be expressed in terms 
of renormalization-group invariant quantities. Introducing
a renormalized magnetization
\begin{equation}
\varphi^2 = {\xi(t,H=0)^2 M(t,H)^2\over \chi(t,H=0)} \,,
\end{equation}
where $t$ is the reduced temperature,
one may write
\begin{equation}
\Delta {\cal F}= {1\over 2} m^2\varphi^2 + 
\sum_{j=2} m^{d-j(d-2)} {1\over (2j)!} g_{2j} \varphi^{2j} .
\label{freeeng}
\end{equation}
Here $m=1/\xi$, $g_{2j}$ are functions of $t$ only, and $d$ is the space 
dimension. In field theory
$\varphi$ is the expectation value of the zero-momentum renormalized
field.  For $t\to 0$ the quantities $g_{2j}$ approach universal
constants (which we indicate with the same symbol) that represent the
zero-momentum $2j$-point renormalized coupling constants.  By
performing a further rescaling
\begin{equation}
\varphi = {m^{(d-2)/2}\over\sqrt{g_4}} z
\label{defzeta}
\end{equation} 
in Eq.\ (\ref{freeeng}), 
the free energy can be  written as
\begin{equation}
\Delta {\cal F} = {m^d\over g_4}A(z),
\label{dAZ}
\end{equation}
where
\begin{equation}
A(z) =   {1\over 2} z^2 + {1\over 4!} z^4 + 
\sum_{j=3} {1\over (2j)!} r_{2j} z^{2j},
\label{AZ}
\end{equation}
and
\begin{equation}
r_{2j} = {g_{2j}\over g_4^{j-1}} \qquad\qquad j\geq 3.
\label{r2j}
\end{equation}
One can show that $z\propto t^{-\beta} M$, and
that the equation of state can be written in the form
\begin{equation}
H\propto t^{\beta\delta} {\partial A(z)\over \partial z}.
\label{eqa}
\end{equation}

The effective potential ${\cal F}(M)$ admits a power-series expansion
also near the coexistence curve, i.e.\ for $t<0$ and $H=0$.  If
$M_0=\lim_{H\to 0^+} M(H)$, for $M>M_0$ (i.e.\ for $H\geq 0$) we have
\begin{equation}
\delta {\cal F}\equiv 
{\cal F} (M) - {\cal F} (M_0) = \sum_{j=2} {1\over j!} a_j (M-M_0)^j.
\end{equation}
In terms of the renormalized magnetization $\varphi$ we can rewrite
\begin{equation}
\delta {\cal F}  ={1\over 2} m^2(\varphi-\varphi_0)^2 + 
\sum_{j=3} m^{d-j(d-2)/2}
{1\over j!} g_{j}^- (\varphi-\varphi_0)^j,
\label{freeengbr}
\end{equation}
where $m\equiv 1/\xi^-$ and $\xi^-$ is the second-moment correlation
length defined in the low-temperature phase.  For $t\to 0^-$, the
quantities $g_j^-$ approach universal constants that represent the
low-temperature zero-momentum $j$-point renormalized coupling
constants.  A simpler parametrization can be obtained if we introduce
\cite{P-V-br-99}
\begin{equation}
u \equiv {M\over M_0},
\label{defu}
\end{equation} 
so that
\begin{equation}
\delta {\cal F} = {m^d \over w^2} B(u),
\label{dAZu}
\end{equation}
where 
\begin{equation}
w^2\equiv \lim_{T\to T_c - }\,  \lim_{H\to 0}{\chi\over M^2 \xi^d} \, .
\label{defw}
\end{equation}
The scaling function $B(u)$ has the following expansion
\begin{equation}
B(u) =   {1\over 2} (u-1)^2 + \sum_{j=3} {1\over j!} v_{j} (u-1)^j,
\label{BZ}
\end{equation}
where
\begin{equation}
v_{j} = {g_j^-\over w^{j-2}} .
\label{uj}
\end{equation}

\subsection{The four-point zero-momentum renormalized coupling}
\label{subef2}

The four-point coupling $g\equiv g_4$ plays an important role in the
field-theoretic perturbative expansion at fixed dimension
\cite{Parisi-80}, which provides an accurate description of the
critical region in the symmetric phase.  In this approach, any
universal quantity is obtained from a series in powers of $g$
($g$-expansion), which is then resummed and evaluated at the
fixed-point value of $g$, $g^*$ (see e.g.\ Refs.\ 
\cite{L-Z-77,B-N-G-M-77}).  The theory is renormalized at zero
momentum by requiring
\begin{eqnarray}
&&\Gamma^{(2)}(p) = 
Z^{-1} \left[ M^2+p^2+O(p^4)\right],
\label{s2e7a}  \\
&&\Gamma^{(4)}(0,0,0,0)=
Z^{-2} \,M\, g. 
\label{s2e7b}
\end{eqnarray}
When $M\rightarrow 0$ the coupling $g$ is driven toward an
infrared-stable zero $g^*$ of the corresponding Callan-Symanzik
$\beta$-function
\begin{equation}
\beta(g)\equiv M \left.\partial g \over \partial M\right|_{g_0,\Lambda}.
\end{equation}
In this context a rescaled coupling is usually introduced
(see e.g.\ Ref.\ \cite{ZJ-book}):
\begin{equation}
\bar{g} = {3\over 16\pi} g.
\end{equation}
An important issue in this field-theoretical approach concerns the
analytic properties of $\beta(g)$, that are relevant for the procedure
of resummation of the $g$-expansion.  General renormalization-group
arguments predict a non-analytic behavior of $\beta(g)$ at $g=g^*$
\cite{Parisi-80}. One expects
a behavior of the form \cite{Nickel-82,Nickel-91}
\begin{equation}
\beta(g) = -\omega  (g^{*} - g) 
   +  b_1 (g^{*} - g)^2 + \ldots 
   +  c_1 (g^{*} - g)^{1+{1\over \Delta}} + \ldots 
  +  d_1 (g^{*} - g)^{\Delta_2\over \Delta} + \ldots 
\label{rg}
\end{equation}
($\Delta=\omega\nu$ and $\Delta_2$ are scaling correction exponents).
In the framework of the $1/N$ expansion of ${\rm O}(N)$ $\phi^4$
models, the analysis \cite{P-V-gr-98} of the next-to-leading order of
the Callan-Symanzik $\beta$-function, calculated in Ref.\ 
\cite{C-P-R-V-96}, shows explicitly the presence of confluent
singularities of the form (\ref{rg}).

In the fixed-dimension field-theoretical approach, a precise
determination of $g^*$ is crucial, since the critical exponents are
obtained by evaluating appropriate (resummed) anomalous dimensions at
$g^*$.  The resummation of the $g$-expansion is usually performed
following the Le Guillou-Zinn-Justin (LZ) procedure \cite{L-Z-77},
which assumes the analyticity of the $\beta$-function.  The presence
of confluent singularities may then cause a slow convergence to the
correct fixed-point value, leading to an underestimate of the
uncertainty derived from stability criteria.  

We have computed $g^*\equiv g_4^*$ from our IHT series by calculating
the critical limit of the quantity $g_4$ defined in Eq.\ 
(\ref{grdef}).  A description of our analysis can be found in App.\ 
\ref{ratioofamp}.  The results are reported in Table \ref{summaryrfc}.
We find good agreement among the results of the three improved
Hamiltonians, that lead to our final estimate:
\begin{equation}
g^*=23.49(4),\qquad\qquad \bar{g}^*=1.402(2).
\label{resg4}
\end{equation} 

\begin{table}[tbp]
\caption{
Results for $g_4^*$, $r_6$, $r_8$, $r_{10}$, $c_2$ and $c_3$
derived from the analysis of the IHT series (see App.\
\ref{seriesanalysis}).  The error is reported as a sum of two terms:
the first one is related to the spread of the approximants; the second
one is related to the uncertainty of the value of $\lambda_4^*$.
}
\label{summaryrfc}
\begin{tabular}{cr@{}lr@{}lr@{}lr@{}lr@{}lr@{}l}
\multicolumn{1}{c}{}&
\multicolumn{2}{c}{$g_4^*$}&
\multicolumn{2}{c}{$r_6$}&
\multicolumn{2}{c}{$r_{8}$}&
\multicolumn{2}{c}{$r_{10}$}&
\multicolumn{2}{c}{$10^4 c_2$}&
\multicolumn{2}{c}{$10^4 c_3$}\\
\tableline \hline
$\lambda_6=0$ & 23&.499(16+20)    & 2&.051(7+2) & 2&.23(5+4) & $-$14&(4) & $-$3&.582(7+6)  & 0&.085(6) \\  
$\lambda_6=1$ & 23&.491(21+40)    & 2&.050(5+4) & 2&.23(5+6) & $-$13&(5) & $-$3&.574(7+20) & 0&.086(4) \\
spin-1       & 23&.487(18+20)     & 2&.046(2+3) & 2&.34(5+3) & $-$8&(25) & $-$3&.568(11+4) & 0&.090(4) \\
\end{tabular}
\end{table}

Table \ref{summarygr} presents a selection of estimates of $\bar{g}^*$
obtained by different approaches.
\begin{table}[tbp]
\squeezetable
\caption{
Estimates of $\bar{g}^*\equiv 3g^*/(16\pi)$.  (sc) and (bcc) in
the HT estimates of Ref.\ \protect\cite{B-C-g-98} denote simple cubic
and bcc lattice respectively.  For values marked with an asterisk,
the error is not quoted explicitly in the reference.
}
\label{summarygr}
\begin{tabular}{r@{}lr@{}lr@{}lr@{}lr@{}lr@{}lr@{}l}
\multicolumn{2}{c}{IHT}&
\multicolumn{2}{c}{HT}&
\multicolumn{2}{c}{$\epsilon$-exp.}&
\multicolumn{2}{c}{$d$=3 $g$-exp.}&
\multicolumn{2}{c}{MC}&
\multicolumn{2}{c}{$d$-exp.}&
\multicolumn{2}{c}{ERG}\\
\tableline \hline
1&.402(2)& 1&.408(7) (sc) \cite{B-C-g-98}& 1&.397(8) \cite{P-V-gr-98} & 1&.411(4) \cite{G-Z-98} 
& 1&.39(3) \cite{Tsypin-94} & 1&.412(14) \cite{B-B-92}  & 1&.23(21) \cite{Morris-97} \\
& & 1&.407(6) (bcc) \cite{B-C-g-98} & 1&.391$^*$ \cite{G-Z-98} & 1&.40$^*$ \cite{M-N-91} & 
1&.408(12) \cite{Kim-99}
& & & 1&.72$^*$ \cite{T-W-94} \\
& & 1&.406(9) \cite{P-V-gr-98} & & & 1&.415$^*$ \cite{S-O-U-K-98} & 1&.49(3) \cite{B-K-96} & & & & \\
& & 1&.414(6) \cite{B-C-g-97} & & & 1&.416(5) \cite{L-Z-77} & 1&.462(12)~\cite{K-L-96} & & & & \\
& & 1&.459(9) \cite{Z-L-F-96} & & &  & &  & & & & & \\
& & 1&.42(9) \cite{Reisz-95} & & &  & &  & & & & & \\
\end{tabular}
\end{table}
The HT estimates of Refs.\ \cite{P-V-gr-98,B-C-g-98,B-C-g-97} were
obtained by using the RT or appropriate biased approximants in order
to handle the leading confluent correction.  The larger result of
Ref.\ \cite{Z-L-F-96} could be explained by an effect of the scaling
corrections.  Field-theoretical estimates are reasonably consistent,
especially those obtained from a constrained analysis of the
$O(\epsilon^4)$ $\epsilon$-expansion \cite{P-V-gr-98}.  In the $d$=3
$g$-expansion approach $g^*$ is determined from the zero of $\beta(g)$
after resumming its available $O(g^7)$ series.  The 
results obtained using the LZ resummation method
\cite{G-Z-98} show a slight discrepancy from our IHT estimates. 
This difference can explain the
apparent discrepancy found in the determination of $\gamma$.  Indeed,
the sensitivity of $\gamma$ to $\bar{g}^*$, quantified in Ref.\ 
\cite{G-Z-98} through $d\gamma/d \bar{g}^*\simeq 0.18$, tells us that
changing the value of $\bar{g}^*$ from 1.411 (which is the value
obtained from the zero of $\beta(g)$) to 1.402 shifts $\gamma$ from
1.2396 to 1.2380, which is much closer to the IHT estimate
$\gamma=1.2371(4)$. Similarly for $\nu$, using $d\nu/d \bar{g}^*\simeq
0.11$ \cite{G-Z-98}, $\nu$ would change from 0.6304 to 0.6294, which
is quite acceptable, since a residual uncertainty due to the
resummation of $\nu(g)$ is still present.  The more general analysis
of the $g$-expansion of Ref.\ \cite{M-N-91} leads to a smaller value
$\bar{g}^*=1.40$, with an uncertainty estimated by the authors to be
about 1\%.  In Table \ref{summarygr} we also report estimates obtained
by approximately solving the exact renormalization group equation
\cite{T-W-94,Morris-97} (ERG), and from a dimensional expansion of the
Green's functions around $d=0$ \cite{B-B-92} ($d$-exp.).  Concerning
Monte Carlo (MC) results, we mention that the result of Ref.\ 
\cite{Tsypin-94} has been obtained by studying the probability
distribution of the average magnetization (see also Ref.\ 
\cite{R-L-J-98} for a work employing a similar approach).  The other
estimates have been obtained from fits to data in the neighborhood 
of $\beta_c$. In
Ref.\ \cite{B-F-M-M-98} Monte Carlo simulations were performed using
the Hamiltonian (\ref{hamiltonian}) with $\lambda_6=0$ and
$\lambda_4=1$, which is close to its optimal value.  A fit to the data
of $g_4$, kindly made available to us by the authors, gives the
estimate $g^*=23.41(24)$ (i.e.\ $\bar{g}^*=1.397(14)$), which is in
agreement with our IHT estimate.  
In Ref.\ \cite{Kim-99} a finite-size scaling technique is used to obtain
data for large correlation lengths, then the estimate of $g^*_4$ is 
extracted by a fit taking into account the leading scaling correction.
The Monte Carlo estimates of Refs.\ \cite{B-K-96,K-L-96} were
larger because the effects of scaling corrections were neglected, as
already observed in Ref.\ \cite{P-V-gr-98}.  A more complete list of
references regarding this issue can be found in Ref.\ 
\cite{P-V-gr-98}.

\subsection{Higher-order zero-momentum renormalized couplings}
\label{subef4}

To compute the HT series of the effective-potential parameters
$r_{2j}$ defined in Eq.\ (\ref{r2j}), we rewrite them in terms of the
zero-momentum connected $2j$-point Green's functions $\chi_{2j}$ as
\begin{eqnarray}
r_6 =&& 10 - {\chi_6\chi_2\over \chi_4^2},\label{r6gr}\\
r_8 =&& 280 - 56{\chi_6\chi_2\over \chi_4^2} + 
        {\chi_8\chi_2^2\over \chi_4^3},\label{r8gr}\\
r_{10} =&& 15400  - 4620 {\chi_6 \chi_2\over \chi_4^2}        
+ 126{\chi_6^2 \chi_2^2\over \chi_4^4}
+ 120 {\chi_8 \chi_2^2\over \chi_4^3} 
-{\chi_{10} \chi_2^3\over \chi_4^4},
\label{r10gr}
\end{eqnarray}
etc...  Details of the analysis of the series are reported in App.\ 
\ref{ratioofamp}. Combining the results reported in Table \ref{summaryrfc},
we obtain the following estimates:
\begin{eqnarray}
r_6 &=& 2.048(5) ,\label{resr6} \\
r_8 &=& 2.28(8) ,\label{resr8} \\
r_{10} &=& -13(4) .\label{resr10} 
\end{eqnarray}
From the results for $r_{2j}$ we can obtain estimates of the
couplings
\begin{eqnarray}
g_6=g^2 r_6 &=& 1130(5), \\
g_8=g^3 r_8 &=& 2.96(11)\times 10^4,\\
g_{10}=g^4 r_{10} &=& -4.0(1.2)\times 10^6.
\end{eqnarray}
In the literature several approaches have been
used for the determination of the couplings $g_{2j}$.  Table
\ref{summaryrj} presents a review of the available estimates of $r_6$,
$r_8$ and $r_{10}$.  
\begin{table}[tbp]
\caption{
Estimates of $r_{2j}$.
When the original reference reports only estimates of $g_{2j}$ 
(see Refs.\ \protect\cite{Tsypin-94,K-L-96,Reisz-95}),
the errors we quote for $r_{2j}$  have been calculated
by considering the estimates of $g_{2j}$ as uncorrelated. 
For values marked with an asterisk,
the error is not quoted explicitly in the reference.
}
\label{summaryrj}
\begin{tabular}{cr@{}lr@{}lr@{}lr@{}lr@{}lr@{}l}
\multicolumn{1}{c}{}&
\multicolumn{2}{c}{IHT}&
\multicolumn{2}{c}{HT}&
\multicolumn{2}{c}{$\epsilon$-exp.}&
\multicolumn{2}{c}{$d$=3 $g$-exp.}&
\multicolumn{2}{c}{MC}&
\multicolumn{2}{c}{ERG}\\
\tableline \hline
$r_6$ & 
2&.048(5) & 1&.99(6) \cite{B-C-g-97}& 2&.058(11) \cite{P-V-ef-98} & 2&.053(8) \cite{G-Z-98} 
& 2&.72(23) \cite{Tsypin-94} & 2&.064(36) \cite{Morris-97} \\
& & & 2&.157(18) \cite{Z-L-F-96} & 2&.12(12) \cite{G-Z-98} & 2&.060$^*$ \cite{S-O-U-K-98} & 3&.37(11) \cite{Kim-99}
& 1&.92$^*$ \cite{T-W-94} \\
& & & 2&.25(9) \cite{L-F-96} & & & &  & 3&.26(26)~\cite{K-L-96}  &  & \\
& & & 2&.5(5) \cite{Reisz-95} & & & & &  &  & &  \\\hline
$r_8$ & 2&.28(8) & 2&.7(4) \cite{B-C-g-97}& 2&.48(28) \cite{P-V-ef-98} & 2&.47(25) \cite{G-Z-98} 
& & & 2&.47(5) \cite{Morris-97} \\
& & & & & 2&.42(30) \cite{G-Z-98} & &
& & & 2&.18$^*$ \cite{T-W-94} \\\hline
$r_{10}$ & 
$-$13&(4) & $-$4&(2) \cite{B-C-g-97}& $-$20&(15) \cite{P-V-ef-98} & $-$25&(18) \cite{G-Z-98} 
& & & $-$18&(4) \cite{Morris-97} \\
& & & & & $-$12&.0(1.1) \cite{G-Z-98} & && & & & \\
\end{tabular}
\end{table}
We also mention the estimate $r_{10}=-10(2)$ we will
obtain in Sec.\ \ref{eqofstate} by studying the equation of state.
The agreement with the field-theoretic calculations based on the
$\epsilon$-expansion \cite{G-Z-98,P-V-ef-98} and on the $d$=3 $g$-expansion
\cite{G-Z-98} is good.  Precise estimates of $r_{2j}$ have also been
obtained in Ref.\ \cite{Morris-97} (see also Ref.\ \cite{T-W-94}) by
ERG, although the estimate of $g_4^*$ by the same method is not as
good.  Additional results have been obtained from HT expansions
\cite{B-C-g-97,Z-L-F-96,Reisz-95} and Monte Carlo simulations
\cite{Tsypin-94,K-L-96} of the Ising model.  The Monte Carlo results
do not agree with the results of other approaches, especially those of
Refs.\ \cite{Kim-99,K-L-96}, which are obtained using finite-size
scaling techniques.  But one should consider the difficulty of such
calculations due to the subtractions that must be performed to compute
the irreducible correlation functions.  A more complete list of
references regarding this issue can be found in Refs.\ 
\cite{G-Z-97,G-Z-98,P-V-ef-98}.

\section{The two-point function}
\label{twopointf}

The critical behavior of the two-point correlation function $G(x)$ of
the order parameter is relevant to the description of critical
scattering phenomena, which can be observed in many experiments, such
as light and X-ray scattering in fluids, magnets...  In Born
approximation the cross section $\Gamma_{fi}$ for particles
of incoming momentum $p_i$ and outgoing momentum $p_f$ is proportional to
the component $k=p_f-p_i$ of the Fourier transform of $G(x)$:
\begin{equation}
\Gamma_{fi}\propto \widetilde{G}(p_f-p_i).
\end{equation}
As a consequence of the critical
behavior of the two-point function $G(x)$ at $T_c$,
\begin{equation}
\widetilde{G}(k)\sim {1\over  k^{2-\eta}},
\label{eq1}
\end{equation}
the cross section for $k\rightarrow 0$ (forward scattering) diverges
as $T\rightarrow T_c$.  When strictly at criticality, Eq.\ (\ref{eq1})
holds for all $k \ll \Lambda$, where $\Lambda$ is a generic cut-off
related to the microscopic structure of the statistical system, e.g.\ 
the inverse lattice spacing in the case of lattice models. In the
vicinity of the critical point, where the relevant correlation length
$\xi$ is large but finite, the behavior (\ref{eq1}) occurs for
$\Lambda \gg k\gg 1/\xi$.  At low momentum, $k\ll 1/\xi$, experiments
show that $G(x)$ is well approximated by a Gaussian
(Ornstein-Zernike) behavior,
\begin{equation}
{\widetilde{G}(0)\over\widetilde{G}(k)}
\simeq 1 + {k^2\over M^2},
\label{gaubeh}
\end{equation}
where $M\sim 1/\xi$ is a mass scale defined at zero momentum (for a
general discussion see e.g.\ Ref.\ \cite{F-B-67}).  Corrections to
Eq.\ (\ref{gaubeh}) are present, and reflect, once more, the
non-Gaussian nature of the Wilson-Fisher fixed point.  The
above-mentioned experimental observations, confirmed by theoretical
studies \cite{C-P-R-V-98,T-F-75}, show that they are small.  In the
following we will improve the determination of the critical two-point
function at low-momentum using IHT series.

In order to study the low-momentum universal critical behavior of the
two-point function $G(x)=\langle \phi(x) \phi(0) \rangle$,
we consider the scaling function
\begin{equation}
g(y) = \chi/\widetilde{G}(k),\qquad\qquad y\equiv k^2/M^2,
\end{equation}
($M\equiv 1/\xi$ and $\xi$ is the second-moment correlation length)
in the critical limit $k,M\to0$ with $y$ fixed.
The scaling function $g(y)$ can be expanded in powers of $y$ around $y=0$:
\begin{equation}
g(y)=1 + y + \sum_{i=2}^\infty c_i y^i.
\label{lexp}
\end{equation}
Other important quantities which characterize the low-momentum behavior of
$g(y)$ are the critical limit of the ratios 
\begin{eqnarray}
S_M&\equiv&M_{\rm gap}^2/M^2,\label{SMdef}\\
S_Z&\equiv& \chi M^2/Z_{\rm gap},\label{SZdef}
\end{eqnarray}
where $M_{\rm gap}$ (the mass gap of the theory) and $Z_{\rm gap}$ 
determine the long-distance behavior of the two-point function: 
\begin{equation}
G(x)\approx  {Z_{\rm gap}\over 4\pi |x|} e^{-M_{\rm gap}|x|}.
\label{largexbehavior}
\end{equation}
The critical limits of $S_M$ and $S_Z$ are related to the negative zero 
$y_0$ of $g(y)$ closest to the origin by 
\begin{eqnarray}
S_M&=&-y_0,\\
S_Z&=& \left.\partial g(y) \over \partial y \right|_{y=y_0}.
\end{eqnarray}
The coefficients $c_i$ can be related to the critical limit of
appropriate dimensionless ratios of spherical moments of $G(x)$ (as
shown explicitly in App.\ \ref{series}) and can be calculated by
analyzing the corresponding HT series.  Some details of the analysis
of our HT series are reported in App.\ \ref{ratioofamp}.  In Table
\ref{summaryrfc} we report the results for $\lambda_6=0,1$ and the spin-1 model.  
We obtain the estimates
\begin{eqnarray}
c_2 &= -&3.576(13) \times 10^{-4},\label{resc2}\\ 
c_3 &= & 0.87(4) \times 10^{-5},\label{resc3}
\end{eqnarray}
and the bound
\begin{equation}
- 10^{-6} \lesssim  c_4 < 0.\label{resc4}
\end{equation}
The constants $c_i$ and $S_M$ can also be calculated by
field-theoretic methods.  They have been computed to $O(\epsilon^3)$
in the framework of the $\epsilon$-expansion \cite{Bray_76}, and to
$O(g^4)$ in the framework of the $d$=3 $g$-expansion
\cite{C-P-R-V-98}.  In Table \ref{ciexpeps} we report the results of
constrained analyses of the $O(\epsilon^3)$ $\epsilon$-expansion of
$c_i$ and $S_M-1$, using exact results in $d=2,1$ ($S_M=1$ and $c_i=0$
in $d=1$; two-dimensional values will be reported in Table \ref{ci2d})
and following the method of Ref.\ \cite{P-V-gr-98}.
\begin{table}[tbp]
\caption{
$c_i$ and $S_M-1$ obtained from $O(\epsilon^3)$ series:
unconstrained analysis (unc) and
analyses constrained in dimensions $d=1,2$.
}
\label{ciexpeps}
\begin{tabular}{cr@{}lr@{}lr@{}lr@{}l}
\multicolumn{1}{c}{}&
\multicolumn{2}{c}{unc}&
\multicolumn{2}{c}{$d=1$}&
\multicolumn{2}{c}{$d=2$}&
\multicolumn{2}{c}{$d=1,2$}\\
\tableline \hline
$10^4(S_M-1)$ & $-$4&.4(1.0) & $-$3&.3(8) &  $-$3&.3(5) &  $-$3&.24(36) \\ 
$ 10^4 c_2$ &  $-$4&.3(9) & $-$3&.2(8) &  $-$3&.3(4) &  $-$3&.30(21) \\ 
$ 10^5 c_3$ &  1&.13(27) & 0&.84(22)  &  0&.76(17)  &  0&.69(10)\\ 
$10^6 c_4$ &  $-$0&.50(13) & $-$0&.37(10)  &  $-$0&.32(8) &  $-$0&.27(5) \\ 
\end{tabular}
\end{table}
Since the constants $c_i$ are of order $O(\epsilon^2)$, we analyzed the
$O(\epsilon)$ series for $c_i/\epsilon^2$.  Errors are indicative
since the series are short.  In Table \ref{summaryci} we
compare the estimates obtained by various approaches: they all agree 
within the quoted errors.

\begin{table}[tbp]
\caption{
Estimates of $S_M$ and $c_{i}$.  (sc) and (bcc) 
denote the simple cubic and the body-centered cubic lattice respectively.
}
\label{summaryci}
\begin{tabular}{cr@{}lr@{}lr@{}lr@{}l}
\multicolumn{1}{c}{}&
\multicolumn{2}{c}{IHT}&
\multicolumn{2}{c}{HT}&
\multicolumn{2}{c}{$\epsilon$-exp.}&
\multicolumn{2}{c}{$d=3$ $g$-exp.}\\
\tableline \hline
$c_2$ & 
$-$3&.576(13)$\times 10^{-4}$ & $-$3&.0(2)$\times 10^{-4}$ \cite{C-P-R-V-98}& 
$-$3&.3(2) $\times 10^{-4}$ & $-$4&.0(5) $\times 10^{-4}$  \\
& && $-$5&.5(1.5)$\times 10^{-4}$ (sc) \cite{T-F-75} & && &\\
& && $-$7&.1(1.5)$\times 10^{-4}$ (bcc) \cite{T-F-75} & && & \\\hline
$c_3$ & 
0&.87(4)$\times 10^{-5}$ & 1&.0(1)$\times 10^{-5}$ \cite{C-P-R-V-98} & 0&.7(1) $\times 10^{-5}$ & 
1&.3(3) $\times 10^{-5}$ \\ 
& && 0&.5(2)$\times 10^{-5}$ (sc) \cite{T-F-75} & && & \\
& && 0&.9(3)$\times 10^{-5}$ (bcc) \cite{T-F-75} & && &  \\\hline
$c_4$ & $-$&$10^{-6} \lesssim   c_4 < 0 $ & & & 
$-$0&.3(1)$\times 10^{-6}$ & $-$0&.6(2)$\times 10^{-6}$\\ \hline
$S_M$ & 0&.999634(4) &   0&.99975(10) \cite{C-P-R-V-98} & 0&.99968(4) & 0&.99959(6)  \\ 
\end{tabular}
\end{table}

As already observed in Ref.\ \cite{C-P-R-V-98}, the coefficients show
the pattern
\begin{equation}
c_i\ll c_{i-1}\ll...\ll c_2 \ll 1\qquad\qquad {\rm for}\qquad i\geq 3.
\label{patternci}
\end{equation}
Therefore, a few terms of the expansion of $g(y)$ in powers of $y$
should be a good approximation in a relatively large region around
$y=0$, larger than $|y|\lesssim 1$. This is in agreement with the
theoretical expectation that the singularity of $g(y)$ nearest to the
origin is the three-particle cut \cite{Ferrel-Scalapino_75,Bray_76}.
If this is the case, the convergence radius $r_g$ of the Taylor
expansion of $g(y)$ is $r_g=9S_M$.  Since, as we shall see, 
$S_M\simeq 1$, at least asymptotically we should have
\begin{equation}
c_{i+1}\simeq {1\over 9}c_i.
\label{pattern-cip1-ci}
\end{equation}
This behavior can be checked explicitly in the large-$N$ limit of the 
$N$-vector model \cite{C-P-R-V-98}.
In two dimensions, the critical two-point function can be written
in terms of the solutions of a Painlev\'e differential equation
\cite{W-M-T-B-76} and it can be verified explicitly that $r_g=9S_M$.  
In Table \ref{ci2d} we report the
values of $S_M$ and $c_i$ for the two-dimensional Ising model.  

\begin{table}[tbp]
\caption{
Values of $S_M$ and $c_i$ for the two-dimensional Ising model in 
the high- and low-temperature phase.
}
\label{ci2d}
\begin{tabular}{r@{$\,$}r@{}lr@{$\,$}r@{}l}
\multicolumn{3}{c}{high temperature \protect\cite{C-P-R-V-96-2}}&
\multicolumn{3}{c}{low temperature}\\
\tableline \hline
$S_M =$&    0&.999196337056               &  $S_M^- =$& 0&.399623590999  \\
$c_2 =$& $-$0&.7936796064$\times 10^{-3}$ &  $c_2^- =$&$-$0&.42989191603 \\
$c_3 =$&    0&.109599108$\times 10^{-4}$  &  $c_3^- =$& 0&.5256121845    \\
$c_4 =$& $-$0&.3127446$\times 10^{-6}$    &  $c_4^- =$&$-$0&.8154613925  \\
$c_5 =$&    0&.126670 $\times 10^{-7}$    &  $c_5^- =$& 1&.422603449     \\
$c_6 =$& $-$0&.62997$\times 10^{-9}$      &  $c_6^- =$&$-$2&.663354573   \\
\end{tabular}
\end{table}

Assuming the pattern (\ref{patternci}), we may estimate $S_M$ and
$S_Z$ from $c_2$, $c_3$, and $c_4$.  Indeed from the equation
$g(y_0)=0$, where $y_0=-S_M$, we obtain
\begin{eqnarray}
S_M &=& 1 + c_2 - c_3  + c_4 + 2 c_2^2 + ... \label{SMest}\\
S_Z &=& 1 - 2 c_2 + 3 c_3 - 4 c_4 - 2 c_2^2 + ... \label{SZest}
\end{eqnarray}
where the ellipses indicate contributions that are negligible with
respect to $c_4$.  In Ref.\ \cite{C-P-R-V-98} the relation
(\ref{SMest}) has been confirmed by a direct analysis of the HT series
of $S_M$.  From Eqs.\ (\ref{SMest}) and (\ref{SZest}) we obtain
$S_M=0.999634(4)$ (from which we can derive an estimate of the ratio
$Q_\xi^+\equiv f^+_{\rm gap}/f^+=1.000183(2)$, cf.\ Eqs.\ 
(\ref{xiamp}) and (\ref{xiosamp})) and $S_Z=1.000741(7)$.

We can also use our results to improve the phenomenological model 
proposed by Bray \cite{Bray_76}. If
we parametrize the large-$y$ behavior of $g(y)$ as
\cite{Fisher-Langer_68}
\begin{equation}
g(y)^{-1} = {A_1\over y^{1 - \eta/2}} 
  \left(1 + {A_2\over y^{(1-\alpha)/(2 \nu)}} +
            {A_3\over y^{1/(2\nu)}}\right),
\label{Bray-param}
\end{equation}
then, by using our estimates of the critical exponents and the 
phenomenological function of Ref.\ \cite{Bray_76}, 
we obtain the following values for the coefficients:
\begin{equation}
A_1 \approx 0.918, \qquad A_2 \approx 2.55, \qquad A_3 \approx - 3.45. 
\label{Bray-vals}
\end{equation}
Estimating reliable errors on these results is practically impossible,
since it is difficult to assess the systematic error due to the many
uncontrolled simplifications that are used.  It is however reassuring
that they are in reasonable agreement the $\epsilon$-expansion
predictions \cite{Bray_76}
\begin{equation}
A_1 \approx 0.92, \qquad A_2 \approx 1.8, \qquad A_3 \approx - 2.7, 
\end{equation}
and with the results of a recent experimental study
\cite{Damay-etal_98}
\begin{equation}
A_1 = 0.915(21),  \qquad A_2 = 2.05(80),  \qquad A_3 = - 2.95(80).
\end{equation}
Bray's phenomenological expression makes also predictions for the
coefficients $c_i$.  The pattern (\ref{pattern-cip1-ci}) is built in
the approach. We find 
$c_2 = - 4.2\cdot 10^{-4}$ and $c_3 = 1.0 \cdot 10^{-5}$, in good
agreement with our IHT estimates.  Therefore, Bray's expression
provides a good description of $g(y)$ for small and large values of
$y$. However, in the intermediate crossover region, as already
observed in Ref.\ \cite{Damay-etal_98}, the agreement is worse: Bray's
interpolation is lower by 20--50\% than the experimental result.
 
In the low-temperature phase, for $y\to 0$, the two-point function
also admits a regular expansion of the form (\ref{lexp}).  However,
the deviation from the Gaussian behavior is much larger. The leading
coefficient $c_2^-$ is larger than $c_2$ by about two orders of
magnitude \cite{C-D-K-74}.  Moreover, by analyzing the low-temperature
series published in Ref.\ \cite{A-T-95} one gets $S_M^-=0.938(8)$ (and
correspondingly $Q^-_\xi\equiv f^-_{\rm gap}/f^-=1.032(4)$). Thus
$S_M^-$ shows a much larger deviation from one (the Gaussian value)
than the corresponding high-temperature phase quantity $S_M$.  The
two-dimensional Ising model shows even larger deviations from Eq.\ 
(\ref{gaubeh}), as one can see from the values of $S_M^-$ and $c_i^-$
reported in Table \ref{ci2d}.  Notice that in the low-temperature
phase of the two-dimensional Ising model the singularity at
$k^2=-M^2_{\rm gap}$ of $\widetilde{G}(k)$ is not a simple pole, but a
branch point \cite{W-M-T-B-76}.  As a consequence, $g(y)$ is not
analytic for $|y|>S_M^-$, and therefore the convergence radius of the
expansion around $y=0$ is $S_M^-$.  For discussions of the analytic
structure of $g(y)$ in the low-temperature phase of the
three-dimensional Ising model, see e.g. Refs.\ 
\cite{C-D-K-74,Bray_76,Ferrel-Scalapino_75,C-H-P-99}.

\section{The critical equation of state}
\label{eqofstate}

\subsection{The parametric representation}
\label{parrepth}

The critical equation of state provides relations among the
thermodynamical quantities in the neighborhood of the critical
temperature, in both phases.  From this equation one can then
derive all the universal ratios of amplitudes involving quantities
defined at zero-momentum (i.e.\ integrated in the volume), such as
specific heat, magnetic susceptibility, etc...

From the analysis of IHT series we have obtained the first few
non-trivial terms of the small-field expansion of the effective
potential in the high-temperature phase. This provides corresponding
information for the equation of state
\begin{equation}
H\propto t^{\beta\delta} F(z),
\label{eqa2}
\end{equation}
where $z \propto M t ^{-\beta}$ and, using Eq.\ (\ref{eqa}), 
\begin{equation}
F(z)={\partial A(z)\over \partial z} = 
z + \case{1}{6}z^3 + \sum_{m=2} F_{2m+1} z^{2m+1}
\label{Fzdef}
\end{equation}
with 
\begin{equation}
F_{2m-1} = {1\over (2m-1)!} r_{2m}.
\end{equation}
The function $H(M,t)$ representing the external field in the critical
equation of state (\ref{eqa2}) satisfies Griffith's analyticity: it is
regular at $M=0$ for $t>0$ fixed and at $t=0$ for $M>0$ fixed.  The
first region corresponds to small $z$ in Eq.\ (\ref{eqa2}), while the
second is related to large $z$, where $F(z)$ can be expanded in the
form
\begin{equation}
F(z) = z^\delta \sum_{n=0} F^{\infty}_n z^{-n/\beta}.
\label{fasymp}
\end{equation}
Of course  $F^\infty_n$ are universal constants.

To reach the coexistence curve, i.e.\ $t<0$ and $H=0$, one should
perform an analytic continuation in the complex $t$-plane
\cite{ZJ-book,G-Z-97}.  The spontaneous magnetization is related to
the complex zero $z_0$ of $F(z)$.  Therefore the description of the
coexistence curve is related to the behavior of $F(z)$ in the
neighbourhood of $z_0$.  In order to obtain a representation of the
critical equation of state that is valid in the whole critical region,
one may use parametric representations, which implement in a simple
way all scaling and analytic properties.  One parametrizes $M$ and $t$
in terms of $R$ and $\theta$
\cite{Schofield-69,S-L-H-69,Josephson-69}:
\begin{eqnarray}
M &=& m_0 R^\beta \theta ,\nonumber \\
t &=& R(1-\theta^2), \nonumber \\
H &=& h_0 R^{\beta\delta}h(\theta),
\label{parrep}
\end{eqnarray}
where $h_0$ and $m_0$ are normalization constants.  The function
$h(\theta)$ is odd and regular at $\theta=1$ and at
$\theta=0$.  The constant $h_0$ can be chosen so that
$h(\theta)=\theta+O(\theta^3)$.  The zero of $h(\theta)$,
$\theta_0>1$, represents the coexistence curve $H=0$, $T<T_c$. The
parametric representation satisfies the requirements of regularity of
the equation of state. One expects at most an essential singularity on
the coexistence curve \cite{singatcoexcurve}.

The relation between $h(\theta)$ and $F(z)$ is given by
\begin{eqnarray}
z &=& \rho \theta \left( 1 - \theta^2\right)^{-\beta},
\label{thzrel} \\
h(\theta) &=& \rho^{-1} 
    \left( 1 - \theta^2\right)^{\beta\delta} F(z(\theta)),
\label{hFrel}
\end{eqnarray}
$\theta>0$, and hyperscaling implies that $\beta\delta = \beta+\gamma$.
Notice that this mapping is invertible only in the region 
$\theta<\theta_{l}$, where $\theta_{l} = (1-2\beta)^{-1/2}$ is the
solution of the equation $z'(\theta)=0$. Thus the values of $\theta$ that
are relevant for the critical equation of state, i.e. 
$0\leq \theta \leq \theta_0$,
must be smaller than $\theta_l$. This fact will not be a real limitation
for us, since the range of values of $\theta$ involved in our calculations
(which will be  $0\leq \theta^2 \leq \theta_0^2\lesssim 1.40$)
will be always far from the limiting value $\theta_{l}^2\simeq 2.88$.

As a consequence of Eqs.\ (\ref{parrep}), (\ref{thzrel}), and
(\ref{hFrel}), we easily obtain the relationships
\begin{equation}
{M \over t^\beta} = \left(m_0 \over \rho\right) z, \qquad
{H \over t^{\beta\delta}} = \left(h_0 \over \rho\right) F(z).
\end{equation}
We can therefore treat $\rho$ as a free parameter, and the scaling
relations between physical variables will not depend on $\rho$,
provided that $m_0$ and $h_0$ are rescaled with $\rho$.  In the
exact parametric equation the value of $\rho$ may be chosen
arbitrarily but, as we shall see, when adopting an approximation
procedure the dependence on $\rho$ is not eliminated, and it may
become important to choose the value of this parameter properly in
order to optimize the approximation.
 
From $\theta_0$ one can obtain the universal rescaled spontaneous
magnetization \cite{G-Z-97}, i.e.\ the complex zero $z_0$ of $F(z)$,
\begin{equation}
z_0 = |z_0|e^{-i\pi\beta},\qquad\qquad |z_0| =
\rho \theta_0 \left(\theta_0^2-1\right)^{-\beta}.
\end{equation}
From the function $h(\theta)$ one can calculate the universal ratios
of amplitudes.  In App.\ \ref{univra} we report the definitions of the
universal ratios of amplitudes that have been introduced in the
literature, and the corresponding expressions in terms of $h(\theta)$.

Expanding  $h(\theta)$ in (odd) powers of $\theta$,
\begin{equation}
h(\theta) = \theta  + \sum_{n=1} h_{2n+1}\theta^{2n+1} ,
\label{hexp}
\end{equation}
and using Eq.\ (\ref{hFrel}), one can find the relations among
$h_{2n+1}$ and the coefficients $F_{2m+1}$ of the expansion of $F(z)$.
The procedure is explained in App.\ \ref{rhotheta}, and the general
result is:
\begin{equation}
h_{2n+1} = \sum_{m=0}^n c_{n,m}\rho^{2m}F_{2m+1},
\label{h2n}
\end{equation}
where
\begin{equation}
c_{n,m}= {1 \over (n-m)!}\prod_{k=1}^{n-m} (2\beta m -\gamma+k-1) ;
\label{cnm}
\end{equation}
notice that $c_{n,n}$ = 1.
In general $h_{2n+1}$ depends on $\gamma$, $\beta$, and on
the coefficients $F_{2m+1}$ with $m \leq n$.

We shall need the explicit form of the first two coefficients:
\begin{eqnarray}
h_3 &=& \case{1}{6}\rho^2 - \gamma,\label{h3}\\
h_5 &=& \case{1}{2}\gamma(\gamma-1) + \case{1}{6} (2\beta-\gamma)\rho^2 
+ F_5 \rho^4 .\label{h5}
\end{eqnarray}

\subsection{Approximation scheme based on stationarity}
\label{stationarity}

In Ref.\ \cite{G-Z-97} Guida and Zinn-Justin use the first few
coefficients of the small-$z$ expansion of $F(z)$ to get polynomial
approximations of $h(\theta)$ which should provide a description that
is reliable in the whole critical region.  The approximations
considered are truncations of the small-$\theta$ expansion of
$h(\rho,\theta)$, i.e.\ 
\begin{equation}
h^{(t)}(\rho,\theta) = 
  \theta  + \sum_{n=1}^{t-1} h_{2n+1}(\rho) \theta^{2n+1},
\label{hexpn}
\end{equation}
where $h_{2n+1}(\rho)$ are given by Eq.\ (\ref{h2n}).
We follow a similar strategy, with a significant difference in the
procedure adopted in order to fix the value of $\rho$ .

By Eqs.\ (\ref{h2n}) and (\ref{cnm}), the coefficients
$h_{2n+1}(\rho)$ included in Eq.\ (\ref{hexpn}) are written in terms
of the $t$ parameters $\gamma$, $\beta$, $F_5$ ... $F_{2t-1}$.  In
practice only the first coefficients of the small-$\theta$ expansion
of $h(\theta)$ are well determined, since we have good estimates only
for the first few $F_{2m+1}$.  Once the order of the truncation has
been decided, one may exploit the freedom of choosing $\rho$ to
optimize the approximation of $h(\theta)$. In this way one may hope to
obtain a good approximation even for small values of $t$.  Ref.\ 
\cite{G-Z-97} proposes to determine the optimal value of $\rho$ by
minimizing the absolute value of $h_{2t-1}(\rho)$, i.e.\ the
coefficient of the highest-order term considered.  The idea underlying
this procedure is to increase the importance of small powers of
$\theta$. Our approach is different.

Our starting point is the independence on $\rho$ of the scaling
function $F(z)$ and, as a consequence, of all universal ratios of
amplitudes that can be extracted from it.  Of course, this property
does not hold anymore when we start from a truncated function
$h^{(t)}(\rho,\theta)$, i.e.\ if we compute universal quantities from
a function $F^{(t)}(\rho,z)$ defined by
\begin{equation}
F^{(t)}(\rho,z) \equiv \widetilde{F}^{\,(t)}(\rho,\theta(\rho,z)),
\label{fntfrel}
\end{equation}
where
\begin{equation}
\widetilde{F}^{\,(t)}(\rho,\theta) = 
  { \rho h^{(t)}(\rho,\theta) \over (1 - \theta^2)^{\beta\delta}}
\label{tt1}
\end{equation}
and $\theta(\rho,z)$ is obtained by inverting Eq.\ (\ref{thzrel}).

In order to optimize $\rho$ for a given truncation
$h^{(t)}(\rho,\theta)$, we propose a procedure based on the physical
requirement of minimal dependence on $\rho$ of the resulting universal
function $F^{(t)}(\rho,z)$.  This can be obtained by assuming $\rho$
to depend on $z$, i.e.\ $\rho = \rho^{(t)}(z)$, and by requiring the
functional stationarity condition
\begin{equation}
{\delta F^{(t)}(\rho^{(t)},z) \over \delta \rho^{(t)}} = 0
\label{functional-stat}
\end{equation}
(see Ref.\ \cite{Kleinert-98} and references therein for a similar
technique applied to the resummation of perturbative power
expansions).  The non-trivial fact, even surprising at first sight, is
that the solution $\rho^{(t)}(z)$ of Eq.\ (\ref{functional-stat}) is
constant.  In other words, for any $t$ there exists a solution
$\rho_t$ independent of $z$ that satisfies the global stationarity
condition
\begin{equation}
\left. \partial F^{(t)}(\rho,z)\over \partial \rho \right|_{\rho=\rho_t}=0.
\label{glstcond}
\end{equation}
This is equivalent to the fact that, for any universal ratio of
amplitudes $R$, its approximation $R^{(t)}(\rho)$ (obtained from
$F^{(t)}(\rho,z)$) satisfies the stationarity condition
\begin{equation}
\left. d R^{(t)}(\rho) \over d\rho \right|_{\rho=\rho_t} = 0.
\end{equation}

The proof of Eq.\ (\ref{glstcond}) is given in App.\ \ref{rhotheta},
where we show that the global stationarity condition amounts to
requiring $\rho_t$ to be a solution of the algebraic equation
\begin{equation}
\left[(2 \beta-1)\rho {\partial \over \partial \rho}-2 \gamma+2 t-2\right]
    h_{2t-1}(\rho) = 0.
\label{stateq}
\end{equation}

The idea behind our scheme of approximation is that, for any
truncation, the stationarity condition enforces the physical request
that the universal ratios of amplitudes be minimally dependent on
$\rho$.  To check the convergence of the approximation, one can repeat
the computation of universal ratios of amplitudes from the truncated
function $h^{(t)}(\rho_t,\theta)$ for different values of $t$, as long
as one has a reliable estimate of $F_{2t-1}$.  We have no a priori
argument in favor of a fast convergence in $t$ of the universal ratios
of amplitudes derived by this procedure towards their exact values.
However, we may appreciate that its lowest-order implementation,
corresponding to $t$=2 in Eq.\ (\ref{hexpn}), reproduces the
well-known formulae of Refs.\ 
\cite{Schofield-69,S-L-H-69,Josephson-69}, which give an effective
optimization of the linear parametric model.  Indeed we obtain from
Eqs.\ (\ref{stateq}) and (\ref{h3}) the $t=2$ solution
\begin{equation}
\rho_2 = \sqrt{ {6 \gamma (\gamma -1) \over \gamma - 2 \beta}} \,.
\end{equation}
In this case the critical equation of state and all critical
amplitudes turn out to be expressible simply in terms of the critical
exponents $\beta$ and $\gamma$. In particular, we found a closed-form
expression for all $F^{(2)}_{2m+1}$ coefficients (see App.\ 
\ref{rhotheta} for a derivation):
\begin{equation}
F^{(2)}_{2m+1}= {(-1)^m \over m!}{\gamma (\gamma-1) \over \rho_2^{2m}} 
\prod_{k=1}^{m-2}(2\beta m-\gamma-k).
\label{coeff2}
\end{equation}

Wallace and Zia \cite{W-Z-74} already noticed that the minimum
condition of Refs.\ \cite{Schofield-69,S-L-H-69,Josephson-69} was
equivalent to a condition of global stationarity for the linear
parametric model.  We have shown that such a global stationarity can
be extended to other parametric models, regardless the linearity
constraint, and can be used to improve the approximation.

The next truncation, corresponding to $t=3$, can also be treated
analytically.  Since it sensibly improves the linear parametric model
in the $3d$ Ising case, we shall present here a few details.  By
applying the stationarity condition (\ref{stateq}) to Eq.\ (\ref{h5}),
we obtain
\begin{equation}
\rho_3 =
  \sqrt{{(\gamma-2\beta)(1-\gamma+2 \beta) \over 12(4 \beta-\gamma)F_5}}
  \left(1-\sqrt{1-{72(2-\gamma)\gamma(\gamma-1)(4\beta-\gamma)F_5 \over 
       (\gamma-2\beta)^2 (1-\gamma+ 2\beta)^2}}\right)^{\!\!{1 \over 2}}.
\end{equation}
Universal ratios of amplitudes may be evaluated in terms of $\rho_3$;
they will now depend only on the parameters $\beta$, $\gamma$ and
$F_5$.  Notice that the predictions of the $t=2$ and $t=3$ models
differ from each other only proportionally to the difference between
the ``experimental'' value of $F_5$ and the value predicted according
to Eq.\ (\ref{coeff2})
\begin{equation}
F_5^{(2)} = {(\gamma-2\beta)^2 \over 72 \, \gamma(\gamma-1)}.
\label{F2}
\end{equation}
If we replace $F_5$ with $F_5^{(2)}$ in the $t=3$ model results, all
the linear parametric model results are automatically reproduced.  In
the $3d$ Ising model, the two values differ by $\sim6\%$, and thus we
expect comparable discrepancies for all universal ratios of
amplitudes.  This can be verified from the numerical results that we
will present in Sec.\ \ref{eqstres} (see Table \ref{eqstdet}).  All
universal ratios of amplitudes obtained from the $t=2$ truncation
(i.e.\ the linear parametric model), using our estimates of $\gamma$
and $\beta$, differ at most by a few per cent from previously
available estimates.  The $t=3$ and higher-order approximations are
consistent with the latter.  The apparent convergence in $t$ of the
results provides a further important support to this scheme.

It is worth noticing that the parametric representation of the equation
of state induces parametric forms for such thermodynamics functions
as the free energy and the susceptibility, as discussed in detail
in App.\ \ref{univparrep}. When we assume a truncated form
of the parametric equation of state, in general only the corresponding
free energy function will admit a polynomial representation.
A peculiar and possibly unique feature of our scheme is the
induced truncation of the function related to the susceptibility, which
turns out to be an even polynomial of degree $2t$ in the variable $\theta$.
App.\ \ref{rhotheta} contains a more extended discussion of these
and other properties of the approximation scheme based on the stationarity
condition.

We have introduced our parametric representation assuming independent
knowledge of $F_5$, ..., $F_{2t-1}$.  It should be noticed that our
results for $t=3$ can also be used as a phenomenological
parametrization, fitting the value of $F_5$ on any known universal
quantity.  As we will show in Sec.\ \ref{eqstres}, the difference with
the linear parametric model of Refs.\ 
\cite{Schofield-69,S-L-H-69,Josephson-69} is not negligible.  On the
other hand, our numerical estimates for $t=4$ show that the difference
from $t=3$ is too small (compared with both theoretical and
experimental precision) to justify the introduction of an additional
phenomenological parameter $F_7$.

\subsection{$\epsilon$-expansion of the parametric representation}
\label{epsilon}

It is interesting to compare our results with the analysis of the
parametric equation of state which can be performed in the context of
the $\epsilon$-expansion, generalizing results presented in
Refs.\ \cite{W-Z-74,B-W-W-72}.

According to Ref.\ \cite{W-Z-74}, within the $\epsilon$-expansion it
is possible to choose a value $\rho_0$ such that, for all $n \geq 2$,
\begin{equation}
h_{2n+1}(\rho_0) = O(\epsilon^{n+1}).
\label{heps}
\end{equation}
The calculation shows that $\rho_0 = \sqrt{2}$.  We proved in App.\ 
\ref{rhotheta} that Eq.\ (\ref{heps}) keeps holding for all choices of
$\rho$ that satisfy the relation $\rho=\rho_0+O(\epsilon)$.  We can
now $\epsilon$-expand our globally stationary solutions for arbitrary
$t$, obtaining
\begin{equation}
\displaystyle\lim_{\epsilon\to0} \rho_t = \rho_0.
\end{equation}
As a consequence, any truncation satisfying the stationarity condition
is an accurate description of the $\epsilon$-expanded parametric
equation of state up to $O(\epsilon^t)$ included.

As a byproduct, we may extract from the linear model relation
(\ref{coeff2}), expanded to $O(\epsilon^2)$, the coefficients of
the $\epsilon$-expansion for $F_{2m+1}$, for $m \geq 2$:
\begin{equation}
F_{2m+1} = \sum_{k=1}^{\infty} f_{mk}\epsilon^k.
\end{equation}
We easily obtained from Eq.\ (\ref{coeff2}) the closed form results
\begin{eqnarray}
f_{m1} &=& {(-1)^m \over 6 m(m-1)}{1 \over 2^m},\\
f_{m2} &=&  f_{m1}\left[{17 \over 27} -{m \over 2} -\left(
{m \over 3}+{1 \over 6}\right)\sum_{k=1}^{m-2}{1 \over k}\right],
\end{eqnarray}
reproducing known results \cite{G-Z-97}.
More generally, knowing the expansion of the coefficients $F_{2m+1}$
to $O(\epsilon^t)$ for $m<t$ is enough to reconstruct all $F_{2m+1}$
for $m \geq t$ to the same accuracy.

\subsection{Results}
\label{eqstres}

As input parameters for the determination of the functions
$h^{(t)}(\rho,\theta)$ we use the results of the IHT expansion:
$\gamma=1.2371(4)$, $\nu=0.63002(23)$, $r_6=2.048(5)$, $r_8=2.28(8)$,
$r_{10}=-13(4)$.  

In Table \ref{eqstdet} we report the universal ratios of amplitude as
derived from truncations corresponding to $t=2,3,4,5$.  We use the
standard notation for the ratios of amplitudes (see e.g.\ Ref.\ 
\cite{F-Z-98}); all definitions can be found in Table \ref{eqst}.  For
comparison, we also report, for $t=3,4$, the results obtained using
the procedure of Ref.\ \cite{G-Z-97}, fixing $\rho$ to the value
$\rho_{{\rm m},t}$ which minimizes the absolute value of the
$O(\theta^{2t-1})$ coefficient\footnote{As already noted in Ref.\ 
\cite{G-Z-97}, for $t=4$ the minimum of $h_7(\rho)$ is zero, while,
for $t=3$, $h_5(\rho)$ never reaches zero.}.  Such results are very
close to those derived from the stationarity condition; this is easily
explained by the fact that the values of $\rho_{{\rm m},t}$ are close
to $\rho_t$.
\begin{table}[tbp]
\squeezetable
\caption{
Universal ratios of amplitudes obtained by taking different
approximations of the parametric function $h(\theta)$. Numbers marked
with an asterisk are inputs, not predictions.  The values $\rho_{{\rm
m},t}$ are obtained as in Ref.\ \protect\cite{G-Z-97}, see text for
details.
}
\label{eqstdet}
\begin{tabular}{cr@{}lr@{}lr@{}lr@{}lr@{}lr@{}l}
\multicolumn{1}{c}{}&
\multicolumn{2}{c}{$h^{(2)}(\rho_2,\theta)$}&
\multicolumn{2}{c}{$h^{(3)}(\rho_3,\theta)$}&
\multicolumn{2}{c}{$h^{(4)}(\rho_4,\theta)$}&
\multicolumn{2}{c}{$h^{(5)}(\rho_5,\theta)$}&
\multicolumn{2}{c}{$h^{(3)}(\rho_{{\rm m},3},\theta)$}&
\multicolumn{2}{c}{$h^{(4)}(\rho_{{\rm m},4},\theta)$}\\
\tableline \hline
$\rho$ 
& 1&.7358(12) & 1&.7407(14) & 1&.7289(83) & 1&.686(51)  &
1&.6889(26) & 1&.651(30) \\
$\theta_0^2$ 
& 1&.3606(11) & 1&.3879(29) & 1&.372(12) & 1&.325(53)  &
1&.3310(13) & 1&.295(27) \\
$F^\infty_0$
& 0&.03280(14) & 0&.03382(18) & 0&.03374(21)  & 0&.03366(26)  
& 0&.03378(18) & 0&.03370(23) \\
$|z_0|$ 
& 2&.825(12) & 2&.7937(17) & 2&.7970(33) & 2&.8012(72)  
& 2&.7955(15) & 2&.7992(45) \\
$U_0$ 
& 0&.5222(16) & 0&.5316(21) & 0&.5295(29) & 0&.5261(60) 
& 0&.5303(19) & 0&.5276(39)  \\
$U_2$ 
& 4&.826(11) & 4&.752(15) & 4&.769(22)  & 4&.797(47)  
& 4&.764(13) & 4&.786(30) \\
$U_4$
& $-$9&.737(41) & $-$8&.918(83) & $-$9&.10(18)  & $-$9&.42(48) 
& $-$9&.061(67) & $-$9&.31(28) \\
$R_c^+$ 
& 0&.05538(13) & 0&.05681(16) & 0&.05644(32)  & 0&.0558(10) 
& 0&.05656(14) & 0&.05606(53)  \\
$R_c^-$ 
& 0&.021976(16) & 0&.022488(30) & 0&.02235(11) & 0&.02211(36) 
& 0&.022387(18) & 0&.02220(19)  \\
$R_4^+$ 
& 7&.9789(64) & 7&.804(10) & 7&.823(18) & 7&.847(40) 
& 7&.8146(85) & 7&.836(25) \\
$R_4^-$ 
& 92&.10(23) & 93&.91(20) & 93&.25(45) & 91&.9(1.7) 
& 93&.25(21) & 92&.27(83) \\
$v_3$ 
& 6&.0116(79) & 6&.0561(68) & 6&.041(11) & 6&.010(39) 
& 6&.0412(73) & 6&.018(20) \\
$v_4$ 
& 16&.320(55) & 16&.121(66) & 16&.21(11)  & 16&.41(24) 
& 16&.239(55) & 16&.38(15) \\
$Q_1^{-\delta}$ 
& 1&.6775(19) & 1&.6588(27) & 1&.6624(44) & 1&.668(10)  
& 1&.6611(25) & 1&.6656(60) \\
$U_2\,R_4^+$ 
& 38&.505(82) & 37&.09(15) & 37&.31(26) & 37&.64(55) 
& 37&.23(13) & 37&.50(35) \\
$R_4^+\,R_c^+$ 
& 0&.4419(13) & 0&.4434(13) & 0&.4416(18) & 0&.4377(56) 
& 0&.4420(13) & 0&.4392(29) \\
$r_6$ 
& 1&.9389(48) & $^*$2&.048(5) &  $^*$2&.048(5) &  $^*$2&.048(5) 
& $^*$2&.048(5) &  $^*$2&.048(5) \\
$r_8$
& 2&.507(31) & 2&.402(39) & $^*$2&.28(8) & $^*$2&.28(8) 
& 2&.365(43) & $^*$2&.28(8) \\ 
$r_{10}$ 
& $-$12&.612(41) & $-$12&.146(60) &  $-$10&.0(1.5) & $^*$$-$13&(4) 
& $-$11&.80(10) &  $-$10&.98(86)  
\end{tabular}
\end{table}
The errors reported in Table \ref{eqstdet} are related to the
uncertainty of the corresponding input parameters (considering them as
independent).  The results for $t=2,3,4$ suggest a good convergence
and give a good support to our analysis.  The results for $t=5$,
although perfectly consistent, are less useful to check convergence,
due to the large uncertainty of $F_9$.  In Table \ref{eqst} we report
our final estimates, obtained using $h^{(4)}(\rho_4,\theta)$; all the
approximations reported in Table \ref{eqstdet} are consistent with
them, except $t=2$.

\begin{table}[tbp]
\caption{
Summary of the results obtained in this paper by our high-temperature
calculations (IHT), by using the parametric representation of the
equation of state (IHT-PR), by analyzing the low-temperature expansion
(LT), and by combining the two approaches (IHT-PR+LT).
Notations are explained in App.\ \protect\ref{univra}.
}
\label{eqst}
\begin{tabular}{lr@{}lr@{}lr@{}lr@{}l}
\multicolumn{1}{l}{}&
\multicolumn{2}{c}{IHT}&
\multicolumn{2}{c}{IHT-PR}&
\multicolumn{2}{c}{LT}&
\multicolumn{2}{c}{IHT-PR+LT}\\
\tableline \hline
$\gamma$ & 1&.2371(4) &&  &&  && \\
$\nu$    & 0&.63002(23) &&  &&  &&  \\
$\alpha$    & 0&.1099(7)  &&  &&  &&  \\
$\eta$   & 0&.0364(4)   &&  &&  && \\
$\beta$   & 0&.32648(18)   &&  &&  && \\
$\delta$   & 4&.7893(22)   &&  &&  && \\
$\sigma$ & 0&.0208(12)  &&  &&  && \\
$r_6$    & 2&.048(5)    &&  &&  && \\
$r_8$    & 2&.28(8)     &&  &&  && \\
$r_{10}$ &$-$13&(4)     & $-$10&(2)  && && \\

$U_0\equiv A^+/A^-$ & &&  0&.530(3) && &&\\
$U_2\equiv C^+/C^-$ & &&  4&.77(2) && && \\
$U_4\equiv C^+_4/C^-_4$ &  && $-$9&.1(2) && && \\
$R_c^+\equiv \alpha A^+C^+/B^2 $ &  && 0&.0564(3)&& && \\
$R_c^-\equiv \alpha A^- C^-/B^2 $ &  && 0&.02235(11) &&&& \\

$R_4^+\equiv - C_4^+B^2/(C^+)^3= |z_0|^2$ &  && 7&.82(2) && &&\\
$R_3\equiv v_3\equiv - C_3^-B/(C^-)^2$  &  && 6&.041(11) && &&\\
$R_4^-\equiv C_4^-B^2/(C^-)^3$ &  && 93&.3(5) && &&\\

$v_4\equiv - R_4^- + 3 R_3^2 $ &  && 16&.21(11) && && \\

$Q_1^{-\delta}\equiv R_\chi \equiv C^+ B^{\delta-1}/(\delta C^c)^\delta$ &  && 1&.662(5) && &&\\
$F^\infty_0$ \quad cf.\ Eq.\ (\ref{fasymp}) & &&   0&.0337(2) && &&\\

$g_4^+\equiv g \equiv -C_4^+/[ (C^+)^2 (f^+)^3] $ & 23&.49(4) && && &&  \\

$w^2\equiv C^- /[ B^2 (f^-)^3]$ & &&  && 4&.75(4) \cite{P-V-gr-98} & &  \\ 

$U_\xi\equiv f^+/f^- = \left( w^2 U_2 R_4^+/g_4^+\right)^{1/3}$ &  && && && 1&.961(7) \\
$Q^+ \equiv \alpha A^+ (f^+)^3 =  R_4^+ R_c^+/g_4^+ $  &   && 0&.01880(8) && && \\

$R^+_\xi\equiv (Q^+)^{1/3} $  &  &&  0&.2659(4) && && \\
$Q^- \equiv \alpha A^- (f^-)^3 =  R_c^-/w^2 $  & && && &&  0&.00471(5) \\
$Q_c \equiv B^2(f^+)^3/C^+=Q^+/R_c^+ = R_4^+/g_4^+$ &  && 0&.3330(10) && &&    \\
$g_3^- \equiv w v_3$ &   && && && 13&.17(6) \\
$g_4^- \equiv w^2 v_4$ &  && &&   && 77&.0(8) \\
$Q^+_\xi\equiv f^+_{\rm gap}/f^+$ & 1&.000183(2) && && && \\
$Q^-_\xi\equiv f^-_{\rm gap}/f^-$ & && && 1&.032(4) \cite{C-P-R-V-98} &&  \\
$U_{\xi_{\rm gap}}\equiv f^+_{\rm gap}/f^-_{\rm gap} =U_\xi Q^+_\xi/Q^-_\xi$ & && && && 1&.901(10) \\
$Q^c_\xi\equiv f^c_{\rm gap}/f^c$ & && && && 1&.024(4) \\
$Q_2\equiv (f^c/f^+)^{2-\eta} C^+/C^c$ & && && && 1&.195(10) \\
\end{tabular}
\end{table}

We should say that the method of Guida and Zinn-Justin to determine
the optimal $\rho$ leads to equivalent results, and shows an apparent
good convergence as well.  However, we believe that the global
stationarity represents a more physical requirement, and it is more
amenable to a theoretical analysis of its convergence properties.
Moreover, as we have shown, it has the linear parametric model of
Refs.\ \cite{Schofield-69,S-L-H-69,Josephson-69,W-Z-74} as the
lowest-order approximation. 

Estimates of other universal ratios of amplitudes can be obtained by
supplementing the above results with the estimates of 
$w^2\equiv C^- /[ B^2 (f^-)^3]$ and 
$Q^-_\xi\equiv f^-_{\rm gap}/f^-$ obtained by an analysis of the
corresponding low-temperature expansion.  The results so obtained are
denoted by IHT-PR+LT in Table \ref{eqst}.  The low-temperature
expansion of $w^2$ can be 
calculated to $O(u^{21})$ on the cubic lattice using the series
published in Refs.\ \cite{A-T-95,Vohwinkel-93}. The results reported in
Table \ref{eqst} were obtained by using the Roskies transform in order
to reduce the systematic effects due to confluent
singularities \cite{P-V-gr-98}.

We also consider a parametric representation of the correlation length.
Following Ref.\ \cite{F-Z-U-98} we write
\begin{eqnarray}
&&\xi^2/\chi = R^{-\eta\nu} a(\theta) ,\label{parrepxi} \\
&&\xi_{\rm gap}^2/\chi = R^{-\eta\nu} a_{\rm gap}(\theta) .\label{parrepxig} 
\end{eqnarray}
We consider the simplest polynomial approximation to $a(\theta)$ and
$a_{\rm gap}(\theta)$:
\begin{eqnarray}
&&a(\theta)\approx a_0 \left( 1 + c\theta^2 \right),\label{appxipar} \\
&&a_{\rm gap}(\theta)\approx 
  a_{{\rm gap},0} \left( 1 + c_{\rm gap}\theta^2 \right),\label{appxigpar}
\end{eqnarray}
where the constants $c$ and $c_{\rm gap}$ can be determined by fitting the
quantities $U_\xi$ and $U_{\xi_{\rm gap}}$. Then, using
Eqs.\ (\ref{parrepxi}), (\ref{parrepxig}), and the parametric
representation of the equation of state, one can estimate the
universal ratios of amplitudes $Q_\xi^c$ and $Q_2$ defined in
Table \ref{eqst}.  Notice that, given the equation of state, the
normalization $a_0$ is not arbitrary, but it may be fixed 
using the zero-momentum four-point coupling $g_4$:
\begin{equation}
a_0 = \left( {h_0/\rho}\right)^{1/3} \left( {m_0/\rho}\right)^{-5/3} 
(g_4^*)^{-2/3},
\end{equation}
where $h_0$, $m_0$ and $\rho$ have been introduced in Eqs.\ 
(\ref{parrep}) and (\ref{thzrel}).  Notice that $a_0$ depends only on
the ratios $h_0/\rho$ and $m_0/\rho$, as it is required of a physical
quantity.  Moreover one has $a_{{\rm gap},0} = (Q^+_\xi)^2\;a_0$.  In
order to check the results obtained from the approximate expressions
(\ref{appxipar}) and (\ref{appxigpar}), we also considered the
following parametric representation \cite{T-F-75}:
\begin{eqnarray}
&&\xi^{-2}  = R^{2\nu} b(\theta) ,\label{parrepxi2} \\
&&\xi_{\rm gap}^{-2} = R^{2\nu} b_{\rm gap}(\theta) ,\label{parrepxig2} 
\end{eqnarray}
and the corresponding polynomial approximations truncated to second
order. The results for $Q_\xi^c$ and $Q_2$ obtained by this second
representation are perfectly consistent with those from the first one.
Our final estimates of $Q_\xi^c$ and $Q_2$ derived by the above
method are reported in Table \ref{eqst}.

In Table \ref{summaryeqst} we compare our results with other
approaches. We find a good overall agreement. 
\begin{table}[tbp]
\squeezetable
\caption{
Estimates of the quantities in Table \ref{eqst} by various
approaches. The experimental data are taken from Ref.\
\protect\cite{P-H-A-91}, unless otherwise stated. ms\ denotes a
magnetic system; bm\ a binary mixture; lv\ a liquid-vapor transition
in a simple fluid. 
For values marked with an asterisk, the error is not quoted
explicitly in the reference.
}
\label{summaryeqst}
\begin{tabular}{cr@{}lr@{}lr@{}lr@{}lr@{}llr@{}l}
\multicolumn{1}{c}{}&
\multicolumn{2}{c}{IHT--PR}&
\multicolumn{2}{c}{HT,LT}&
\multicolumn{2}{c}{MC}&
\multicolumn{2}{c}{$\epsilon$-exp.}&
\multicolumn{2}{c}{$d$=3 exp.}&
\multicolumn{3}{c}{experiments}\\
\tableline \hline
$U_0$& 0&.530(3)& 0&.523(9) \cite{L-F-89}& 0&.560(10) \cite{H-P-97}&0&.527(37) \cite{G-Z-98}&
0&.537(19) \cite{G-Z-98} & bm & 0&.56(2) \\
&& & 0&.51$^*$ \cite{B-H-K-75} &0&.550(12) \cite{H-P-97} &0&.524(10) \cite{N-A-85,Bervillier-86}&
0&.540(11) \cite{L-M-S-D-98} & lv & 0&.50(3) \\
&& & & & 0&.567(16) \cite{H-P-97}& & & 0&.541(14) \cite{B-B-M-N-87}
& ms & 0&.51(3) \\
&& & & & & & & & & & lv \cite{H-S-99} & 0&.53$^{+8}_{-7}$ \\
&& & & & & & & & & & lv \cite{S-N-93} & 0&.538(17) \\\hline
$U_2$ &  4&.77(2) & 4&.95(15) \cite{L-F-89} & 4&.75(3) \cite{C-H-97}& 4&.73(16) \cite{G-Z-98} &
4&.79(10) \cite{G-Z-98} & bm & 4&.4(4) \\
      &   &       & 5&.01$^*$ \cite{T-F-75}     &  4&.72(11) \cite{E-S-99}  &  4&.9$^*$ \cite{N-A-85} & 
4&.77(30) \cite{B-B-M-N-87} & lv & 4&.9(2) \\
      &   &       &  &                      &  &                    &  4&.8$^*$ \cite{B-L-Z-74,A-H-76} & 
4&.72(17) \cite{G-K-M-96} & ms & 5&.1(6)\\
      &   &       &  &                      &  & &  && && ms\cite{B-Y-87} & 4&.6(2)\\\hline
$U_4$ &  $-$9&.1(2)& $-$&9.0(3) \cite{Z-L-F-96} & && $-$8&.6(1.5) \cite{G-Z-98} &
$-$9&.1(6) \cite{G-Z-98} & & \\\hline
$R_c^+$ &  0&.0564(3) & 0&.0581(10) \cite{Z-L-F-96}& && 0&.0569(35) \cite{G-Z-98} &
0&.0574(20) \cite{G-Z-98} &  bm & 0&.050(15) \\
&  & & & & && & & 
0&.0594(10) \cite{B-B-M-N-87} &  lv & 0&.047(10) \\\hline
$R_4^+$ & 7&.82(2) & 7&.94(12) \cite{F-Z-98} & & & 8&.24(34) \cite{G-Z-97} & 7&.84$^*$ \cite{G-Z-98} & & \\\hline

$R_3\equiv v_3$ & 6&.041(11) & 6&.44(30) \cite{F-Z-98,Z-L-F-96} & & & 5&.99(5) \cite{P-V-br-99} & 
6&.08(6) \cite{G-Z-98} & & \\
  &  & &  & & & & 6&.07(19) \cite{G-Z-98} & && & \\\hline
$R_4^-$ & 93&.3(5) & 107&(13) \cite{Z-L-F-96,F-Z-98} &  &  &  &  & && &\\\hline
$v_4$ & 16&.21(11) & & &  &  & 15&.8(1.4) \cite{P-V-br-99} & && &\\\hline
$Q_1^{-\delta}$ &  1&.662(5) & 1&.57(23) \cite{Z-F-96,F-Z-98}& && 1&.648(36) \cite{G-Z-98} &
1&.669(18) \cite{G-Z-98} &  bm & 1&.75(30) \\
&  & & & & && 1&.67$^*$ \cite{N-A-85,Bervillier-86}& 
1&.7$^*$ \cite{B-B-M-N-87} &  lv & 1&.69(14) \\\hline
$w^2$ & &  & 4&.75(4) \cite{P-V-gr-98} & 4&.77(3) \cite{C-H-97} & & & 4&.73$^*$ \cite{G-K-M-96} &&  \\
 & &  & 4&.71(5) \cite{Z-L-F-96,Fisher-pv} & & & & & && & \\\hline
$U_\xi$ & 1&.961(7)& 1&.96(1) \cite{L-F-89} & 1&.95(2) \cite{C-H-97}& 1&.91$^*$ \cite{B-L-Z-74} &
2&.013(28) \cite{G-K-M-96} & bm & 1&.93(7)  \\
        &  &        &  1&.96$^*$ \cite{T-F-75} & 2&.06(1) \cite{R-Z-W-94}& & &
&& ms & 1&.92(15)  \\\hline
$Q^+$ & 0&.01880(8) & 0&.0202(9) \cite{B-C-99} & 0&.0193(10) \cite{H-P-97}& 0&.0197$^*$ \cite{B-G-80,Bervillier-86} &
0&.01968(15) \cite{B-B-85} & lv \cite{Edwards-84} & 0&.0174(32) \\
& & & 0&.01880(15) \cite{L-F-89} & & &&  & & & lv\cite{H-S-99} & 0&.023(4) \\\hline
$Q^-$ & 0&.00471(5) & 0&.00477(20) \cite{F-Z-98} & 0&.0463(17)\cite{H-P-97} & & & && & \\\hline
$Q_c$ & 0&.3330(10) & 0&.324(6) \cite{F-Z-98} & 0&.328(5) \cite{C-H-97} & && 0&.331(9) \cite{B-B-M-N-87}  
& bm & 0&.33(5) \\
& &  & & & &  & & & & & lv & 0&.35(4) \\\hline
$g_3^-$ & 13&.17(6) & 13&.9(4) \cite{Z-L-F-96} & 13&.6(5) \cite{Tsypin-97} & 13&.06(12)\cite{P-V-br-99}  & 
&& & \\\hline
$g_4^-$ & 77&.0(8) & 85&$^*$ \cite{Z-L-F-96} & 108&(7) \cite{Tsypin-97} & 75&(7)\cite{P-V-br-99} & && &\\\hline
$Q_\xi^+$ & 1&.000183(2) & 1&.0001$^*$ \cite{F-Z-98} & & & 1&.00016(2) & 
1&.00021(3) && \\\hline
$Q_\xi^c$ & 1&.024(4) & 1&.007(3) \cite{F-Z-98}  & &&  & &&&& \\\hline
$Q_\xi^-$ & & & 1&.032(4) \cite{C-P-R-V-98} & 1&.031(6) \cite{C-H-P-99,A-C-C-H-97} &  & &&&& \\
& & & 1&.037(3) \cite{F-Z-98} & & &  & &&&& \\\hline
$Q_2$ & 1&.195(10) & 1&.17(2) \cite{Z-F-96,F-Z-98} & &&  1&.13$^*$ \cite{B-L-Z-74} &&&& \\
\end{tabular}
\end{table}
Our results appear to substantially improve the estimates of most of
the universal ratios considered.  In Table \ref{summaryeqst} we have
collected results obtained by high-temperature and low-temperature
expansions (HT,LT), Monte Carlo simulations (MC), field-theoretical
methods such as $\epsilon$-expansion and various kinds of expansions
at fixed dimension $d=3$, and experiments.  Concerning the HT,LT
estimates, we mention the recent Ref.\ \cite{F-Z-98} where a review of
such results is presented.  The agreement with the most recent Monte
Carlo simulations is good, especially with the results reported in
Ref.\ \cite{C-H-97}, which are quite precise. However we note that the
estimates of $U_0$ reported in Ref.\ \cite{H-P-97} are slightly larger.
Moreover there is an apparent discrepancy with the estimate of $g_4^-$
of Ref.\ \cite{Tsypin-97}.  It is worth mentioning that the result of
Ref.\ \cite{E-S-99} was obtained simulating a four-dimensional $SU(2)$
lattice gauge model at finite temperature, whose phase transition is
expected to be in the $3d$ Ising universality class.
Field-theoretical estimates are in general less precise, although
perfectly consistent. We mention that the results denoted by ``$d=3$
exp.'' are obtained from different kinds of expansions: $g$-expansion
\cite{G-Z-97,G-Z-98,B-B-M-N-87,B-B-85}, minimal renormalization
without $\epsilon$-expansion \cite{L-M-S-D-98,S-D-89}, expansion in
the coupling $u\equiv 3 w^2$ defined in the low-temperature phase
\cite{G-K-M-96}.  In Refs.\ \cite{G-Z-97,G-Z-98} Guida and Zinn-Justin
used the $d=3$ $g$- and $\epsilon$-expansion to calculate the
small-field expansion of the effective potential and a parametric
representation of the critical equation of state.  We also mention the
results (not included in this table) of an approach based on
approximate solution of exact renormalization equations (see e.g.\ 
Refs.\ \cite{Morris-97,T-W-94,B-T-W-96}).  Some results can be found
in Ref.\ \cite{B-T-W-96}: $U_2\cong4.29$, $g_3^-\cong15.24$,
$Q_1^{-\delta}\cong1.61$ and $U_\xi\cong1.86$.

We report in Table \ref{summaryeqst} experimental results for three
interesting physical systems exhibiting a critical point belonging to
the $3d$ Ising universality class: binary mixtures, liquid-vapor
transitions and uni-axial antiferromagnetic systems. A review of
experimental data can be found in Ref.\ \cite{P-H-A-91}. Most of the
results shown in Table \ref{summaryeqst} were reported in Refs.\ 
\cite{G-Z-97,C-H-97}. They should give an overview of the level of
precision reached by experiments.

For sake of comparison, in Table \ref{eqstd2} we report the universal
ratios of amplitudes for the two-dimensional Ising model.  The purely
thermal results are taken from Ref.\ \cite{W-M-T-B-76}, where the
exact two-point function has been written in terms of the solution of
a Painlev\'e equation.  $Q^+$ and $Q^-$ have been computed by us
solving numerically the differential equations reported in Ref.\ 
\cite{W-M-T-B-76}.  The ratios involving amplitudes along the critical
isotherm can be obtained using the results reported in Ref.\ 
\cite{Delfino-98}.  For the quantities that are not known exactly, we
report estimates derived from the high- and low-temperature expansion.
Such estimates are quite accurate and should be reliable because the
leading correction to scaling is analytic, since the subleading
exponent $\Delta$ is expected to be larger than one (see e.g.\ Ref.\ 
\cite{B-F-85} and references therein).  In particular the available
exact calculations \cite{W-M-T-B-76} for the square-lattice Ising
model near criticality have shown only analytic corrections to the
leading power law.  Therefore the traditional methods of series
analysis should work well.

\begin{table}[tbp]
\caption{
Universal ratios of amplitudes for the two-dimensional Ising model.
Since the specific heat diverges logarithmically in the
two-dimensional Ising model, the specific heat amplitudes $A^\pm$ are
defined by $C_H\approx - A^\pm \ln t$.
}
\label{eqstd2}
\begin{tabular}{lr@{}l}
$\gamma$ &  7&/4 \\
$\nu$ &  1& \\
$U_0\equiv A^+/A^-$ & 1&  \\
$U_2\equiv C^+/C^-$ & 37&.69365201    \\
$R_c^+\equiv A^+C^+/B^2 $ & 0&.31856939 \\
$R_c^-\equiv A^- C^-/B^2 $ & 0&.00845154 \\
$Q_1^{-\delta}\equiv R_\chi \equiv C^+ B^{\delta-1}/(\delta C^c)^\delta$ & 6&.77828502  \\
$w^2\equiv C^- /[ B^2 (f^-)^2]$ &  0&.53152607 \\ 
$U_\xi\equiv f^+/f^- $ &  3&.16249504 \\
$U_{\xi_{\rm gap}}\equiv f^+_{\rm gap}/f^-_{\rm gap}$ & 2& \\
$Q^+ \equiv A^+ (f^+)^2$  &  0&.15902704  \\
$Q^- \equiv A^- (f^-)^2 $  &  0&.015900517 \\
$Q^+_\xi\equiv f^+_{\rm gap}/f^+$ & 1&.000402074 \\
$Q^c_\xi\equiv f^c_{\rm gap}/f^c$&  1&.0786828  \\
$Q^-_\xi\equiv f^-_{\rm gap}/f^-$&  1&.581883299 \\
$Q_2\equiv (f^c/f^+)^{2-\eta} C^+/C^c$ &  2&.8355305\\\hline 
$g_4^+\equiv g \equiv -C_4^+/[ (C^+)^2 (f^+)^2] $ & 14&.694(2) \cite{B-C-96,P-V-gr-98}  \\
$r_6$ & 3&.678(2) \cite{P-V-ef-98} \\
$r_8$ & 26&.0(2) \cite{P-V-ef-98} \\
$r_{10}$ & 275&(15) \cite{P-V-ef-98} \\
$v_3\equiv - C_3^-B/(C^-)^2$  & 33&.011(6) \cite{Z-L-F-96,P-V-br-99} \\
$v_4 \equiv -C_4^-B^2/(C^-)^3 + 3 v_3^2 $ & 48&.6(1.2) \cite{P-V-br-99} \\
\end{tabular}
\end{table}

\acknowledgments

We gratefully acknowledge useful discussions with Riccardo Guida and 
Alan Sokal, and
correspondence with Victor Mart\'{\i}n-Mayor and Giorgio Parisi.
We thank Martin Hasenbusch for making available to us
the estimate of $D^*$ for the spin-1 model,
that allowed us to revise our original manuscript and
improve the final results.

\appendix

\section{Generation and analysis of the high-temperature expansion for
improved Hamiltonians}
\label{seriesanalysis}

\subsection{Definitions}
\label{series}

Before discussing the series computation, let us define all the
quantities we are interested in and fix the notation.

Starting from the two-point function 
$G(x)\equiv \langle \phi(0) \phi(x)\rangle$, 
we define its spherical moments 
\begin{equation}
m_{2j} = \sum_x (x^2)^j G(x)
\end{equation}
($\chi\equiv m_0$) and the first non-spherical moments
\begin{equation}
q_{4,2j} = \sum_x (x^2)^j \left[ x^4 - \case{3}{5}(x^2)^2\right]  G(x)
\label{q4}
\end{equation}
(where $x^n \equiv \sum_i x^n_i$).

The second-moment correlation length is defined  by
\begin{equation}
\xi^2 \equiv M^{-2} =  {m_2\over 6 \chi}.
\end{equation}
The coefficients $c_i$ of the low-momentum expansion of the function
$g(y)$ introduced in Sec.\ \ref{twopointf} can be related to the
critical limit of appropriate dimensionless ratios of spherical
moments, or of the corresponding weighted moments
$\overline{m}_{2j}\equiv m_{2j}/\chi$.  Introducing the quantities
\begin{equation}
u_{2j}={1\over (2j+1)!} \, \overline{m}_{2j}M^{2j},
\label{vj}
\end{equation}
one can define combinations of $u_{2j}$ (that we will still call $c_i$
to avoid introducing new symbols) having $c_i$ as critical limit:
\begin{eqnarray}
c_2&=& 1-u_4 + \case{1}{20} M^2  ,\label{c2j}\\
c_3&=& 1-2u_4+u_6 - \case{1}{840}M^4,\label{c3j}\\
c_4&=& 1-3u_4+  u_4^2 +2u_6 - u_8 + \case{1}{4725}M^4 + 
\case{1}{60480}M^6  ,
\label{c4j}
\end{eqnarray}
etc... Notice that the terms proportional to powers of $M^2$ do not
contribute in the critical limit $M\rightarrow 0$, but they allow us
to define improved estimators \cite{C-P-R-V-98}.  Indeed in the
lattice Gaussian limit, defined by the two-point function
\begin{equation}
\widetilde{G}(k) = {1\over \hat{k}^2 + M^2},\qquad\qquad \hat{k}^2 
= \sum_i 4 \sin^2 (k_i/2),
\end{equation}
$c_i=0$ independently of $M$, and not only in the critical limit
$M\rightarrow 0$.

The zero-momentum connected Green's functions are defined by
\begin{equation}
\chi_{2j} = \sum_{x_2,...,x_{2j}}\langle \phi(0) \phi(x_2)
...\phi(x_{2j-1}) \phi(x_{2j})\rangle_c ;
\end{equation}
in particular, $\chi_2\equiv \chi$.

\subsection{Linked cluster expansion}
\label{LCE}

We computed the high-temperature expansion by the linked cluster
expansion (LCE) technique.  A general introduction to the LCE can be
found in Ref.\ \cite{Wortis}.  We modeled the application
of the LCE to ${\rm O}(N)$-symmetric models after Ref.\ \cite{L-W-88}.

In order to perform the LCE for the most general model described by
the Hamiltonian (\ref{hamiltonian}), we parametrize the potential
$V(\phi^2)$ in terms of the ``single-site moments'' \cite{L-W-88}
\begin{equation}
\label{m02k}
\mpalla_{2k} = 
{\Gamma\!\left({\textstyle{1\over2}} N\right) \over 
 2^k\Gamma\!\left({\textstyle{1\over2}} N + k\right)}\,
{J_{N-1+2k} \over J_{N-1}}\,, \quad
J_k = \int_0^\infty dx\,x^k \exp\!\left(-V(x^2)\right).
\end{equation}
We compute our series for fixed $N$, leaving all $\mpalla_{2k}$ as
free parameters: each term of the series is a polynomial in
$\mpalla_{2k}$ with rational coefficients.

With the aim of computing as many terms of the series as possible, we
adopted all the technical developments of Ref.\ \cite{Reisz-95a}, and
we introduced more improvements of our own; in this Appendix, we will
only describe these; readers not familiar with technical details of
the LCE should consult Refs.\ \cite{L-W-88,Reisz-95a}.

As discussed in Ref.\ \cite{Reisz-95a}, Sec.\ 3, the LCE requires a
unique representation of graphs; it is convenient to implement this by
defining a canonical form for the incidence matrix.  The reduction to
canonical form of a graph with $V$ vertices requires in principle the
comparison of the $V!$ incidence matrices obtained by permutation of
vertices, which is clearly unmanageable for large graphs; even with
the introduction of the ``extended vertex ordering'' of Ref.\ 
\cite{Reisz-95a}, this operation remains the dominant factor in the
computation time; therefore, we devoted a large effort to the
optimization of this aspect of the computation.  On one hand, we have
perfected the extended vertex ordering, and we are able to recognize
inequivalent vertices much more often.  On the other hand, we search
for (a subgroup of) the symmetry group of the incidence matrix, which
allows us to perform explicitly only one permutation for each
equivalence class.  Altogether, the largest sets of vertices which are
explicitly permuted are of size 5 or less (except for a few hundred
diagrams requiring 6, and a handful requiring permutations of 7 or 8
elements).

The next most computer-intensive operation is the computation of
embedding numbers and color factors; it is optimized by
``remembering'' each computed value in a table, compatibly with
available memory.  This is crucial for color factors, which are
computed recursively, and very effective for embedding numbers.

The problem of handling integer and rational quantities which do not
fit into machine precision is solved by using the GNU multiprecision
(gmp) library.  Neither multiprecision nor polynomials in
$\mpalla_{2k}$ are necessary for the most expensive sections of the
computation; therefore they have little impact on the computation
time.

In order to speed up the handling of search and insertion into ordered
sets of data, we make extensive use of AVL trees (height-balanced
binary trees) (cfr.\ e.g.\ Ref.\ \cite{Knuth}, Chapt.\ 6.2.3).  AVL
trees are used to manipulate graph sets, multivariate polynomials,
and tables of embedding numbers and of color factors.

The LCE is dramatically simplified by restricting actual computations 
to the set of one-particle irreducible graphs.  One must however
establish the relationship between the usual moments and
susceptibilities and their irreducible parts.  To this purpose, we
found it convenient to define a generating functional of irreducible
moments (irreducible momentum-space 2-point functions):
\begin{equation}
G^{\rm 1pi}(\vec p\,) = 
\sum_{\vec x} \exp(i \vec p\cdot\vec x) \, G^{\rm 1pi}(\vec x),
\end{equation}
where $G^{\rm 1pi}(\vec x)$ is the irreducible-graph contribution to
the field-field correlation.  One may then prove the relationship
\begin{equation}
\bigl[G(\vec p\,)\bigr]^{-1} = 
\bigl[G^{\rm 1pi}(\vec p\,)\bigr]^{-1} - 
2\beta \sum_{i=1}^d \cos p_i .
\label{G1pi}
\end{equation}
By expanding both sides of Eq.\ (\ref{G1pi}) in powers of $p^2_i$, it
is trivial to establish all desired relationships, both for spherical
and for non-spherical moments.

We have calculated $\chi$ and $m_2$ to 20th order, and the other
moments of the two-point functions to 19th order.  We have calculated
$\chi_4$ to 18th order, $\chi_6$ to 17th order, $\chi_8$ to 16th
order, and $\chi_{10}$ to 15th order.  Using Eqs.\ (\ref{grdef}),
(\ref{r6gr}), (\ref{r8gr}), and (\ref{r10gr}), one can obtain the
HT series corresponding to zero-momentum four-point
coupling $g_4$ and the quantities $r_{2j}$ we have introduced to
parametrize the effective potential.

It is useful to factorize out the leading dependence on $\beta$:
\begin{equation}
{\cal O} = \beta^r \sum_{i=0}^n a_i \beta^i;
\label{series-normalization}
\end{equation}
the values of $r$ and $n$ are summarized in Table \ref{series-summary}.
In the following we will analyze the series
normalized to start with $O(\beta^0)$, i.e.\ $a_0 + \beta a_1 +...$

\begin{table}[tbp]
\caption{Summary of normalization and length of our IHT series.
}
\label{series-summary}
\begin{tabular}{ccl}
\multicolumn{1}{c}{$\cal O$}&
\multicolumn{1}{c}{$r$}&
\multicolumn{1}{c}{$n$} \\
\tableline 
$\chi$ & 0 & 20 \\
$\xi^2$ & 1 & 19 \\
$g_4$ & $-{3\over2}$ & 17 \\
$r_6$ & 0 & 17 \\
$r_8$ & 0 & 16 \\
$r_{10}$ & 0 & 15 \\
$c_2$ & 4 & 13 \\
$c_3$ & 3 & 13 \\
$c_4$ & 2 & 13 \\
\end{tabular}
\end{table}

We have checked our series for $\chi_{2n}$ and $m_2$ against the
available series of the standard Ising model (see e.g.\ Refs.\ 
\cite{B-C-97,B-C-g-97}); in this special case our only new result is
the 18th order coefficient of the expansion of $\chi_4$:
\begin{equation}
a_{18}(\chi_4)=-{171450770247965944104542584\over 32564156625}.
\end{equation}
We have checked the (new) series for $m_4$ and $m_6$ by changing
variables to $v=\tanh\beta$ and verifying that all coefficients
become integer numbers.

It would be pointless to present here the full results for an
arbitrary potential: the resulting expressions are only fit for
further computer manipulation.  For the three potentials we are
interested in, we computed $\mpalla_{2k}$ by numerical integration (to
32 digit precision or higher).  The coefficients $a_i$ of the HT
series for $\lambda_6=0$ and $\lambda_4=1.1$ and for $\lambda_6=1$ and
$\lambda_4=1.9$ are reported in Tables \ref{seriesl60} and
\ref{seriesl61} respectively.  The series for the spin-1 model defined
by Eq.\ (\ref{spin1action}), with $D=0.641$, are reported in
Table \ref{seriesspin1}.

\begin{table}[tbp]
\squeezetable
\caption{IHT series for $\lambda_4 = 1.1$, $\lambda_6 = 0$,
in the notation of Eq.\ (\ref{series-normalization}).
}
\label{seriesl60}
\begin{tabular}{ccccc}
$i$ & $\chi$ & $m_2$ & $m_4$ & $m_6$ \\
\tableline 
0  & 0.5308447611308816674  &  0  &  
     0  &  0 \\ 
1  & 1.6907769625206168952  &  1.6907769625206168952  &  
     1.6907769625206168952  &  1.6907769625206168952 \\ 
2  & 4.8619910172171602715  &  10.770481113538254496  &  
     28.721282969435345322  &  86.163848908306035967 \\ 
3  & 13.927143445449825611  &  48.231864289335148202  &  
     219.75546850876176116  &  1168.8527451895890195 \\ 
4  & 38.903779467013842036  &  188.05200645833592066  &  
     1229.8912709035843796  &  9662.7116928978662527 \\ 
5  & 108.53608309082300208  &  675.87104272683156690  &  
     5832.6111092485591617  &  61434.000941321774177 \\ 
6  & 299.26769419241406575  &  2309.0702869335329515  &  
     24922.163253442953617  &  332091.05589205489865 \\ 
7  & 824.59140866646260666  &  7605.2927810004027352  &  
     99136.392920539417813  &  1607441.8553355166225 \\ 
8  & 2256.9464691160956608  &  24394.253637363046531  &  
     374199.26312351789406  &  7181423.6410152125984 \\ 
9  & 6174.4168460205032479  &  76627.209216394784414  &  
     1356978.6661855495156  &  30189095.008610024020 \\ 
10 & 16819.879385593690953  &  236799.07667260657765  &  
     4767392.0099162457155  &  120979394.99942695905 \\ 
11 & 45803.222727034040360  &  721928.56294418965891  &  
     16324368.332600723787  &  466443860.83571669282 \\ 
12 & 124363.26835432977776  &  2176629.7412550662641  &  
     54721422.909210908059  &  1742031434.2158738860 \\ 
13 & 337573.74963124787949  &  6500509.9999124014627  &  
     180180889.59217314660  &  6334414835.2189134577 \\ 
14 & 914347.51881247417342  &  19258195.825678901432  &  
     584286774.86828839115  &  22515209536.395375901 \\ 
15 & 2476042.0363235919004  &  56653368.820414865245  &  
     1869882856.0802601457  &  78474598792.492084546 \\ 
16 & 6694111.3933182426254  &  165647959.63348214176  &  
     5915646880.4853995649  &  268883672555.38373998 \\ 
17 & 18094604.163418613844  &  481713425.26839816320  &  
     18526280962.623843359  &  907568914837.87389317 \\ 
18 & 48847832.893538572297  &  1394159442.3995129568  &  
     57500007423.420861038  &  3022858145406.1028395 \\ 
19 & 131848611.02423050678  &  4017559436.5586218326  &  
     177034534120.13444060  &  9949496764882.0179674 \\ 
20 & 355511932.47075480765  &  11532862706.754267638  &  
       &   \\ 
\hline
\hline
$i$ & $\chi_4$ & $\chi_6$ & $\chi_8$ & $\chi_{10}$ \\
\hline
0  & $-$0.3285640660980093563  &  1.0162413020868264428  &  
     $-$6.8394260743547676540  &  79.348906365011205720 \\ 
1  & $-$4.1859963164157358115  &  25.898050097125655128  &  
     $-$286.46285381596139097  &  4926.0498824266722100 \\ 
2  & $-$30.741724996686361292  &  334.99400966656728040  &  
     $-$5647.7911946128472205  &  136420.74020456605398 \\ 
3  & $-$176.53992137927974557  &  3085.9534765513322166  &  
     $-$75010.539622248383529  &  2448228.6617818973385 \\ 
4  & $-$873.19795113342604018  &  22962.676420970454675  &  
     $-$772240.19869798230118  &  33066114.972427672856 \\ 
5  & $-$3914.7539681033628945  &  147391.51074600633293  &  
     $-$6637987.3112014615524  &  364115447.75387624436 \\ 
6  & $-$16340.897140343714854  &  848040.97611583569625  &  
     $-$49819302.751247368336  &  3432806859.5766444770 \\ 
7  & $-$64653.470284596876205  &  4484314.0227663949556  &  
     $-$336218289.35412445864  &  28624021619.321695067 \\ 
8  & $-$245234.96688614286659  &  22168058.403461618364  &  
     $-$2082668486.5197392127  &  215985650524.12674877 \\ 
9  & $-$899257.84241359785957  &  103719954.32133484536  &  
     $-$12019618575.926765056  &  1499822923734.9247399 \\ 
10 & $-$3206654.1029660245146  &  463530204.01159412237  &  
     $-$65362330202.998798196  &  9707913378518.1866160 \\ 
11 & $-$11170819.137408319309  &  1992634695.4221763266  &  
     $-$337846725947.59204461  &  59157394345479.522579 \\ 
12 & $-$38148679.051940544866  &  8285182135.0991054145  &  
     $-$1671339685352.6240713  &  342085313031114.78303 \\ 
13 & $-$128071730.18983471414  &  33466763808.875455279  &  
     $-$7957523515361.0083551  &  1889271366312617.3703 \\ 
14 & $-$423602975.57444761466  &  131799942528.27775884  &  
     $-$36629909644439.962983  &  10018120044594804.748 \\ 
15 & $-$1382909952.8269485885  &  507558776082.59290672  &  
     $-$163635571877509.11631  &  51230076380872446.008 \\ 
16 & $-$4462746050.5347000940  &  1915992506452.6475673  &  
     $-$711670345253605.66031  &   \\ 
17 & $-$14253923929.146690502  &  7104558940304.9779228  &  
  &  \\
18 & $-$45107295178.923296542  &    &  
  &  \\
\end{tabular}
\end{table}

\begin{table}[tbp]
\squeezetable
\caption{IHT series for $\lambda_4 = 1.9$, $\lambda_6 = 1$,
in the notation of Eq.\ (\ref{series-normalization}).
}
\label{seriesl61}
\begin{tabular}{ccccc}
$i$ & $\chi$ & $m_2$ & $m_4$ & $m_6$  \\
\tableline 
0  & 0.4655662671465330507  &  0  &  
     0  &  0 \\ 
1  & 1.3005116946285418792  &  1.3005116946285418792  &  
     1.3005116946285418792  &  1.3005116946285418792 \\ 
2  & 3.2867727127185864813  &  7.2656925005834656853  &  
     19.375180001555908494  &  58.125540004667725483 \\ 
3  & 8.2759367806706747301  &  28.571904795057895903  &  
     130.05174486699400177  &  691.57352659837378755 \\ 
4  & 20.312529070243777317  &  97.872800052605172528  &  
     638.95885613696143199  &  5016.1872942503349323 \\ 
5  & 49.794483203882242272  &  309.06347996559814197  &  
     2661.2142140348739306  &  27994.182028926323346 \\ 
6  & 120.62506581503040408  &  927.75225884375746610  &  
     9988.4118850888212117  &  132873.57808445335775 \\ 
7  & 292.01512515747113432  &  2684.8185172507997345  &  
     34904.461272349558702  &  564842.72150668529468 \\ 
8  & 702.16648194884216675  &  7566.3506306935978232  &  
     115746.99820735924226  &  2216511.1678444057633 \\ 
9  & 1687.6428772292658916  &  20882.094002198352809  &  
     368765.50096064887178  &  8184889.2748500168622 \\ 
10 & 4038.7968304000570954  &  56696.707453278695709  &  
     1138240.4288922759401  &  28813837.691500628877 \\ 
11 & 9662.2729936573649464  &  151863.26982182684635  &  
     3424267.0781090006844  &  97595910.383064147131 \\ 
12 & 23047.113364799224307  &  402271.59171959125975  &  
     10084793.611992545198  &  320214529.50558023009 \\ 
13 & 54959.286082391555970  &  1055489.7357595885681  &  
     29174001.507844433518  &  1022942183.3309332937 \\ 
14 & 130774.53710516575464  &  2747201.8540323841754  &  
     83116951.940376136421  &  3194364390.8578671172 \\ 
15 & 311110.31236133784486  &  7100121.4902629495116  &  
     233696696.59215911285  &  9781470144.1262582176 \\ 
16 & 738903.79800753751923  &  18238478.826399953589  &  
     649550962.77873855718  &  29444730149.930149105 \\ 
17 & 1754633.3818777485475  &  46596282.414619468624  &  
     1787187905.0885303101  &  87315557903.862730209 \\ 
18 & 4161219.5431591307916  &  118476705.93261076006  &  
     4873253901.0331542961  &  255504682223.76959562 \\ 
19 & 9867152.2571694621736  &  299943150.21579862108  &  
     13181855628.291951390  &  738841191899.23449896 \\ 
20 & 23372660.203375142541  &  756429264.37368967452  &  
       &   \\ 
\hline
\hline
$i$ & $\chi_4$ & $\chi_6$ & $\chi_8$ & $\chi_{10}$ \\
\hline
0  & $-$0.2477796481363361878  &  0.6474598784982264035  &  
     $-$3.7217906767025756640  &  37.013405312193356664 \\ 
1  & $-$2.7685883005851708908  &  14.535362477178409902  &  
     $-$137.07512921231932765  &  2014.8224919025960131 \\ 
2  & $-$17.877560582981223820  &  165.70820207079800267  &  
     $-$2380.8220701405889901  &  49093.219551382446508 \\ 
3  & $-$90.322805654272940178  &  1345.3663489182523069  &  
     $-$27871.670743361006737  &  776123.99513244411557 \\ 
4  & $-$393.02065426992610848  &  8820.7362245901247130  &  
     $-$252935.23058597306885  &  9238516.9347113237042 \\ 
5  & $-$1549.8790048885508754  &  49871.943864090224636  &  
     $-$1916250.8876237147160  &  89671059.994338257912 \\ 
6  & $-$5689.6695707018226942  &  252682.48564411688205  &  
     $-$12673128.761295226766  &  745143260.15631620580 \\ 
7  & $-$19795.290447377859541  &  1176295.0381597851008  &  
     $-$75349625.332312049631  &  5475799689.5566930641 \\ 
8  & $-$66017.086587684140210  &  5118124.4639866386541  &  
     $-$411110139.91029709454  &  36408447077.502609221 \\ 
9  & $-$212822.79826473961810  &  21072833.518843848267  &  
     $-$2089380048.3140434305  &  222743859156.05535943 \\ 
10 & $-$667124.90202267593913  &  82859426.257839273325  &  
     $-$10003697691.834647146  &  1270011540667.3854746 \\ 
11 & $-$2042817.2956666846385  &  313350700.53757124062  &  
     $-$45518090240.849158501  &  6816107886501.6215858 \\ 
12 & $-$6131769.6806946007770  &  1146009747.0203113140  &  
     $-$198194900838.32752342  &  34708877742973.050540 \\ 
13 & $-$18092493.782248051554  &  4071313753.6304055865  &  
     $-$830439274834.30354664  &  168778204360271.32357 \\ 
14 & $-$52592383.442563299219  &  14100289090.570869350  &  
     $-$3363665155462.8191575  &  787889916506321.20775 \\ 
15 & $-$150889630.61165148373  &  47747821664.976757633  &  
     $-$13220617132779.560155  &  3546559365689180.8887 \\ 
16 & $-$427911822.28715837917  &  158482857336.16489875  &  
     $-$50583038860319.049950  &    \\ 
17 & $-$1201047651.7022285867  &  516675010346.30966717  &  
  &  \\
18 & $-$3339910306.5273359851  &     &  
  &  \\
\end{tabular}
\end{table}

\begin{table}[tbp]
\squeezetable
\caption{IHT series for the spin-1 model at $D=0.641$,
in the notation of Eq.\ (\ref{series-normalization}).
}
\label{seriesspin1}
\begin{tabular}{ccccc}
$i$ & $\chi$ & $m_2$ & $m_4$ & $m_6$  \\
\tableline 
0  & 0.5130338416658140921  &  0  &  
     0  &  0 \\ 
1  & 1.5792223361663016211  &  1.5792223361663016211  &  
     1.5792223361663016211  &  1.5792223361663016211 \\ 
2  & 4.4354864269385182067  &  9.7223340236143130564  &  
     25.926224062971501484  &  77.778672188914504451 \\ 
3  & 12.419491590434972647  &  42.346809834993589030  &  
     191.98340105778667094  &  1019.9725391572417242 \\ 
4  & 33.791261055416279831  &  160.98579902461367310  &  
     1043.4445460560890431  &  8166.3561104899782974 \\ 
5  & 91.841711514904201711  &  564.00666776024395394  &  
     4815.3070263215122748  &  50394.717412597175181 \\ 
6  & 246.45252550898370464  &  1877.9601316327756011  &  
     20036.716263907461491  &  264816.64944314665340 \\ 
7  & 661.02891687705666766  &  6026.3998915743002481  &  
     77637.764412960183453  &  1247156.1972913768310 \\ 
8  & 1760.2485042097784968  &  18829.052334178552758  &  
     285474.20654850401533  &  5423869.9597014869631 \\ 
9  & 4685.9023162918237050  &  57600.716340522523798  &  
     1008432.9506590528288  &  22201293.939743630952 \\ 
10 & 12417.403239568002677  &  173324.21633191405061  &  
     3450900.8811683331178  &  86641778.653884117489 \\ 
11 & 32897.780017066145091  &  514450.09422679607285  &  
     11508712.418722055563  &  325332157.49685048174 \\ 
12 & 86884.987650751268743  &  1509919.6022647962218  &  
     37570398.804037029475  &  1183310452.1936542300 \\ 
13 & 229424.66023796502223  &  4389267.7826683406013  &  
     120463565.05045829114  &  4190401181.7453112798 \\ 
14 & 604434.19630362500628  &  12656102.176997207352  &  
     380359458.94391796446  &  14504931859.074136829 \\ 
15 & 1592166.9939650416411  &  36234119.703394609364  &  
     1185138499.8785854689  &  49231114687.307644210 \\ 
16 & 4186778.2089201191352  &  103100174.19850927716  &  
     3650157580.5925683557  &  164257098550.90866386 \\ 
17 & 11008100.036706697979  &  291755921.33251277078  &  
     11128154630.539032580  &  539841179613.24402666 \\ 
18 & 28904025.427069749972  &  821639224.91959656471  &  
     33620312241.342947618  &  1750686524137.2717781 \\ 
19 & 75884596.302083003892  &  2303832207.2589619187  &  
     100754983048.94595013  &  5610143692354.6457755 \\ 
20 & 199011100.35574405792  &  6434727599.1288912159  &  
       &   \\ 
\hline
\hline
$i$ & $\chi_4$ & $\chi_6$ & $\chi_8$ & $\chi_{10}$ \\
\hline
0  & $-$0.2765773264173367185  &  0.6159505110893571460  &  
     $-$2.9991913236448423261  &  25.639540414004216349 \\ 
1  & $-$3.4054446789491065718  &  15.965825500576724169  &  
     $-$131.09722874868945636  &  1673.3057480542363748 \\ 
2  & $-$24.570159964990933901  &  209.17712832988124972  &  
     $-$2673.1181642081994199  &  48487.693538927944393 \\ 
3  & $-$139.04496175368819963  &  1941.9314061871553985  &  
     $-$36418.744278635469309  &  903571.05878070985095 \\ 
4  & $-$676.69722203050291907  &  14481.161465713991310  &  
     $-$381670.70958440882592  &  12572456.840428388939 \\ 
5  & $-$2980.2850555040833295  &  92716.594093707793992  &  
     $-$3318236.6956045245656  &  141625856.39478899275 \\ 
6  & $-$12201.170328998898251  &  530071.30088368454939  &  
     $-$25054434.002489226202  &  1357653370.1441467390 \\ 
7  & $-$47288.798001661947371  &  2776591.5196504503986  &  
     $-$169364792.32951428214  &  11451941832.665412740 \\ 
8  & $-$175521.23292975181789  &  13563137.626324266820  &  
     $-$1047051748.2193837515  &  87035901992.179110029 \\ 
9  & $-$629296.76861756838811  &  62579486.626554750976  &  
     $-$6012877960.8045829709  &  606512902870.52199480 \\ 
10 & $-$2192523.1579334804197  &  275333754.56012219662  &  
     $-$32454249850.587780052  &  3927249840355.5223369 \\ 
11 & $-$7458569.3308517637566  &  1163638939.7290930525  &  
     $-$166147643973.94469398  &  23876098793430.398990 \\ 
12 & $-$24861185.382016374834  &  4751112394.9532757501  &  
     $-$812614050679.29384404  &  137426012654323.52351 \\ 
13 & $-$81432521.735317351753  &  18826990715.719661290  &  
     $-$3819167751639.5573175  &  753928052790611.90171 \\ 
14 & $-$262699929.49158922336  &  72675832066.845002904  &  
     $-$17330670355620.376973  &  3964188356224407.7875 \\ 
15 & $-$836235938.80657455801  &  274128146369.26786810  &  
     $-$76232334051550.698537  &  20070120847424617.843 \\ 
16 & $-$2630659142.3082426455  &  1012932475761.7342902  &  
     $-$326121067148253.92416  &   \\ 
17 & $-$8189047799.9075180970  &  3674558299661.5473376  &  
  &  \\
18 & $-$25252383565.446882139  &    &  
  &  \\
\end{tabular}
\end{table}

\subsection{Critical exponents}
\label{exponents}

In order to determine the critical exponent $\gamma$ from the $n$th
order series of $\chi$ ($n=20$ in our case) we used quasi-diagonal
first, second and third order integral approximants (IA1's, IA2's and
IA3's respectively).

IA1's are solutions of the first-order linear differential equation
\begin{equation}
P_1(x)f^\prime (x)+ P_0(x)f(x)+R(x)= 0,
\label{IA1def}
\end{equation}
where the functions $P_i(x)$ and $R(x)$ are polynomials that are
determined by the known $n$th order small-$x$ expansion of $f(x)$.  We
considered $[m_1/m_0/k]$ IA1's with
\begin{eqnarray}
&&m_1+m_0+k+2\geq n-p,\nonumber \\
&&{\rm Max}\left[\lfloor (n-p-2)/3 \rfloor -q, 2\right]
  \leq m_1,m_0,k \leq \lceil (n-p-2)/3\rceil +q,
\label{gaapprox1}
\end{eqnarray}
where $m_1,m_0,k$ are the orders of the polynomial $P_1$, $P_0$ and
$R$ respectively.  The parameter $q$ determines the degree of
off-diagonality allowed.  Since the best approximants are expected to
be those diagonal or quasi-diagonal, we considered sets of
approximants corresponding to $q=3$. For a given integer number $p$,
only approximants using $\bar{n}$ terms with $n\geq \bar{n}\geq n-p$
are selected by (\ref{gaapprox1}). In our analysis we considered the
values $p=0,1$.

IA2's are solutions of the second-order linear differential equation
\begin{equation}
P_2(x) f^{\prime\prime}(x) + P_1(x)f^\prime (x)+ P_0(x)f(x)+R(x)= 0.
\end{equation}
We considered $[m_2/m_1/m_0/k]$ IA2's  with 
\begin{eqnarray}
&&m_2+m_1+m_0+k+4\geq n-p,\nonumber \\
&&{\rm Max}\left[\lfloor (n-p-4)/4 \rfloor -q, 2\right]\leq m_2,m_1,m_0,k 
    \leq \lceil (n-p-4)/4\rceil +q,
\label{gaapprox}
\end{eqnarray}
where $m_2,m_1,m_0,k$ are the orders of the polynomial $P_2$, $P_1$, 
$P_0$ and $R$ respectively.
Again the parameter $q$ determines the degree of off-diagonality
allowed, and  we consider sets of approximants 
corresponding to $q=2$. 

IA3's are solutions of the third-order linear differential equation
\begin{equation}
P_3(x) f^{\prime\prime\prime}(x) +
P_2(x) f^{\prime\prime}(x) +  P_1(x)f^\prime (x)+ P_0(x)f(x)+R(x)= 0.
\end{equation}
We considered $[m_3/m_2/m_1/m_0/k]$ IA3's  with 
\begin{eqnarray}
&&m_3+m_2+m_1+m_0+k+6\geq n-p,\nonumber \\
&&{\rm Max}\left[\lfloor (n-p-6)/5 \rfloor -q, 2\right]
  \leq m_3,m_2,m_1,m_0,k \leq \lceil (n-p-6)/5\rceil +q.
\label{gaapprox3}
\end{eqnarray}
We considered sets of approximants with $q=2$. 

As estimate of $\beta_c$ and $\gamma$ from each set of IA's, we
took the average of the values corresponding to all non-defective
approximants listed above.  The error we quote is the standard
deviation.  Approximants are considered defective
when they present spurious singularities close to the real axis for
${\rm Re}\ \beta \lesssim \beta_c$.  More precisely we considered
defective the approximants with spurious singularities in the
rectangle
\begin{equation}
x_{\rm min} \leq {\rm Re} z \leq x_{\rm max},\qquad 
| {\rm Im} z | \leq y_{\rm max}
\label{filter}
\end{equation}
where $z\equiv \beta/\beta_c$.  The special values of $x_{\rm min}$,
$x_{\rm max}$ and $y_{\rm max}$ are fixed essentially by 
stability criteria, and may differ in the various analysis.  One
should always check that, within a reasonable and rather wide range of
values, the results depend very little on the values of $x_{\rm min}$,
$x_{\rm max}$, and $y_{\rm max}$.  The condition (\ref{filter}) cannot be
too strict, otherwise only few approximants are left. In this case the
analysis would be less robust and therefore less reliable.  In the
analysis of the critical exponents we fixed $x_{\rm min} = 0.5$,
$x_{\rm max} = 1.5$ and $y_{\rm max} = 0.5$.  Sometimes we also
eliminated seemingly good approximants whose results were very far
from the average of the other approximants (more than three standard
deviations).

As a further check of our analysis we used the fact that $\chi$ must
present an antiferromagnetic singularity at 
$\beta_c^{\rm af}=-\beta_c$ with exponent $1-\alpha$
\cite{Fisher-62}, cf.\ Eq.\ (\ref{fisheraf}).  We verified the
existence of a singularity at $\beta\simeq -\beta_c$ and
calculated the associated exponent. In some analyses we selected the
approximants with a pair of singularities $\beta^{\rm af}_c$ and
$\beta_c$ such that 
$\beta_c + \beta^{\rm af}_c\leq \varepsilon \beta_c$, and extracted
the estimates of $\beta_c$, $\gamma$ and $\gamma^{\rm af}$ from them.
As in Ref.\ \cite{F-C-85}, we also considered IA's where the
polynomial associated with the highest derivative of $f(x)$ is even,
i.e.\ it is a polynomial in $x^2$. We will denote them by
b$_{\rm af}$IA's.  This ensures that if $x_c$ is a singularity of an
approximant then $-x_c$ will also be a singular point.

In Table \ref{gammares} we present the results for some values of the
parameters $p,q,\varepsilon$ introduced above (when the value of
$\varepsilon$ is not explicitly shown it means that the corresponding
constraint was not implemented).  We quote the ``ratio of
approximants'' r-app $(l-s)/t$, where $t$ is the total number of
approximants in the given set, $l$ is the number of non-defective
approximants (passing the test (\ref{filter})), and $s$ is the number
of seemingly good approximants which are excluded because their
results are very far from the other approximants; $l-s$ is the number
of ``good'' approximants used in the analysis; notice that $s \ll l$,
and $l-s$ is never too small.  We found the IA2 analysis to give the
most stable results, especially with respect to the change of the
number of terms of the series considered.  Therefore we consider the
IA2 results to be the most reliable.  Moreover, IA1's are not
completely satisfactory in reproducing the antiferromagnetic
singularity when its presence is not biased.  In the biased analyses
where $\beta_c$ is forced, IA1's, IA2's and IA3's give almost
equivalent results.

\begin{table}[tbp]
\squeezetable
\caption{
Results of various analyses of the 20th order IHT series for
$\chi$. r-app is explained in the text.  In the biased analyses
forcing the value of $\beta_c$ the error is reported as a sum: the
first term is related to the spread of the approximants at the central
value of $\beta_c$, while the second one is related to the uncertainty
of the value of $\beta_c$ and it is estimated by varying $\beta_c$.
}
\label{gammares}
\begin{tabular}{clcr@{}lr@{}lr@{}lr@{}l}
\multicolumn{1}{c}{}&
\multicolumn{1}{c}{approx}&
\multicolumn{1}{c}{r-app}&
\multicolumn{2}{c}{$\beta_c$}&
\multicolumn{2}{c}{$\gamma$}&
\multicolumn{2}{c}{$\beta_c^{\rm af}$}&
\multicolumn{2}{c}{$\gamma^{\rm af}$}\\
\tableline \hline
$\lambda_6=0,\;\lambda_4=1.08$
&IA2$_{q=2,p=0}$                       &   $(78-2)/85$ & 0&.3760701(22) & 1&.23717(36) & $-$0&.376(6)  & $-$1&.0(7)  \\
&IA3$_{q=2,p=0}$                       &   $(62-1)/65$ & 0&.3760699(21) & 1&.23713(37) & $-$0&.377(4)  & $-$1&.0(3)  \\
&b$_{\rm af}$IA3$_{q=2,p=0}         $  & $(63-1)/65$   & 0&.3760703(18) & 1&.23719(30) & & & $-$0&.903(10) \\ 

$\lambda_6=0,\;\lambda_4=1.10$
&IA1$_{q=3,p=0}$                       &   $30/37$     & 0&.3750945(27) & 1&.23684(42) & $-$0&.375(3)  & $-$0&.9(3)  \\
&IA1$_{q=3,p=0,\varepsilon=10^{-2}}$   &   $(26-1)/37$ & 0&.3750955(18) & 1&.23699(26) & $-$0&.3760(8) & $-$0&.77(8)  \\
&IA1$_{q=3,p=0,\varepsilon=10^{-3}}$   &   $1/37$      & 0&.3750956     & 1&.23701     & $-$0&.3754    & $-$0&.81  \\
&b$_{\rm af}$IA1$_{q=3,p=0}         $  & $(24-2)/37$   & 0&.3750937(46) & 1&.23673(69) & & & $-$0&.887(11) \\ 

&IA2$_{q=2,p=0}$                       &   $(78-2)/85$ & 0&.3750975(18) & 1&.23734(29) & $-$0&.376(6)  & $-$1&.0(6)  \\
&IA2$_{q=2,p=0,\varepsilon=10^{-2}}$   & $(69-1)/85$   & 0&.3750974(21) & 1&.23733(35) & $-$0&.375(2)  & $-$0&.9(2)  \\
&IA2$_{q=2,p=0,\varepsilon=10^{-3}}$   &   $16/85$     & 0&.3750975(19) & 1&.23735(31) & $-$0&.3751(2) & $-$0&.90(4)\\
&b$_{\rm af}$IA2$_{q=2,p=0}            $  & $(54-1)/85$& 0&.3750989(31) & 1&.23753(46) & & & $-$0&.906(18) \\ 
&IA2$_{q=2,p=1}$                       & $(151-3)/165$ & 0&.3750974(26) & 1&.23731(39) & $-$0&.376(4)  & $-$1&.0(5) \\
&IA2$_{q=2,p=1,\varepsilon=10^{-3}} $  & $25/165$      & 0&.3750975(16) & 1&.23735(25) & $-$0&.3751(2) & $-$0&.90(3) \\

&IA3$_{q=2,p=0}$                       & $(62-1)/65$   & 0&.3750971(21) & 1&.23728(36) & $-$0&.375(4)  & $-$1&.0(5)  \\
&IA3$_{q=2,p=0,\varepsilon=10^{-3}}$   & $21/65$       & 0&.3750983(12) & 1&.23749(19) & $-$0&.3750(2) & $-$0&.92(2)\\
&b$_{\rm af}$IA3$_{q=2,p=0}         $  & $(63-1)/65$   & 0&.3750976(18) & 1&.23734(30) & & & $-$0&.902(11) \\ 
&IA3$_{q=2,p=1}$                       & $(99-3)/100$  & 0&.3750957(55) & 1&.23704(98) & $-$0&.375(3)  & $-$1&.0(4)  \\

&b$_{\beta_c^{\rm MC}}$IA1$_{q=3,p=0}$ & $38/48$ & 0&.3750966(4) \cite{Hasenbusch-99}     & 1&.23718(7+7) & & & \\
&b$_{\beta_c^{\rm MC}}$IA2$_{q=2,p=0}$ & $(114-3)/115$ & 0&.3750966(4) \cite{Hasenbusch-99}     & 1&.23720(2+7) & & & \\
&b$_{\pm\beta_c^{\rm MC}}$IA2$_{q=2,p=0}$ & $(90-2)/100$ & 0&.3750966(4) \cite{Hasenbusch-99}& 1&.23719(3+7) & & & \\
&b$_{\pm\beta_c^{\rm MC}}$IA3$_{q=2,p=0}$ & $(61-1)/63$ & 0&.3750966(4) \cite{Hasenbusch-99}& 1&.23719(8+7) & & & \\
$\lambda_6=0,\;\lambda_4=1.12$
&IA2$_{q=2,p=0}$           &   $(77-2)/85$ & 0&.3741203(16) & 1&.23748(26) & $-$0&.374(4)  & $-$1&.0(6)\\
&IA3$_{q=2,p=0}$           &   $(62-1)/65$ & 0&.3741199(21) & 1&.23742(36) & $-$0&.375(4)  & $-$1&.0(3)\\
&b$_{\rm af}$IA3$_{q=2,p=0}         $  & $(62-1)/65$   & 0&.3741203(17) & 1&.23747(29) & & & $-$0&.904(9) \\\hline 

$\lambda_6=1,\;\lambda_4=1.86$  
& IA2$_{q=2,p=0}$ &  $(82-3)/85$  & 0&.4307605(22) & 1&.23676(30) & $-$0&.431(5) & $-$1&.1(5)  \\
$\lambda_6=1,\;\lambda_4=1.90$  
& IA1$_{q=3,p=0}$ &  $37/37$          & 0&.4269723(81)             & 1&.2363(11)  & $-$0&.429(7) & $-$0&.7(5)  \\
&b$_{\rm af}$IA1$_{q=3,p=0}  $  & $(29-1)/37$   & 0&.426972(7) & 1&.2363(9) & & & $-$0&.890(16) \\ 
& IA2$_{q=2,p=0}$ &  $(83-2)/85$      & 0&.4269779(24)             & 1&.23711(34) & $-$0&.428(5) & $-$1&.0(5)  \\
& IA2$_{q=2,p=0,\varepsilon=10^{-3}}$ &  $23/85$  & 0&.4269779(32) & 1&.23710(45) & $-$0&.4271(2) & $-$0&.90(3)  \\
&b$_{\rm af}$IA2$_{q=2,p=0}$ &  $(69-2)/85$  & 0&.4269791(32) & 1&.23725(48) & & & $-$0&.903(16)  \\
& IA2$_{q=2,p=1}$ &  $(156-4)/165$    & 0&.4269777(32)             & 1&.23707(42) & $-$0&.428(5) & $-$0&.9(5)  \\
& IA2$_{q=2,p=1,\varepsilon=10^{-3}}$ &  $31/165$  & 0&.4269777(30)& 1&.23708(41) & $-$0&.4271(2) & $-$0&.89(4)  \\
& IA3$_{q=2,p=0}$ &  $(63-1)/65$      & 0&.4269782(24)             & 1&.23714(38) & $-$0&.429(6) & $-$1&.0(4)  \\
&b$_{\rm af}$IA3$_{q=2,p=0}$ &  $(63-1)/65$  & 0&.4269786(25) & 1&.23719(36) & & & $-$0&.906(10)  \\
$\lambda_6=1,\;\lambda_4=1.94$  
& IA2$_{q=2,p=0}$ &  $(82-3)/85$  & 0&.4232606(21) & 1&.23738(30) & $-$0&.423(6) & $-$1&.1(9)  \\\hline

spin-1, $D=0.633$ 
& IA2$_{q=2,p=0}$ &$(82-4)/85$ & 0&.3845065(27) & 1&.23683(34) & $-$0&.384(8) & $-$1&.1(1.2)\\
& b$_{\rm af}$IA2$_{q=2,p=0}$ &$(73-4)/85$ & 0&.3845076(17) & 1&.23698(25) & & & $-$0&.905(12)\\

spin-1, $D=0.641$ 
& b$_{\rm af}$IA1$_{q=3,p=0}$ &$(30-3)/37$ & 0&.3856634(41) & 1&.2360(6) & & & $-$0&.887(11)\\
& IA2$_{q=2,p=0}$ &$(73-2)/85$ & 0&.3856681(33) & 1&.23674(38) & $-$0&.386(3) & $-$0&.9(4)\\
& IA2$_{q=2,p=0,\varepsilon=10^{-3}}$ &$(37-2)/85$ & 0&.3856685(17) & 1&.23678(23) & $-$0&.3860(3) & $-$0&.87(3)\\
&b$_{\rm af}$IA2$_{q=2,p=0}$  & $(73-4)/85$ & 0&.3856691(16) & 1&.23687(25) & & & $-$0&.905(12)\\
& IA2$_{q=2,p=1}$ &$(146-3)/165$ & 0&.3856669(53) & 1&.23655(69) & $-$0&.385(10) & $-$1&.2(2.0)\\
& IA3$_{q=2,p=0}$ &$(62-5)/65$ & 0&.3856686(36) & 1&.23678(58) & $-$0&.385(4) & $-$1&.1(5)\\
&b$_{\rm af}$IA3$_{q=2,p=0}$  & $(61-2)/65$ & 0&.3856682(38) & 1&.23673(60) & & & $-$0&.911(13)\\

spin-1, $D=0.649$ 
& IA2$_{q=2,p=0}$ &$(82-3)/85$ & 0&.3868365(32) & 1&.23660(36) & $-$0&.386(7) & $-$1&.2(1.0)\\
&b$_{\rm af}$IA2$_{q=2,p=0}$  & $(73-4)/85$ & 0&.3868377(16) & 1&.23676(24) & & & $-$0&.905(12)\\
\end{tabular}
\end{table}

The results are quite stable, and the value of $\gamma^{\rm af}$ is
always consistent with $1-\alpha\simeq 0.89$.  
From the results of Table \ref{gammares}, combining the results of the IA2
and IA3 analyses (selecting the results denoted by IA2$_{q=2,p=0}$, 
b$_{\rm af}$IA2$_{q=2,p=0}$,  IA3$_{q=2,p=0}$
and b$_{\rm af}$IA3$_{q=2,p=0}$) we obtain the following estimates
\begin{equation}
\beta_c = 0.3750973(14),\qquad\gamma = 1.23732(24) 
        \qquad {\rm for} \quad \lambda_6=0,\quad \lambda_4=1.10,
\label{betacl60}
\end{equation}
\begin{equation}
\beta_c = 0.4269780(18),\qquad\gamma = 1.23712(26) 
        \qquad {\rm for} \quad \lambda_6=1,\quad \lambda_4=1.90,
\label{betacl61}
\end{equation}
and
\begin{equation}
\beta_c = 0.3856688(20),\qquad\gamma = 1.23680(30) 
        \qquad {\rm for} \quad \hbox{\rm spin-1},\quad D=0.641.
\label{betacspin1}
\end{equation}
Taking also into account the uncertainty of $\lambda_4^*$ and $D^*$,
we arrive at
the estimates of Table \ref{summaryexp}.  Notice that the value of
$\beta_c$ at $\lambda_4=1.10$ and $\lambda_6=0$ is in agreement with
the Monte Carlo estimate of Ref.\ \cite{Hasenbusch-99}, i.e.\ 
$\beta_c=0.3750966(4)$ (where according to the author the error does
not include possible systematic errors).  From the analysis of the
antiferromagnetic singularity using the b$_{\rm af}$IA's we obtain the
following estimate for $\alpha$:
\begin{equation}
\alpha=0.105(10),  
\end{equation}
which is in good agreement with the much more precise
estimate (\ref{resalpha}) obtained assuming
 hyperscaling.

In order to determine $\nu$ from the analysis of the HT series of
$\xi^2$, we followed the suggestion of Ref.\ \cite{Guttmann-87}, i.e.\ 
to use the estimate of $\beta_c$ derived from the analysis of $\chi$
in order to bias the analysis of the series of $\xi^2$.  We analyzed
the 19th order series of $\xi^2/\beta$ and employed biased integral
approximants (b$_{\beta_c}$IA). For instance, biased IA2's can be
obtained from the solutions of the equation
\begin{equation}
\left( 1 - {x/\beta_c}\right)
P_2(x) f^{\prime\prime}(x) + P_1(x)f^\prime (x)+ P_0(x)f(x)+R(x)= 0.
\end{equation}
In this case we considered the approximants satisfying the conditions
\begin{eqnarray}
&&m_2+m_1+m_0+k+3\geq n-p,\nonumber\\
&&{\rm Max}\left[\lfloor (n-p-3)/4 \rfloor -q, 2\right]
  \leq m_2,m_1,m_0,k \leq \lceil (n-p-3)/4\rceil +q,
\end{eqnarray}
where, as before, $m_i$ and $k$ are the order of the polynomials $P_i$
and $R$ respectively.  We also tried doubly-biased IA2
(b$_{\pm\beta_c}$IA2) where also a singularity at $-\beta_c$ is forced
using solutions of the equation
\begin{equation}
\left( 1 - x^2/\beta_c^2\right)
P_2(x) f^{\prime\prime}(x) + 
P_1(x)f^\prime (x)+ P_0(x)f(x)+R(x)= 0.
\end{equation}

In Table \ref{nures} we report the results of some of the analyses we
performed.  In the case of the b$_{\pm\beta_c}$IA2 analysis we also
report the exponent at the antiferromagnetic singularity which turned
out to be always consistent with $1-\alpha$.  The error of $\nu$ is
given as a sum of two terms: the first one is computed from the spread
of the approximants at $\beta_c$, the second one is related to the
uncertainty of $\beta_c$. 
\begin{table}[tbp]
\squeezetable
\caption{
Results of the analysis of the 19th order IHT  series for
$\xi^2/\beta$.  r-app is explained in the text.  The error is reported
as a sum of two terms.  The first term is related to the spread of the
approximants at the central value of $\beta_c$, while the second one
is related to the uncertainty of the value of $\beta_c$ and it is
estimated by varying $\beta_c$.
}
\label{nures}
\begin{tabular}{clcr@{}lr@{}l}
\multicolumn{1}{c}{}&
\multicolumn{1}{c}{approx}&
\multicolumn{1}{c}{r-app}&
\multicolumn{2}{c}{$\nu$}&
\multicolumn{2}{c}{$\gamma^{\rm af}$}\\
\tableline \hline
$\lambda_6=0,\;\lambda_4=1.08$& b$_{\beta_c}$IA2$_{q=2,p=0}$ & $(54-2)/70$     & 0&.63004(2+11) &  &  \\
$\lambda_6=0,\;\lambda_4=1.10$
  & b$_{\beta_c}$IA1$_{q=3,p=0}$ & $35/37$     & 0&.63012(2+10) & &  \\
  & b$_{\beta_c}$IA2$_{q=2,p=0}$ & $51/70$     & 0&.63016(3+9) & &  \\
  & b$_{\pm\beta_c}$IA2$_{q=2,p=0}$ & $(53-3)/55$ & 0&.63015(5+10) & $-$0&.88(5) \\
  & b$_{\beta_c}$IA3$_{q=2,p=0}$ & $28/35$     & 0&.63009(14+10) & &  \\
  & b$_{\beta_c}$IA2$_{q=2,p=1}$ & $(86-4)/132$     & 0&.63016(3+9) & &  \\
  & b$_{\beta_c^{\rm MC}}$IA2$_{q=2,p=0}$ & $(47-2)/55$ & 0&.63012(3+3) & &  \\
$\lambda_6=0,\;\lambda_4=1.12$& b$_{\beta_c}$IA2$_{q=2,p=0}$ & $(49-2)/70$     & 0&.63027(3+9) &  &  \\\hline
$\lambda_6=1,\;\lambda_4=1.86$ & b$_{\beta_c}$IA2$_{q=1,p=0}$ & $(60-1)/70$     & 0&.62978(1+11) &  & \\
$\lambda_6=1,\;\lambda_4=1.90$
  & b$_{\beta_c}$IA1$_{q=3,p=0}$ & $34/37$     & 0&.63000(2+12) & &  \\
  & b$_{\beta_c}$IA2$_{q=2,p=0}$ & $(61-2)/70$     & 0&.63003(2+11) & &  \\
  & b$_{\pm\beta_c}$IA2$_{q=2,p=0}$ & $(55-2)/55$ & 0&.63003(7+11) & $-$0&.87(16) \\
  & b$_{\beta_c}$IA2$_{q=2,p=1}$ & $(113-4)/132$     & 0&.63003(3+10) & &  \\
  & b$_{\beta_c}$IA3$_{q=2,p=0}$ & $(27-1)/34$     & 0&.62988(17+14) & &  \\
$\lambda_6=1,\;\lambda_4=1.94$ & b$_{\beta_c}$IA2$_{q=2,p=0}$ & $(55-2)/70$     & 0&.63023(2+11) & & \\\hline

${\rm spin}-1,\; D=0.633$ 
& b$_{\beta_c}$IA2$_{q=2,p=0}$ & $(67-1)/70$ & 0&.62998(2+13) & &  \\ 

${\rm spin}-1,\; D=0.641$ 
& b$_{\beta_c}$IA1$_{q=3,p=0}$ & $33/37$ & 0&.62988(5+15) & &  \\ 
& b$_{\beta_c}$IA2$_{q=2,p=0}$ & $(66-1)/70$ & 0&.62990(2+13) & &  \\ 
& b$_{\beta_c}$IA3$_{q=2,p=0}$ & $24/35$ & 0&.62981(12+19) & &  \\ 

${\rm spin}-1,\; D=0.649$ 
& b$_{\beta_c}$IA2$_{q=2,p=0}$ & $(66-1)/70$ & 0&.62982(2+13) & &  \\ 
\end{tabular}
\end{table}
We quote as our final estimates:
\begin{equation}
\nu = 0.63015(12) \qquad {\rm for} \quad \lambda_6=0,\quad \lambda_4=1.10,
\end{equation}
\begin{equation}
\nu = 0.63003(13) \qquad {\rm for} \quad \lambda_6=1,\quad \lambda_4=1.90.
\end{equation}
and
\begin{equation}
\nu = 0.62990(15)         \qquad {\rm for} \quad \hbox{\rm spin-1},\quad D=0.641.
\end{equation}
Taking also into account the uncertainty of $\lambda_4^*$ we arrive at
the estimates of Table \ref{summaryexp}.  We mention that unbiased IA
analyses of the 19th series of $\xi^2/\beta$ give consistent but less
precise estimates of $\beta_c$ (cf.\ Eqs.\ (\ref{betacl60}) and
(\ref{betacl61})) and $\nu$.

As a check of our results, we performed a biased analysis of $\chi$
and $\xi^2$ at $\lambda_6=0$ and $\lambda_4=1.10$, using the value
$\beta_c=0.3750966(4)$ obtained in Ref.\ \cite{Hasenbusch-99} by
Monte Carlo simulations based on finite-size scaling techniques.
Although the author of Ref.\ \cite{Hasenbusch-99} says that the error
on $\beta_c$ does not include systematic errors, we used it as a check
and found (see Tables \ref{gammares} and \ref{nures})
$\gamma=1.23720(2+7)$ and $\nu=0.63012(3+3)$ (the first error is
related to the spread of the approximants at $\beta_c=0.3750966$ and
the second one to the error on $\beta_c$), which are perfectly
consistent with our final estimates reported in Table
\ref{summaryexp}.

In order to obtain an estimate of $\eta$ without using the scaling
relation $\gamma=(2-\eta)\nu$, we employed the so-called critical
point renormalization method (CPRM).  The idea of the CPRM is that
from two series $D(x)$ and $E(x)$ which are singular at the same point
$x_0$, $D(x)=\sum_i d_i x^i\sim (x_0-x)^{-\delta}$ and 
$E(x)=\sum_i e_i x^i\sim (x_0-x)^{-\epsilon}$, one constructs a new
series $F(x)=\sum_i (d_i/e_i)x^i$.  The function $F(x)$ is singular at
$x=1$ and for $x\to 1$ behaves as $F(x)\sim (1-x)^{-\phi}$, where
$\phi = 1+ \delta - \epsilon$.  Therefore the analysis of $F(x)$
provides an unbiased estimate of the difference between the critical
exponents of the two functions $D(x)$ and $E(x)$.  The series $F(x)$
may be analyzed by employing biased approximants with a singularity at
$x_c=1$.  In order to check for possible systematic errors, we applied
the CPRM to the series of $\xi^2/\beta$ and $\chi$ (analyzing the
corresponding 19th order series) and to the series of $\xi^2$ and
$\chi$ (analyzing the corresponding 20th order series).  We used IA's
biased at $x_c=1$.  In Table \ref{etares} we present the results of
the analysis for some values of the parameters $q,p$.  We obtain
$\eta\nu = 0.02294(20)$ at $\lambda_6=0$ and $\lambda_4=1.1$,
$\eta\nu = 0.02287(20)$ at $\lambda_6=1$ and $\lambda_4=1.9$,
and $\eta\nu = 0.02305(20)$ for spin-1 and $D=0.641$.
Taking again into account the uncertainty of $\lambda_4^*$ and $D^*$
we then obtain the
estimate reported in Table \ref{summaryexp}.  

\begin{table}[tbp]
\squeezetable
\caption{
Results for $\eta$ obtained using the CPRM:
$(a)$ applied to $\xi^2$ and $\chi$ (20 orders);
$(b)$ applied to $\xi^2/\beta$ and $\chi$ (19 orders).
r-app is explained in the text.
}
\label{etares}
\begin{tabular}{clcr@{}l}
\multicolumn{1}{c}{}&
\multicolumn{1}{c}{approx}&
\multicolumn{1}{c}{r-app}&
\multicolumn{2}{c}{$\eta\nu$}\\
\tableline \hline
$\lambda_6=0,\;\lambda_4=1.08$ & (a) bIA2$_{q=2,p=0}$ & $(95-1)/115$  & 0&.02274(13) \\
  & (b) bIA2$_{q=2,p=0}$ & $38/70$             & 0&.02294(8) \\
$\lambda_6=0,\;\lambda_4=1.10 $ 
  & (a) bIA2$_{q=2,p=0}$ & $95/115$  & 0&.02280(14) \\
  & (a) bIA2$_{q=2,p=1}$ & $(150-1)/185$         & 0&.02280(16) \\
  & (a) bIA3$_{q=2,p=0}$ & $(47-1)/61$  & 0&.02280(37) \\
  & (b) bIA1$_{q=3,p=0}$ & $28/37$             & 0&.02300(8) \\
  & (b) bIA2$_{q=2,p=0}$ & $36/70$             & 0&.02301(8) \\
  & (b) bIA2$_{q=2,p=1}$ & $86/132$       & 0&.02309(12) \\
  & (b) bIA3$_{q=2,p=0}$ & $31/34$             & 0&.02311(22) \\
$\lambda_6=0,\;\lambda_4=1.12$ 
& (a) bIA2$_{q=2,p=0}$ & $97/115$   & 0&.02285(15) \\
& (b) bIA2$_{q=2,p=0}$ & $35/70$             & 0&.02308(9) \\\hline
$\lambda_6=1,\;\lambda_4=1.86$
& (a) bIA2$_{q=2,p=0}$ & $(94-5)/115$     & 0&.02267(12) \\
& (b) bIA2$_{q=2,p=0}$ & $39/70$     & 0&.02285(12) \\
$\lambda_6=1,\;\lambda_4=1.90$
  & (a) bIA2$_{q=2,p=0}$ & $(90-2)/115$      & 0&.02278(12) \\
  & (b) bIA2$_{q=2,p=0}$ & $37/70$           & 0&.02298(11) \\
$\lambda_6=1,\;\lambda_4=1.94$
& (a) bIA2$_{q=2,p=0}$ & $(92-2)/115$   & 0&.02288(13) \\
& (b) bIA2$_{q=2,p=0}$ & $32/70$   & 0&.02312(10) \\\hline
spin-1, $D=0.633$
& (a) bIA2$_{q=2,p=0}$ & $(84-1)/115$            & 0&.02292(40) \\
& (b) bIA2$_{q=2,p=0}$ & $36/70$                 & 0&.02316(22) \\
spin-1, $D=0.641$
& (a) bIA2$_{q=2,p=0}$ & $(85-2)/115$            & 0&.02288(40) \\
& (b) bIA2$_{q=2,p=0}$ & $37/70$                 & 0&.02312(22) \\
spin-1, $D=0.649$
& (a) bIA2$_{q=2,p=0}$ & $(84-1)/115$            & 0&.02285(43) \\
& (b) bIA2$_{q=2,p=0}$ & $37/70$                 & 0&.02307(22) \\
\end{tabular}
\end{table}

The CPRM was also employed in order to estimate the exponent $\sigma$.
It was applied to the 18th order series of $\chi\xi^2/\beta$ and
$q_{4,0}/\beta$. The results are displayed in Table \ref{sigmares}.
We find $\sigma\nu=0.0134(8)$ for $\lambda_6=0$ and $\lambda_4=1.1$,
$\sigma\nu=0.0134(9)$ for $\lambda_6=1$ and $\lambda_4=1.9$,
and $\sigma\nu=0.0127(6)$ for spin-1 at $D=0.641$.

\begin{table}[tbp]
\squeezetable
\caption{
Results for $\sigma$ obtained using the CPRM
applied to $m_2/\beta$ and $q_{4,0}/\beta$ (18 orders).
Here we used $x_{\rm min} = 0.9$,
$x_{\rm max} = 1.1$, $x_{\rm max} = 0.1$.
r-app is explained in the text.
}
\label{sigmares}
\begin{tabular}{clcr@{}l}
\multicolumn{1}{c}{}&
\multicolumn{1}{c}{approx}&
\multicolumn{1}{c}{r-app}&
\multicolumn{2}{c}{$\sigma\nu$}\\
\tableline \hline
$\lambda_6=0,\;\lambda_4=1.08 $&  bIA2$_{q=1,p=0}$ & $29/34$             & 0&.0134(8) \\
$\lambda_6=0,\;\lambda_4=1.10 $&  bIA2$_{q=1,p=0}$ & $29/34$             & 0&.0134(8) \\
  &  bIA2$_{q=2,p=0}$ & $53/62$             & 0&.0133(10) \\
$\lambda_6=0,\;\lambda_4=1.12 $ &  bIA2$_{q=1,p=0}$ & $28/34$             & 0&.0135(9) \\\hline
$\lambda_6=1,\;\lambda_4=1.86 $&  bIA2$_{q=1,p=0}$ & $29/34$                 & 0&.0133(9) \\
$\lambda_6=1,\;\lambda_4=1.90 $&  bIA2$_{q=1,p=0}$ & $29/34$                 & 0&.0134(9) \\
&  bIA2$_{q=2,p=0}$ & $52/62$                 & 0&.0132(12) \\
$\lambda_6=1,\;\lambda_4=1.94 $&  bIA2$_{q=1,p=0}$ & $28/34$                 & 0&.0135(9) \\\hline
spin-1, $D=0.633$&  bIA2$_{q=1,p=0}$ & $21/34$                 & 0&.0127(5) \\
spin-1, $D=0.641$&  bIA2$_{q=1,p=0}$ & $21/34$                 & 0&.0127(5) \\
&  bIA2$_{q=2,p=0}$ & $37/62$                 & 0&.0128(6) \\
spin-1, $D=0.649$&  bIA2$_{q=1,p=0}$ & $21/34$                 & 0&.0126(5) \\
\end{tabular}
\end{table}

\subsection{Ratios of amplitudes}
\label{ratioofamp}

In the following we describe the analysis we employed in order to
evaluate universal ratios of amplitudes, such as $g_4$, $r_{2j}$ and
$c_i$, from the corresponding HT series.  In the case of
$g$, $c_2$, $c_3$ and $c_4$ we analyzed the series 
$\beta^{3/2} g_4 =\sum_{i=0}^{17} a_i\beta^i$, 
$\beta^{-4}c_2 = \sum_{i=0}^{13} a_i \beta^i$, 
$\beta^{-3}c_3 = \sum_{i=0}^{13} a_i \beta^i$, and 
$\beta^{-2}c_4 = \sum_{i=0}^{13} a_i \beta^i$. In order
to obtain estimates of the universal critical limit of $g_4$,
$r_{2j}$, and $c_i$, we evaluated the approximants of the
corresponding HT series at $\beta_c$ (as determined from the
analysis of the magnetic susceptibility), and multiplied by the
appropriate power of $\beta_c$.

For an $n$th order series we considered three sets of approximants:
Pad\'e approximants (PA's), Dlog-Pad\'e approximants (DPA's) and
first-order integral approximants (IA1's).

(I) $[l/m]$ PA's with 
\begin{eqnarray}
&&l+m \geq n-2, \\
&&{\rm Max}\left[ n/2-q,4\right] \leq l,m \leq  n/2 +q,
\label{paapprox}
\end{eqnarray}
where $l,m$ are the orders of the polynomials respectively in the
numerator and denominator of the PA.  The parameter $q$ determines the
degree of off-diagonality allowed. The best approximants should be
those diagonal or quasi-diagonal.  So we considered PA's selected
using $q=3$.  As estimate from the PA's we take the average of the
values at $\beta_c$ of the non-defective approximants using all the
available terms of the series and satisfying the condition
(\ref{paapprox}) with $q=2$.  The error we quote is the standard
deviation around the estimate of the results from all the
non-defective approximants listed above.  We considered defective PA's
with spurious singularities in the rectangle defined in Eq.\ 
(\ref{filter}) with $x_{\rm min} = 0.9$ ($x_{\rm min}=0$ only in the
case of $r_{10}$) $x_{\rm max} = 1.01$ and $y_{\rm max} = 0.1$.

(II) $[l/m]$ DPA's with
\begin{eqnarray}
&&l+m \geq n-2,\nonumber\\
&&{\rm Max}\left[(n-1)/2-q,4\right] \leq l,m \leq  (n-1)/2+q,
\end{eqnarray}
where $l,m$ are the orders of the polynomials respectively in the
numerator and denominator of the PA of the series of its logarithmic
derivative.  We again fixed $q=3$.  The estimate with the
corresponding error is obtained as in the case of PA's.  We considered
defective DPA's with spurious singularities in the rectangle with
$x_{\rm min} = 0$, $x_{\rm max} = 1.01$ and $y_{\rm max} = 0.1$, cf.\ 
Eq.\ (\ref{filter}).

(III) $[m_1/m_0/k]$ IA1's given by Eq.\ (\ref{gaapprox1}).  The
off-diagonality parameter was fixed to be $q=3$, and $p=1$.
As estimate we take
the average of the values at $\beta_c$ of all non-defective
approximants listed above.  The error we quote is the standard
deviation around the estimate.  We considered defective IA1's with
spurious singularities in the rectangle and $x_{\rm min} = 0$, 
$x_{\rm max} = 1.001$ and $y_{\rm max} = 0.1$.

As in the case of the critical exponents, sometimes we also eliminated
seemingly good approximants whose results were very far from the
average of the other approximants.  In order to arrive at a final
estimate, the results from PA's, DPA's, and IA's were then combined
taking also into account the relative number of non-defective
approximants (before combining the results we divided the apparent
error of each set of approximants by the square root of the ratio
between the number of non-defective and the total number of
approximants).  Of course all the above procedure to arrive at a final
estimate is rather subjective. But we believe it provides reasonable
estimates of the quantity at hand and its uncertainty. We report in
Table \ref{detailfrc} the results of each set of approximants, so that
the readers can judge the reliability of our final estimates.  The
second error in the combined estimate is related to the uncertainty of
the value of $\lambda_4^*$ and $D^*$; it is estimated by varying $\lambda_4$ in
the range $1.08\hbox{--}1.12$ for $\lambda_6=0$, 
$1.86\hbox{--}1.94$ for $\lambda_6=1$, and
$D$ in the range $0.633\hbox{--}0.649$ for the spin-1 model. 
  Errors due to the uncertainty
of $\beta_c$ are negligible.

\begin{table}[tbp]
\squeezetable
\caption{
Results of PA, DPA, and IA1 analyses of the series for $g_4$, $r_{2j}$
and $c_i$. When results are not reported, it means that for that
quantity no acceptable results were obtained from that class of
approximants. The fraction at subscript is the number of non-defective
approximants over the total number of approximants. The last column
contains the estimates obtained by combining the three classes of
approximants.
}
\label{detailfrc}
\begin{tabular}{ccr@{}lr@{}lr@{}lr@{}l}
\multicolumn{1}{c}{}&
\multicolumn{1}{c}{}&
\multicolumn{2}{c}{PA}&
\multicolumn{2}{c}{DPA}&
\multicolumn{2}{c}{IA1}&
\multicolumn{2}{c}{combined}\\
\tableline \hline
$g_4^*$& $\lambda_6=0\;,\lambda_4=1.10$ & 
23&.500(60)$_{17/21}$ & 23&.491(25)$_{16/18}$ & 23&.504(18)$_{49/73}$ & 23&.499(16+20)\\
&$\lambda_6=1\;,\lambda_4=1.90$ & 
23&.487(45)$_{17/21}$ & 23&.474(46)$_{17/18}$ & 23&.498(24)$_{57/73}$ & 23&.491(21+40)\\
&spin-1, $D=0.641$ & 
23&.486(19)$_{20/21}$ & 23&.492(88)$_{17/18}$ & 23&.491(52)$_{53/73}$ & 23&.487(18+20) \\ \hline

$r_6$& $\lambda_6=0\;,\lambda_4=1.10$ & 
2&.051(13)$_{19/21}$  & 2&.058(12)$_{11/18}$ & 2&.048(7)$_{32/73}$ & 2&.051(7+2) \\
     & $\lambda_6=1\;,\lambda_4=1.90$ & 
2&.052(12)$_{18/21}$  & 2&.063(14)$_{11/18}$ & 2&.048(4)$_{33/73}$ & 2&.050(5+4) \\
&spin-1, $D=0.641$ & 
2&.0493(65)$_{20/21}$ & 2&.0461(24)$_{16/18}$ & 2&.0456(16)$_{23/73}$ & 2&.046(2+3) \\ \hline

$r_8$& $\lambda_6=0\;,\lambda_4=1.10$ & 
2&.24(9)$_{17/18}$    & 2&.21(13)$_{17/21}$  & 2&.23(5)$_{37/69}$  & 2&.23(5+4) \\
     & $\lambda_6=1\;,\lambda_4=1.90$ & 
2&.23(11)$_{18/18}$   & 2&.23(9)$_{17/21}$   & 2&.23(5)$_{36/69}$  & 2&.23(5+6) \\
&spin-1, $D=0.641$ & 
2&.40(8)$_{16/18}$ & 2&.31(5)$_{17/21}$ & 2&.42(13)$_{26/69}$ & 2&.34(5+3) \\ \hline

$r_{10}$&$\lambda_6=0\;,\lambda_4=1.10$ & 
$-$14&(5)$_{15/21}$ & &&  $-$13&.3(1.3)$_{6/61}$ & $-$14&(4+0) \\
        &$\lambda_6=1\;,\lambda_4=1.90$ & 
$-$14&(5)$_{14/21}$ & &&  $-$12&(4)$_{10/61}$ & $-$13&(5+0) \\
&spin-1, $D=0.641$ & 
$-$10&(21)$_{13/21}$ & & & 4&(36)$_{14/61}$&$-$8&(25+0) \\\hline

$10^4c_2$& $\lambda_6=0\;,\lambda_4=1.10$ &
$-$3&.582(8)$_{15/15}$ & $-$3&.580(29)$_{12/12}$&$-$3&.586(24)$_{24/33}$&$-$3&.582(7+6)\\
         & $\lambda_6=1\;,\lambda_4=1.90$ &
$-$3&.574(7)$_{14/15}$ & $-$3&.574(26)$_{12/12}$&$-$3&.585(38)$_{24/33}$&$-$3&.574(7+20) \\
&spin-1, $D=0.641$ & 
$-$3&.570(12)$_{15/15}$ & $-$3&.562(36)$_{11/12}$&$-$3&.554(28)$_{25/33}$&$-$3&.568(11+4) \\\hline

$10^4c_3$& $\lambda_6=0\;,\lambda_4=1.10$ & 
0&.087(8)$_{12/15}$& 0&.080(10)$_{3/12}$ & 0&.084(7)$_{26/36}$& 0&.085(6+0) \\
            &$\lambda_6=1\;,\lambda_4=1.90$ & 
0&.086(5)$_{11/15}$& 0&.078(12)$_{2/12}$ & 0&.086(5)$_{26/36}$& 0&.086(4+0) \\
&spin-1, $D=0.641$ & 
0&.095(14)$_{14/15}$& 0&.100(12)$_{2/12}$ & 0&.090(4)$_{30/36}$& 0&.090(4+0) \\
\end{tabular}
\end{table}

We have also performed analyses of the series of $g_4$ for
$\lambda_6=0$ and several values of $\lambda_4$ by employing the
Roskies transform \cite{Roskies-81}.  The idea of the Roskies
transform (RT) is to perform biased analyses which take into account
the leading confluent singularity. For the Ising model, where
$\Delta\simeq 1/2$, one replaces the variable $\beta$ in the original
expansion with a new variable $z$, defined by 
$1-z = (1 - \beta/\beta_c)^{1/2}$.  If the original series has
square-root scaling correction terms, the transformed series has
analytic correction terms, which can be handled by standard PA's or
DPA's.  Note that in principle IA1's should be able to detect the
first non-analytic correction to scaling, but they probably need
many more terms of the series, and practically need to be explicitly
biased as in the case of PA's and DPA's. Indeed the IA1 results
without the RT turn out to be substantially equivalent to those
obtained using PA's and DPA's.  In Table \ref{detailfRT} we report the
details of the analysis without and with RT for some values of
$\lambda_4$ and $\lambda_6=0$.  These results are plotted in Fig.\ 
\ref{figg}.

\begin{table}[tbp]
\squeezetable
\caption{
Details of the analysis of the 17th order series for
$\beta^{-3/2}g_4(\beta)$ with and without the use of the RT for some
values of $\lambda_4$ and $\lambda_6=0$.  
In the PA, DPA and IA analyses with RT we used 
$q=2$ (other approximants turned out to be much less stable). 
We fixed $x_{\rm min} = 0$,
$x_{\rm max} = 1.1$ and $y_{\rm max} = 0.25$ for PA and DPA, 
and $x_{\rm min} = 0$,
$x_{\rm max} = 1.01$ and $y_{\rm max} = 0.25$ for IA.
In order to perform a homogeneous comparison we used the same
procedure for the direct analysis without RT (except that we used 
$x_{\rm max} = 1.01$ and $y_{\rm max} = 0.1$). 
The fraction at subscript is the number of non-defective
approximants over the total number of approximants.
}
\label{detailfRT}
\begin{tabular}{ccr@{}lr@{}lr@{}lr@{}l}
\multicolumn{1}{c}{$\lambda_4$}&
\multicolumn{1}{c}{}&
\multicolumn{2}{c}{PA}&
\multicolumn{2}{c}{DPA}&
\multicolumn{2}{c}{IA}&
\multicolumn{2}{c}{combined}\\
\tableline \hline
0.5     & direct & 22&.62(58)$_{10/15}$  & 22&.43(16)$_{9/12}$  & 22&.55(19)$_{19/37}$ & 22&.48(15)\\
         & RT     & 23&.75(27)$_{8/15}$  & 23&.48(25)$_{8/12}$  & 23&.29(50)$_{15/37}$ & 23&.56(23)\\
0.7     & direct  & 23&.04(28)$_{10/15}$ & 22&.92(14)$_{9/12}$  & 22&.95(13)$_{15/37}$ & 22&.94(12)\\
         & RT     & 23&.62(31)$_{13/15}$ & 23&.54(26)$_{9/12}$  & 23&.35(35)$_{17/37}$ & 23&.54(20)\\
1.0     & direct & 23&.58(23)$_{9/15}$   & 23&.40(6)$_{8/12}$   & 23&.38(19)$_{15/37}$ & 23&.41(7)\\
        & RT     & 23&.56(27)$_{13/15}$  & 23&.57(15)$_{9/12}$  & 23&.45(21)$_{14/37}$ & 23&.55(14)\\
1.1     & direct & 23&.500(64)$_{12/15}$ & 23&.491(23)$_{11/12}$  & 23&.494(16)$_{17/37}$ & 23&.493(16)\\
        & RT     & 23&.56(22)$_{13/15}$  & 23&.59(16)$_{9/12}$    & 23&.44(17)$_{14/37}$ & 23&.55(13)\\
1.2     & direct & 23&.63(4)$_{10/15}$  & 23&.613(9)$_{11/12}$  & 23&.612(9)$_{20/37}$ & 23&.613(8)\\
        & RT     & 23&.56(20)$_{13/15}$ & 23&.54(34)$_{11/12}$  & 23&.43(16)$_{12/37}$ & 23&.52(15)\\
1.5     & direct & 23&.93(4)$_{14/15}$  & 23&.92(6)$_{7/12}$  & & & 23&.93(4)\\
        & RT     & 23&.55(27)$_{12/15}$ & 23&.53(31)$_{11/12}$  & 23&.41(13)$_{11/37}$ & 23&.48(16)\\
2.0     & direct & 24&.14(28)$_{15/15}$  & 24&.07(17)$_{3/12}$  & 24&.15(9)$_{31/37}$ & 24&.15(9)\\
        & RT     & 23&.61(22)$_{12/15}$ & 23&.52(18)$_{10/12}$  & 23&.44(12)$_{17/37}$ & 23&.50(12)\\
3.0     & direct & 24&.42(14)$_{15/15}$  & 24&.14(39)$_{6/12}$  & 24&.40(19)$_{15/37}$ & 24&.40(12)\\
        & RT     & 23&.61(23)$_{14/15}$ & 23&.47(14)$_{11/12}$  & 23&.34(18)$_{16/37}$ & 23&.48(11)\\
$\infty$ & direct & 24&.78(10)$_{15/15}$   & 24&.57(19)$_{10/12}$  & 24&.81(16)$_{9/37}$ & 24&.75(9)\\
& RT     & 23&.59(20)$_{13/15}$  & 23&.47(11)$_{10/12}$  & 23&.48(16)$_{13/37}$ & 23&.50(10)\\
\end{tabular}
\end{table}

\section{Universal ratios of amplitudes}
\label{univra}

\subsection{Notations}
\label{notationseqst}

Universal ratios of amplitudes characterize the behavior in the
critical domain of thermodynamical quantities that do not depend on
the normalizations of the external (e.g.\ magnetic) field, order
parameter (e.g.\ magnetization) and temperature.  Amplitude ratios of
zero-momentum quantities can be derived from the critical equation of
state. We consider several amplitudes derived from the singular
behavior of: the specific heat
\begin{equation}
C_H = A^{\pm} |t|^{-\alpha},
\label{sphamp}
\end{equation}
the magnetic susceptibility
\begin{equation}
\chi = C^{\pm} |t|^{-\gamma},
\label{chiamp}
\end{equation}
the spontaneous magnetization on the coexistence curve
\begin{equation}
M = B |t|^{-\beta},
\label{magamp}
\end{equation}
the zero-momentum connected $n$-point correlation functions
\begin{equation}
\chi_n = C_n^\pm |t|^{-\gamma-(n-2)\beta\delta}.
\label{chinamp}
\end{equation}
We complete our list of amplitudes by
considering the second-moment correlation length 
\begin{equation}
\xi = f^{\pm} |t|^{-\nu},
\label{xiamp}
\end{equation}
and the true (on-shell) correlation length, 
describing the large distance behavior of the two-point function,
\begin{equation}
\xi_{\rm gap} = f_{\rm gap}^{\pm} |t|^{-\nu}.
\label{xiosamp}
\end{equation}
One can also define amplitudes along the critical isotherm, e.g.\ 
\begin{eqnarray}
\chi &=& C^c |H|^{-{\gamma/\beta\delta}}, \label{chicris}\\
\xi &=& f^c |H|^{-{\nu/\beta\delta}}, \label{xicris}\\
\xi_{\rm gap} &=& f_{\rm gap}^c |H|^{-{\nu/\beta\delta}}. \label{xigapcris}
\end{eqnarray}

\subsection{Universal ratios of amplitudes from the parametric representation} 
\label{univparrep}

In the following we report the expressions of the universal ratios of
amplitudes in terms of the parametric representation (\ref{parrep}) of
the critical equation of state.

The singular part of the free energy per unit volume can be written as
\begin{equation}
{\cal F} = h_0 m_0 R^{2-\alpha} g(\theta),
\end{equation}
where $g(\theta)$ is the solution of the first-order differential
equation
\begin{equation}
(1-\theta^2) g'(\theta) + 2(2-\alpha)\theta g(\theta) = 
(1-\theta^2 + 2\beta\theta^2)h(\theta)
\label{pp1}
\end{equation}
that is regular at $\theta=1$.
One may also write
\begin{eqnarray}
&\displaystyle\chi^{-1} = 
 {h_0\over m_0} R^\gamma g_2(\theta),\qquad\qquad
&g_2(\theta) = { 2\beta\delta \theta h(\theta) + (1-\theta^2) h'(\theta)\over 
(1-\theta^2 + 2\beta\theta^2)},\label{pp2}\\
&\displaystyle\chi_3 = 
 {m_0\over h_0^2} R^{-2\gamma-\beta} g_3(\theta),\qquad\qquad
&g_3(\theta) = - { (1-\theta^2)g_2'(\theta) +  2\gamma\theta g_2(\theta)\over 
g_2(\theta)^3 (1-\theta^2 + 2\beta\theta^2)},\\
&\displaystyle\chi_4 =
 {m_0\over h_0^3} R^{-3\gamma-2\beta} g_4(\theta),\qquad\qquad
&g_4(\theta) = 
{ (1-\theta^2)g_3'(\theta) -  2(2\gamma+\beta)\theta g_3(\theta)\over 
 g_2(\theta) (1-\theta^2 + 2\beta\theta^2)}.
\end{eqnarray}
Using the above formulae, one can then calculate the universal ratios of
amplitudes:
\begin{eqnarray}
&&U_0\equiv {A^+\over A^-} = 
(\theta_0^2 - 1 )^{2-\alpha} {g(0)\over g(\theta_0)},\\
&&U_2\equiv {C^+\over C^-} = 
(\theta_0^2 - 1 )^{-\gamma} {g_2(0)\over g_2(\theta_0)},\\ 
&&u_4\equiv {C_4^+\over C_4^-} = 
(\theta_0^2 - 1 )^{-3\gamma-2\beta} {g_4(0)\over g_4(\theta_0)},\\ 
&& R_c^+ \equiv {\alpha A^+ C^+\over B^2} =
- \alpha (1-\alpha)(2-\alpha) (\theta_0^2 - 1 )^{2\beta} \theta_0^{-2} g(0),\\
&& R_c^- \equiv {\alpha A^- C^-\over B^2} = {R_c^+\over U_0U_2},\\ 
&& v_3\equiv R_3 \equiv - {C_3^-B\over (C^-)^2}= 
- \theta_0 g_2(\theta_0)^2 g_3(\theta_0),\\
&& v_4 \equiv  -  {C_4^-B^2\over (C^-)^3} + 3v_3^2 = 
\theta_0^2  g_2(\theta_0)^3 
  \left[ 3g_2(\theta_0)g_3(\theta_0)^2 -  g_4(\theta_0)\right] ,\\
&& R_4^+ \equiv  - {C_4^+ B^2\over (C^+)^3} = |z_0|^2 ,\\
&& Q_1^{-\delta}\equiv R_\chi
   \equiv {C^+ B^{\delta-1}\over (\delta C^c)^\delta}=
(\theta_0^2-1)^{-\gamma} \theta_0^{\delta-1} h(1),\\
&& F^\infty_0 \equiv  \lim_{z\rightarrow \infty} z^{-\delta} F(z) = 
        \rho^{1-\delta} h(1).
\label{thetaform}
\end{eqnarray}
Using Eq.\ (\ref{hFrel}) one can compute $F(z)$ and obtain the
small-$z$ expansion coefficients of the effective potential $r_{2j}$
in terms of the critical exponents and the coefficients $h_{2l+1}$ of
the expansion of $h(\theta)$.

\section{Approximation scheme for the parametric representation of the
equation of state based on stationarity}
\label{rhotheta}

The parametric form of the critical equation of state, described by
Eqs.\ (\ref{parrep}), (\ref{thzrel}), and (\ref{hFrel}), shows a
formal dependence on the auxiliary parameter $\rho$.

However all physically relevant amplitude ratios are independent of
$\rho$, because they may be expressed in terms of the invariant
function $F(z)$ and its derivatives, evaluated at such special values
of $z$ as $z=0$, $z=\infty$ and $z=z_0$, where $F(z_0)=0$.  Notice
that, despite the apparent dependence generated by the relation 
$z_0 \equiv z(\rho,\theta_0(\rho))$, from the definition it follows
that $z_0$ must necessarily be independent of $\rho$.

We can exploit these facts in order to set up an approximation
procedure in which the function $h(\rho, \theta)$, entering the
scaling equation of state, is truncated to some simpler (polynomial)
function $h^{(t)}(\rho,\theta)$ and the value of $\rho$ is properly
fixed to optimize the approximation.

We found that, at any given order in the truncation, it is possible
and convenient to choose $\rho$ in such a way that all the (truncated)
universal amplitude ratios be simultaneously stationary against
infinitesimal variations of $\rho$ itself.

Starting from $h^{(t)} (\rho, \theta)$ we may reconstruct the function
\begin{equation}
\widetilde F^{(t)}(\rho,\theta)=
{\rho h^{(t)}(\rho,\theta) \over (1-\theta^2)^{\beta+\gamma}}.
\end{equation}
In order that all truncated amplitudes be simultaneously stationary in
$\rho$, it is necessary that the function 
$F^{(t)}(\rho,z) \equiv \widetilde F^{(t)}(\rho,\theta(\rho,z))$ be
stationary with respect to variations of $\rho$ for any value of $z$.

We shall prove that for any polynomial truncation
$h^{(t)}(\rho,\theta)$ it is possible to find a value $\rho_t$,
independent of $z$, such that
\begin{equation}
\left. \partial F^{(t)}(\rho,z) \over \partial \rho \right|_{\rho=\rho_t}= 0,
\end{equation}
a property which we shall term ``global stationarity''.

In order to prove our statement, let us rephrase the above condition
into the form
\begin{equation}
{\partial \widetilde F^{(t)}(\rho,\theta) \over \partial \rho} + 
{\partial \widetilde F^{(t)}(\rho,\theta) \over \partial \theta}
{\partial \theta \over \partial \rho} = 0,
\label{Ftildestat}
\end{equation}
where the implicit function theorem allows us to write
\begin{equation}
{\partial \theta \over \partial \rho}= - 
{\partial z/\partial \rho \over \partial z/\partial \theta} \, .
\end{equation}
The definitions (\ref{thzrel}) and (\ref{hFrel}) imply
\begin{equation}
{\partial z \over \partial \rho}={z \over \rho}, \qquad \quad
{\partial z \over \partial \theta}
  = z\left({1 \over \theta}+ {2 \beta \theta \over 1- \theta^2}\right).
\end{equation}
Moreover it is trivial to show that
\begin{eqnarray}
{\partial \widetilde F^{(t)}(\rho,\theta) \over \partial \rho} &=& 
\widetilde F^{(t)}(\rho,\theta) \left({1 \over \rho} +
     {1 \over h^{(t)}(\rho,\theta)}{\partial h^{(t)}(\rho,\theta) \over 
      \partial \rho}\right), \\
{\partial \widetilde F^{(t)}(\rho,\theta) \over \partial \theta} &=& 
\widetilde F^{(t)}(\rho,\theta) 
\left({2(\beta+\gamma)\theta \over 1-\theta^2}+ 
    {1 \over h^{(t)}(\rho,\theta)}{\partial h^{(t)}(\rho,\theta) \over 
     \partial \theta}\right).
\end{eqnarray}
Substitution of these expressions into Eq.\ (\ref{Ftildestat}) leads
to the following form of the global stationarity condition:
\begin{equation}
[1-(1+2 \gamma)\theta^2]h^{(t)}(\rho,\theta)+ [1+(2 \beta-1)\theta^2]\rho 
{\partial h^{(t)}(\rho,\theta) \over \partial \rho}- [1-\theta^2]\theta 
{\partial h^{(t)}(\rho,\theta) \over \partial \theta}= 0.
\end{equation}
Let us now write down $h^{(t)}(\rho,\theta)$ as a power series in the
odd powers of $\theta$:
\begin{equation}
h^{(t)}(\rho,\theta)= \theta+ \sum_{n=1}^{t-1} h_{2n+1}(\rho)\theta^{2n+1}.
\end{equation}
The series-expanded stationarity condition then takes the form
\begin{equation}
\sum_{n=1}^{t-1} 
\left\{\left[\rho {\partial \over \partial \rho}-2 n\right]
        h_{2n+1}(\rho) +
        \left[(2 \beta-1)\rho {\partial \over \partial \rho} -
            2 \gamma+2 n-2\right] h_{2n-1}(\rho)\right\}\theta^{2n+1} = 0,
\label{hstation}
\end{equation}
with the convention $h_1=1$.

Let us now notice that, in the absence of truncations, the above
equation must be identically true, since the original function $F(z)$
is totally independent of $\rho$.  This fact implies that the
coefficients of the above power-series expansion must vanish
individually, and this gives us an infinite set of recursive
differential equations for the functions $h_{2n+1}(\rho)$, which must
be automatically satisfied when the coefficients $h_{2n+1}(\rho)$ are
properly defined.

A truncation corresponds to arbitrarily suppressing all coefficients
starting from $h_{2t+1}(\rho)$.  Hence the global stationarity
condition simply amounts to requiring
\begin{equation}
\left\{\left[(2 \beta-1)\rho 
        {\partial \over \partial \rho}-2 \gamma+2 t-2\right]
    h_{2t-1}(\rho)\right\} \theta^{2t+1}=0,
\label{stationh}
\end{equation}
because all other terms vanish.  The resulting equation can be solved
by choosing $\rho_t$ such that the term in curly brackets vanishes,
independent of $\theta$. This concludes our proof.

The effectiveness of this scheme is beautifully illustrated by its
lowest-order implementation, corresponding to the so-called ``linear
parametric model'', in the context of the three-dimensional Ising
model.

Let us truncate the exact scaling function $h(\rho, \theta)$ to its
cubic approximation
\begin{equation}
h^{(2)} (\rho, \theta)= \theta + h_3(\rho) \theta^3,
\end{equation}
where $h_3(\rho)$ is taken from Eq.\ (\ref{h3}).
Substituting $h_3$ into the stationarity condition (\ref{stationh})
for $t=2$ we find
\begin{equation}
{1 \over 3}(2 \beta - \gamma)\rho^2-2 \gamma(1-\gamma)=0,
\end{equation}
which leads to
\begin{equation}
\rho_2 = \sqrt{ {6 \gamma (\gamma -1) \over \gamma - 2 \beta}}.
\end{equation}

The truncated scaling function vanishes at the value $\theta_0$:
\begin{equation}
 \theta_0^2 \equiv -{1 \over h_3 (\rho_2)}  = 
 {\gamma -2 \beta \over \gamma (1- 2 \beta)} \, .
\end{equation}

In this approximation the scaling equation of state turns out to be
expressible simply in terms of the critical exponents $\beta$ and
$\gamma$.  As a consequence, all the universal ratios may then be
approximated to lowest order by appropriate algebraic combinations of
the critical exponents.

The above results reproduce the old formulae by Schofield, Litster,
and Ho \cite{S-L-H-69}, who obtained expressions for critical
amplitudes in terms of critical exponents from a minimum condition
imposed on the predictions extracted from a parametric scaling
equation of state.

In the case of the linear parametric model, the globality of the
stationarity property introduced by the above authors was shown by
Wallace and Zia \cite{W-Z-74}, who adopted a slightly different, but
essentially equivalent, formulation of the above model.

As we showed above, global stationarity can be imposed on
parametric models regardless of the linearity constraint.
The next truncation, corresponding to $t=3$, can also be treated
analytically.  Our starting point will be
\begin{equation}
h^{(3)} (\rho, \theta)= \theta + h_3(\rho) \theta^3+ h_5(\rho) \theta^5.
\end{equation}
The coefficients $h^{(3)}(\rho)$ and $h^{(5)}(\rho)$ are reported in
Eqs.\ (\ref{h3}) and (\ref{h5}).  By applying the stationarity
condition (\ref{stationh}) to $h_5(\rho)$, we obtain
\begin{equation}
\rho_3 =\sqrt{{(\gamma-2\beta)(1-\gamma+2 \beta) \over 
    12(4 \beta-\gamma)F_5}}
    \left(1-\sqrt{1-{72(2-\gamma)\gamma(\gamma-1)(4\beta-\gamma)F_5 \over 
        (\gamma-2\beta)^2 (1-\gamma+ 2\beta)^2}}\right)^{\!\!{1 \over 2}}.
\end{equation}

The truncated scaling function vanishes when $\theta$ takes the value
$\theta_0$, which is now given by the relation
\begin{equation}
\theta_0^2 = {h_3(\rho_3) \over 2 h_5(\rho_3)}
    \left(\sqrt{1-{4 h_5(\rho_3) \over h_3^2(\rho_3)}}-1\right).
\label{theta0-t3}
\end{equation}

A general feature of truncated parametric models is the possibility of
making a prediction on higher-order coefficients $F_{2n+1}$, for 
$n \geq t$, in terms of lower-order coefficients. This is a natural
consequence of having included by the parametrization some information
on the asymptotic behavior of $F(z)$ for large $z$.  In practice we
observe that each $F_{2n+1}$ appears first in the coefficient
$h_{2n+1}$, in the form of a free constant of integration in the
solution of the recursive differential equation relating $h_{2n+1}$ to
$h_{2n-1}$.  Since a truncation corresponds to setting $h_{2n+1}=0$
starting from $n=t$, this fixes the (truncated) value of all
$F_{2n+1}$ starting from $F_{2t+1}$.

As an important consequence of this mechanism we observe that
truncated models deviate from the exact solution only proportionally
to the difference between exact and predicted coefficients, and this
difference may be quite small even for very low-order truncations.

In order to turn the above considerations into quantitative estimates,
we need to get further insight into the properties of the functions
$h_{2n+1}(\rho)$, especially in the vicinity of the stationary point
$\rho_t$.  To this purpose we introduce the expansion
\begin{equation}
h_{2n+1}(\rho)= \sum_{m=0}^n c_{n,m}\rho^{2m}F_{2m+1}.
\label{hexpan}
\end{equation}
Substituting this expansion as an Ansatz into the recursive
differential equations, we check that $F_{2m+1}$ act as free
parameters (integration constants), while the coefficients $c_{n,m}$
must obey the following algebraic recursive equations:
\begin{equation}
(n-m)c_{n,m}= [(2\beta-1)m-\gamma+n-1]c_{n-1,m},
\label{recurscnm}
\end{equation}
for all $n>m$, subject to the initial conditions $c_{m,m}=1$.  It is
possible to find a closed-form solution to Eq.\ (\ref{recurscnm}):
\begin{equation}
c_{n,m}= {1 \over (n-m)!}\prod_{k=1}^{n-m} (2\beta m -\gamma+k-1) ,
\end{equation}
but for our purposes the recursive equations will sometimes be more
useful than their explicit solutions.

Let us define the coefficients of the $z$-expansion of the truncated
scaling function evaluated at the stationary point:
\begin{equation}
F^{(t)}(\rho_t,z) = 
z + \case{1}{6}z^3 + \sum_{m=2} F_{2m+1}^{(t)} z^{2m+1} .
\label{Ftdef}
\end{equation}
By definition, $F_{2m+1}^{(t)}$ coincide with their exact value
$F_{2m+1}$ for all $m<t$, while for $m \geq t$ they are determined by
the condition $h_{2m+1}(\rho_t)=0$, which, according to Eq.\ 
(\ref{hexpan}), implies
\begin{equation}
\sum_{m=0}^n  c_{n,m}\rho_t^{2m} F_{2m+1}^{(t)} = 0
\label{sumzero}
\end{equation}
for all $n \geq t$.

We can now prove the following lemma:
\begin{equation}
\sum_{m=1}^n m \, c_{n,m}\rho_t^{2m} F_{2m+1}^{(t)} = 0
\label{lemma}
\end{equation}
holds for all $n \ge t$.

The proof is by induction.  Let us assume the lemma to hold for a
given value $n$; then, as a consequence of Eqs.\ (\ref{lemma}) and
(\ref{sumzero}), we obtain
\begin{equation}
\sum_{m=0}^n [(2\beta-1)m-\gamma+n]c_{n,m}\rho_t^{2m} F_{2m+1}^{(t)}=0.
\end{equation}
Notice that the above equation holds also for the initial value
$n=t-1$, since in that case it coincides with the global stationarity
condition.

By use of the recursion equations (\ref{recurscnm}) we now obtain
\begin{equation}
\sum_{m=0}^n (n+1-m) c_{n+1,m}\rho_t^{2m} F_{2m+1}^{(t)}=0.
\end{equation}
Because of the factor $(n+1-m)$ the sum can trivially be extended up
to $n+1$, hence:
\begin{equation}
(n+1) \sum_{m=0}^{n+1} c_{n+1,m}\rho_t^{2m} F_{2m+1}^{(t)} =
\sum_{m=1}^{n+1}m \, c_{n+1,m}\rho_t^{2m} F_{2m+1}^{(t)}.
\end{equation}
The l.h.s.\ vanishes by definition (cf.\ Eq.\ (\ref{sumzero})), hence
the r.h.s.\ vanishes and the proof is completed.

The above lemma is instrumental in evaluating the difference between
the predictions originated by two subsequent truncations.  By applying
once more the definition of $F_{2m+1}^{(t)}$ one can easily show that
\begin{equation}
\sum_{m=0}^n  c_{n,m}
\Bigl[\rho_{t+1}^{2m}(F_{2m+1}^{(t+1)}-F_{2m+1}^{(t)})+ 
      (\rho_{t+1}^{2m}-\rho_t^{2m})F_{2m+1}^{(t)}\Bigr] = 0 
\end{equation}
for all $n > t$.  Let us now expand the equation to first order in the
difference $\rho_{t+1}^2-\rho_t^2$, and make explicit use of the lemma
to obtain
\begin{equation}
\sum_{m=t}^n  c_{n,m}\rho_t^{2m}
\Bigl(F_{2m+1}^{(t+1)}-F_{2m+1}^{(t)}\Bigr) \cong 0
\label{cSumF}
\end{equation}
for all $n>t$. It is crucial that $F_{2m+1}^{(t+1)}-F_{2m+1}^{(t)}=0$
for all $m<t$.

The equation we obtained allows us to express (within the
approximation) all differences $F_{2m+1}^{(t+1)}-F_{2m+1}^{(t)}$ in
terms of the single quantity
\begin{equation}
\delta F_t \equiv F_{2t+1}-F_{2t+1}^{(t)}.
\end{equation}
Knowledge of the $c_{n,m}$ and some ingenuity lead to the explicit
solution of Eq.\ (\ref{cSumF}):
\begin{equation}
F_{2m+1}^{(t+1)}-F_{2m+1}^{(t)} \cong d_{t,m}{\delta F_t \over \rho_t^{2(m-t)}}
\end{equation}
where, for all $m>t$,
\begin{equation}
d_{t,m}= {(-1)^{m-t} \over (m-t)!}
    (2\beta t-\gamma) \prod_{k=1}^{m-t-1} (2\beta m -\gamma-k),
\end{equation}
and obviously $d_{t,t}=1$.

As a corollary of this result, by comparing Eq.\ (\ref{cSumF}) when
$t=1$ to Eq.\ (\ref{lemma}) when $t=2$,
we may write down a closed-form expression for all $F_{2m+1}^{(2)}$
coefficients $(m \geq 1)$:
\begin{equation}
F_{2m+1}^{(2)} = {1 \over 6m}{d_{1,m} \over \rho_2^{2(m-1)}},
\end{equation}
which completes our analysis of the linear parametric model.
 
One may also show that $\delta F_t$ is related to the variation of
$\rho_t$ by the (linearized) relation
\begin{equation}
\delta F_t \cong {1-2\beta \over 2\beta t-\gamma}
\left(\sum_{m=0}^t m^2 c_{t,m} \rho_t^{2(m-t-1)}F_{2m+1}\right)
(\rho_{t+1}^2-\rho_t^2).
\end{equation}
Our numerical estimates, presented in Table \ref{eqstdet}, show that
$\rho_{t+1}^2-\rho_t^2$ is indeed small ($\lesssim 0.01$).

In order to evaluate amplitude ratios, as shown in App.\ \ref{univra},
we must also reconstruct the functions $g(\theta)$ and $g_2(\theta)$,
by solving Eqs.\ (\ref{pp1}) and (\ref{pp2}) respectively.  These
functions may be expanded in even powers of $\theta$, with
coefficients that are functions of $\rho$ satisfying the same
differential equations as $h_{2n+1}$, Eqs.\ (\ref{hstation}). One may
show that, for any given truncation $h^{(t)}(\rho,\theta)$ and
arbitrary values of $\rho$,
\begin{equation}
g^{(t)}(\rho,\theta)=
\sum_{n=0}^\infty \left( \sum_{m=0}^n c_{nm} \rho^{2m} {F^{(t)}_{2m+1}\over 2m+2}\right)
\theta^{2n+2} + A(1-\theta^2)^{2\beta+\gamma},
\label{pp3}
\end{equation}
where $A$ is an integration constant reflecting the arbitrariness in
the zero-field value of the free energy. One may also show that for
$n\geq t$
\begin{equation}
\sum_{m=0}^n c_{nm} \rho^{2m} {F^{(t)}_{2m+1}\over 2m+2} 
\sim {(-1)^n\over (n+1)!} \prod_{k=0}^n(2\beta+\gamma-k)
\label{pp4}
\end{equation}
where the terms in the r.h.s.\ are the coefficients of the Taylor
expansion for $(1-\theta^2)^{2\beta+\gamma}$. As a consequence the
constant $A$ may always be chosen such that $g^{(t)}(\rho,\theta)$ is
truncated to $O(\theta^{2t})$ for any arbitrary choice of $\rho$.

In turn one may also prove that, for any $h^{(t)}(\rho,\theta)$,
\begin{equation}
g_2^{(t)}(\rho,\theta)=
\sum_{n=0}^\infty \left[ \sum_{m=0}^n c_{nm} \rho^{2m} (2m+1) F^{(t)}_{2m+1}\right]
\theta^{2n}.
\label{pp5}
\end{equation}
Now, according to Eqs.\ (\ref{sumzero}) and (\ref{lemma}), when we
choose for $\rho$ the globally stationary value $\rho_t$, the
coefficients in square brakets vanish for all $n\geq t$. As a
consequence for any $t$ the value $\rho_t$ insures the truncation of
$g_2^{(t)}(\rho_t,\theta)$ to $O(\theta^{2t-2})$.  Thus a unique
feature of $\rho_t$ is the simultaneous and consistent truncation of
$h(\theta)$ and $g_2(\theta)$.

Notice that we might start by imposing a global stationarity
condition directly on a truncated $g_2(\rho,\theta)$, obtaining a
different stationary value for $\rho$, and make use of Eq.\ 
(\ref{pp2}) in order to reconstruct the corresponding $h(\theta)$.
However in this case, since $h(\theta)$ must be an odd function of
$\theta$, there is no arbitrary integration constant (which is
physically a trivial consequence of the definition of a reduced
temperature) and therefore $h(\theta)$ cannot be truncated. The
resulting parametric model is mathematically consistent, but in
practice unappealing, because the calculation of $\theta_0$ from the
equation $h(\theta_0)=0$ and the evaluation of universal amplitude
ratios becomes quite cumbersome.

The above described formalism can be usefully employed in the context
of the $\epsilon$-expansion of the critical equation of state.
Comparison with $\epsilon$-expansion results will also shed further
light on the meaning and relevance of the results derived by the
prescription of global stationarity.

Our starting point will be the result of Wallace and Zia
\cite{W-Z-74}, who showed that, when appropriate conditions are
imposed on the zeroth order approximation, the parametric form of the
critical equation of state is automatically truncated in the powers of
$\theta^2$ when expanded in the parameter $\epsilon =4-d$.  For easier
comparison, notice that the parameter $b$ introduced by Schofield
\cite{Schofield-69} and used by Wallace and Zia is the same as our
$\theta_0$, and the variable change from $\theta_0$ to $\rho$ poses no
conceptual problem.

In our reformulation one may state that, within the
$\epsilon$-expansion, it is possible to choose to lowest order a value
$\rho_0$ in such a way that, expanding the parametric equation of
state in $\theta$ and $\epsilon$, one finds, for all $n \geq 2$,
\begin{equation}
h_{2n+1}(\rho_0)= O(\epsilon^{n+1}),
\label{hnrho0}
\end{equation}
and this property should survive the replacement 
$\rho_0 \rightarrow \rho_0 + O(\epsilon)$.

As a first application of our formalism, we can verify the consistency
of the above statements, by checking that, for all $n \geq 2$, the
condition
\begin{equation}
\sum_{m=0}^n c_{n,m} \rho_0^{2m} F_{2m+1} =  O(\epsilon^{n+1})
\end{equation}
implies
\begin{equation}
\sum_{m=1}^n  m \, c_{n,m} \rho_0^{2m} F_{2m+1} =  O(\epsilon^{n}).
\end{equation}
The proof is by induction. Assuming the property to hold for a given
$n$, and exploiting the fact that $2 \beta -1 = O(\epsilon)$, we
obtain
\begin{equation}
\sum_{m=0}^n [(2 \beta-1)m-\gamma+n] c_{n,m} \rho_0^{2m} F_{2m+1}
= O(\epsilon^{n+1}).
\label{induc}
\end{equation}
The initial condition, corresponding to the case $n=1$, has the
explicit form 
\begin{equation}
\gamma (\gamma-1)+ {1 \over 6}(2 \beta- \gamma)\rho_0^2 = O(\epsilon^2),
\end{equation}
and is a definition of $\rho_0$. Notice that
$\rho_0 =\displaystyle\lim_{\epsilon\to0} \rho_2$, and in the Ising
model $\rho_0^2 = 2$.

By applying the recursion equations we then obtain
\begin{equation}
\sum_{m=0}^n (n+1-m) c_{n+1,m} \rho_0^{2m} F_{2m+1} =  O(\epsilon^{n+1}).
\end{equation}
The sum can trivially be extended to $n+1$ and, recalling the
hypothesis, we obtain
\begin{equation}
\sum_{m=1}^{n+1}  m \, c_{n+1,m} \rho_0^{2m} F_{2m+1} =  O(\epsilon^{n+1}),
\end{equation}
thus completing the proof.  Along the same lines it is
straightforward to prove that, for all $n \geq 2$,
\begin{equation}
\sum_{m=1}^n  m^k c_{n,m} \rho_0^{2m} F_{2m+1} =  O(\epsilon^{n-k+1})
\end{equation}
for all integers $k \leq n$. The initial condition ($n=k$) is trivially 
satisfied for all $n \geq 2$:
\begin{equation}
\sum_{m=1}^n  m^n c_{n,m} \rho_0^{2m} F_{2m+1} =  O(\epsilon).
\end{equation}
As a consequence the more general statement
\begin{equation}
h_{2n+1}(\rho)= O(\epsilon^{n+1})
\end{equation}
holds for all $\rho$ admitting an $\epsilon$-expansion and possessing the
limit $\displaystyle\lim_{\epsilon\to0} \rho = \rho_0$.

This relation implies in turn that, expanding in $\epsilon$ the 
coefficients $F_{2m+1}$ for $m \geq 2$ according to
\begin{equation}
F_{2m+1} = \sum_{k=1}^{\infty} f_{mk}\epsilon^k,
\end{equation}
when the $f_{mk}$ for $m<k$ are known then all $f_{mk}$ for $m \geq k$
are fully determined.

As a simple application of the above we obtained the following closed
form result:
\begin{equation}
f_{m1}= {(-1)^m \over m(m-1)} \rho_0^{-2m}
  \displaystyle\lim_{\epsilon\to0} {\gamma -1 \over \epsilon},
\end{equation}
where $\gamma \cong 1 + {1\over6}\epsilon$ and $\rho_0=\sqrt{2}$.

Let us now consider the linear parametric model with global
stationarity in the context of the $\epsilon$-expansion: $\rho_2$
satisfies the condition $\rho_2 = \rho_0 + O(\epsilon)$, though it
does not coincide (and is not expected to) with the
$\epsilon$-expanded $\rho$ value adopted by Guida and Zinn-Justin
\cite{G-Z-97}.

Now notice that for any higher-order truncation the stationarity
condition is still solved by $\rho_t = \rho_0 + O(\epsilon)$, as shown
explicitly by the above derived Eq.\ (\ref{induc}).  As a consequence,
any stationary truncation is an accurate description of the
$\epsilon$-expanded parametric equation of state up to
$O(\epsilon^t)$ included.  Actually the freedom of choosing $\rho$
leaves such an expansion highly under-determined, and many other
prescriptions might work, including that of fixing $\rho$ (or
alternatively $\theta_0$) to its zeroth order value.  It is however
certainly pleasant to recognize that our approach based on stationarity
falls naturally into
the set of consistent truncations. As a side remark, notice that all
the coefficients of the $\epsilon$-expansion of $\rho_t$ will in
general be changed order by order in $t$, and will also in general be
complex numbers. This fact will by no means affect the real character
of the expanded physical amplitudes, and will not even prevent $\rho$
and $\theta_0$ from taking real values in the actual three-dimensional
calculations.



\begin{references}

\bibitem{ZJ-book} 
J.~Zinn-Justin,
{\em Quantum Field Theory and Critical Phenomena},
third edition (Clarendon Press, Oxford, 1996).

\bibitem{Nickel-82}
B.~G.~Nickel, in {\em Phase Transitions},
M.~L\'evy, J.~C.~Le~Guillou, and J.~Zinn-Justin eds.,
(Plenum, New York and London, 1982).

\bibitem{Gaunt-82}
D.~S.~Gaunt, in {\em Phase Transitions},
M.~L\'evy, J.~C.~Le~Guillou, and J.~Zinn-Justin eds.,
(Plenum, New York and London, 1982).

\bibitem{ZJ-79} 
J.~Zinn-Justin,
J.\ Physique {\bf 40}, 969 (1979);
J.\ Physique {\bf 42}, 783 (1981).

\bibitem{C-F-N-82} 
J.-H.~Chen, M.~E.~Fisher, and B.~G.~Nickel,
Phys.\ Rev.\ Lett.\ {\bf 48}, 630  (1982).

\bibitem{Adler-83} 
J.~Adler, J.\ Phys.\ {\bf A~16}, 3585 (1983).

\bibitem{G-R-84} M.~J.~George and J.~J.~Rehr,
Phys.\ Rev.\ Lett.\ {\bf 53}, 2063 (1984).

\bibitem{F-C-85} 
M.~E.~Fisher and J.~H.~Chen,
J.\ Physique {\bf 46}, 1645 (1985).

\bibitem{int-appr-ref}
D.~L.~Hunter and G.~A.~Baker, Jr.,
Phys.\ Rev.\ {\bf B~7}, 3346 (1973); {\bf B~7}, 3377 (1973); 
{\bf B~19}, 3808 (1979);
M.~E.~Fisher and H.~Au-Yang,
J.\ Phys.\ {\bf A~12}, 1677 (1979); Erratum {\bf A~13}, 1517 (1980);
A.~J.~Guttmann and G.~S.~Joyce, J.\ Phys.\ {\bf A~5}, L81 (1972);
J.~J.~Rehr, A.~J.~Guttmann, and G.~S~Joyce, 
J.\ Phys.\ {\bf A~13}, 1587 (1980).

\bibitem{Guttrev} 
A.~J.~Guttmann, in {\em Phase Transitions and Critical Phenomena},
vol.\ 13, C.~Domb and J.~Lebowitz eds.\ 
(Academic Press, New York, 1989).

\bibitem{Roskies-81} 
R.~Z.~Roskies,
Phys.\ Rev.\ {\bf B~24}, 5305 (1981).

\bibitem{A-M-P-82} 
J.~Adler, M.~Moshe, and V.~Privman,
Phys.\ Rev.\ {\bf B~26}, 1411 (1982); 
Phys.\ Rev.\ {\bf B~26}, 3958 (1982).

\bibitem{B-C-97} 
P.~Butera and M.~Comi, 
Phys.\ Rev.\ {\bf B~56}, 8212 (1997).

\bibitem{P-V-gr-98} 
A.~Pelissetto and E.~Vicari,
Nucl.\ Phys.\ {\bf B~519}, 626 (1998).

\bibitem{B-C-g-98}
 P.~Butera and M.~Comi, 
Phys.\ Rev.\ {\bf B~58}, 11552 (1998).

\bibitem{N-R-90}
B.~G.~Nickel and J.~J.~Rehr,
J.\ Stat.\ Phys.\ {\bf 61}, 1 (1990).

\bibitem{H-P-V-98} 
M.~Hasenbusch, K.~Pinn, and S.~Vinti,
Phys.\ Rev.\ {\bf B 59}, 11471 (1999).

\bibitem{B-F-M-M-98}
H.~G.~Ballesteros, L.~A.~Fern\'andez, V.~Mart\'{\i}n-Mayor,
and A.~Mu\~noz Sudupe, Phys.\ Lett.\ {\bf B~441}, 330 (1998).

\bibitem{B-F-M-M-P-R-99}
H.~G.~Ballesteros, L.~A.~Fern\'andez, V.~Mart\'{\i}n-Mayor,
A.~Mu\~noz Sudupe, G.~Parisi, and J.~J.~Ruiz-Lorenzo,
J.\ Phys.\ {\bf A~32}, 1  (1999).

\bibitem{Hasenbusch-99} 
M.~Hasenbusch, 
``A Monte Carlo study of leading order scaling corrections of $\phi^4$
theory on a three dimensional lattice'',
preprint {\tt hep-lat/9902026},
J.\ Phys.\ {\bf A}, in press.

\bibitem{B-N-97} 
P.~Belohorec and B.~G.~Nickel,
``Accurate universal and two-parameter model results
from a Monte-Carlo renormalization group study'',
Guelph University report (1997).

\bibitem{Hasenbusch-99-2}
M.~Hasenbusch, 
private communication,
to be published in the Habilitationsschrift.

\bibitem{G-Z-97} 
R.~Guida and J.~Zinn-Justin,
Nucl.\ Phys.\ {\bf B~489}, 626 (1997).

\bibitem{Schofield-69}
P.~Schofield, Phys.\ Rev.\ Lett.\ {\bf 22}, 606 (1969).

\bibitem{S-L-H-69}
P.~Schofield, J.~D.~Lister and J.~T.~Ho,
Phys.\ Rev.\ Lett.\ {\bf 23}, 1098 (1969).

\bibitem{Josephson-69}
B.~D.~Josephson, 
J.\ Phys.\ {\bf C}: Solid State Phys.\ {\bf 2}, 1113 (1969).

\bibitem{W-Z-74} 
D.~J.~Wallace and R.~P.~K.~Zia,
J.\ Phys.\ {\bf C~7}, 3480 (1974).

\bibitem{Wegner-76} 
F.~J.~Wegner, in
{\em Phase Transitions and Critical Phenomena},
vol.\ 6, C.~Domb and M.~S.~Green eds.
(Academic Press, New York, 1976).

\bibitem{A-F-83}
A.~Aharony and M.~F.~Fisher,
Phys.\ Rev.\ {\bf B~27}, 4394 (1983).

\bibitem{Kong-etal_86}
X.-P. Kong, H. Au-Yang, and J. H. H. Perk,
Phys. Lett. {\bf A 116}, 54 (1986);
{\bf A 118}, 336 (1986).

\bibitem{Gartenhaus-McCullough_88}
S. Gartenhaus and W. S. Mc Cullough,
Phys. Rev. {\bf B 38}, 11688 (1988).

\bibitem{Salas-Sokal_99}
J. Salas and A. D. Sokal,
``Universal Amplitude Ratios in the Critical Two-Dimensional Ising Model
on a Torus", {\tt cond-mat/9904038}.


\bibitem{G-R-76}
G.~R.~Golner and E.~K.~Riedel,
Phys.\ Lett.\ {\bf A~58}, 11 (1976).

\bibitem{N-R-84}
K.~E.~Newman and E.~K.~Riedel,
Phys.\ Rev.\ {\bf B~30}, 6615 (1984).

\bibitem{Brower-Tamayo_89}
R. C. Brower and P. Tamayo,
Phys. Rev. Lett. {\bf 62}, 1087 (1989).


\bibitem{B-C-g-97}
P.~Butera and M.~Comi, Phys.\ Rev.\ {\bf E~55}, 6391  (1997).

\bibitem{Fisher-62}
M.~E.~Fisher, Philos.\ Mag.\ {\bf 7}, 1731 (1962).

\bibitem{Guttmann-87} 
A.~J.~Guttmann, J.\ Phys.\ {\bf A~20}, 1839, 1855 (1987).

\bibitem{C-P-R-V-98} 
M.~Campostrini, A.~Pelissetto, P.~Rossi, and E.~Vicari,
Europhys.\ Lett.\ {\bf 38}, 577 (1997); 
Phys.\ Rev.\ {\bf E~57}, 184 (1998).

\bibitem{H-P-97} 
M.~Hasenbusch and K.~Pinn,
J.\ Phys.\ {\bf A~31}, 6157 (1998).

\bibitem{T-B-96}
A.~L.~Talapov and H.~W.~J.~Bl\"ote,
J.\ Phys.\ {\bf A~29}, 5727 (1996).

\bibitem{G-T-96} 
R.~Gupta and P.~Tamayo, 
Int.\ J.\ Mod.\ Phys.\ {\bf C~7}, 305 (1996).

\bibitem{B-L-H-95}
H.~W.~J.~Bl\"ote, E.~Luijten, and J.~R.~Heringa, 
J.\ Phys.\ {\bf A~28}, 6289 (1995).

\bibitem{G-Z-98} R.~Guida and J.~Zinn-Justin, 
J.~Phys.~{\bf A~31}, 8103 (1998).

\bibitem{Kleinert-98}
H.~Kleinert,
Phys.\ Rev.\ {\bf D~57}, 2264 (1998);
Addendum to paper ``Strong-Coupling Behavior of $\phi^4$-Theories
and Critical Exponents'',
preprint {\tt cond-mat/9803268}; 
``Critical Exponents from Seven-Loop Strong-Coupling $\phi^4$-Theory
in Three Dimensions'',
preprint {\tt hep-th/9812197}.

\bibitem{M-N-91} D.~B.~Murray and B.~G.~Nickel,
``Revised estimates for critical exponents for the continuum
$n$-vector model in 3 dimensions'', unpublished
Guelph University report (1991).

\bibitem{L-Z-77} J.~C.~Le Guillou and J.~Zinn-Justin,
Phys.\ Rev.\ Lett.\ {\bf 39}, 95 (1977); 
Phys.\ Rev.\ {\bf B~21}, 3976 (1980).

\bibitem{Morris-97}
T.~Morris, Nucl.\ Phys.\ {\bf B~495}, 477 (1997).

\bibitem{T-W-94}
N.~Tetradis and C.~Wetterich, Nucl.\ Phys.\ {\bf B~422}, 541 (1994).

\bibitem{B-N-G-M-77} 
G.~A.~Baker, Jr.,  B.~G.~Nickel, M.~S.~Green, and D.~I.~Meiron,
Phys.\ Rev.\ Lett.\ {\bf 36}, 1351 (1977);
G.~A.~Baker, Jr., B.~G.~Nickel, and D.~I.~Meiron,
Phys.\ Rev.\ {\bf B~17}, 1365 (1978).

\bibitem{C-G-L-T-83} 
K.~G.~Chetyrkin, S.~G.~Gorishny, S.~A.~Larin, and F.~V.~Tkachov,
Phys.\ Lett.\ {\bf B~132},  351 (1983).

\bibitem{K-N-S-C-L-93} 
H.~Kleinert, J.~Neu, V.~Schulte-Frohlinde, K.~G.~Chetyrkin,
and S.~A.~Larin,
Phys.\ Lett.\ {\bf B~272}, 39 (1991);
Erratum {\bf B~319}, 545 (1993).

\bibitem{L-Z-89}
J.~C.~Le Guillou and J.~Zinn-Justin,
J.~Physique {\bf 48}, 19 (1987).


\bibitem{Anisimov}
M.~A.~Anisimov, 
{\em Critical Phenomena in Liquids and Liquid Crystals\/} 
(Gordon and Breach, New York, 1991).

\bibitem{P-H-A-91} 
V.~Privman, P.~C.~Hohenberg, and A.~Aharony, 
in {\em Phase Transitions and Critical Phenomena}, vol.\ 14,
C.~Domb and J.~L.~Lebowitz eds., (Academic Press, New York, 1991).

\bibitem{H-S-99} 
A.~Haupt and J.~Straub,
Phys.\ Rev.\ {\bf E~59}, 1795 (1999).

\bibitem{S-N-93}
J.~Straub and K.~Nitsche,
Fluid Phase Equilibria {\bf 88}, 183 (1993).

\bibitem{Edwards-84}
T.~J.~Edwards, thesis,
University of Western Australia, 1984 (unpublished),
cited in Ref.\ \cite{H-S-99}.

\bibitem{A-P-B-94} 
I.~M.~Abdulagatov, N.~G.~Polikhsonidi, and R.~G.~Batyrova, 
J.\ Chem.\ Therm.\ {\bf 26}, 1031 (1994).

\bibitem{K-A-S-W-95}
S.~Kuwabara, H.~Aoyama, H.~Sato, and K.~Watanabe,
J.~Chem.~Eng.~Data {\bf 40}, 112 (1995)

\bibitem{Damay-etal_98}
P. Damay, F. Leclercq, R. Magli, F. Formisano, and P. Lindner,
Phys. Rev. {\bf B 58}, 12038 (1998).


\bibitem{R-J-98} P.~F.~Rebillot and D.~T.~Jacobs, 
J.\ Chem.\ Phys.\ {\bf 109}, 4009 (1998).

\bibitem{P-C-84}
M.~W.~Pestak and M.~H.~W.~Chan,
Phys.\ Rev.\ {\bf B~30}, 274 (1984).

\bibitem{K-M-T-O-94}
S.~Kawase, K.~Maruyama, S.~Tamaki, and H.~Okazaki,
J.\ Phys.\ Condens.\ Matter {\bf 6}, 10237 (1994).

\bibitem{Jacobs-86}
D.~T.~Jacobs, Phys.\ Rev.\ {\bf A~33}, 2605 (1986).

\bibitem{H-K-K-K-85}
K.~Hamano, T.~Kawazura, T.~Koyama, and N.~Kuwahara,
J.\ Chem.\ Phys.\ {\bf 82}, 2718 (1985).

\bibitem{W-G-W-92}
U.~W\"urz, M.~Grubi\'c, and D.~Woermann, 
Ber.\ Bunsenges.\ Phys.\ Chem.\ {\bf 96}, 1460 (1992).

\bibitem{A-S-94} 
X.~An and W.~Shen,
J.\ Chem.\ Therm.\ {\bf 26}, 461 (1994).

\bibitem{A-S-W-Z-94} 
X.~An, W.~Shen, H.~Wang, and G.~Zheng,
J.~Chem.\ Therm.\ {\bf 25}, 1373 (1994).

\bibitem{S-W-W-93}
W.~Schr\"oer, S.~Wiegand, and H.~Weing\"artner,
Ber.\ Bunsenges.\ Phys.\ Chem.\ {\bf 97}, 975 (1993).

\bibitem{B-W-W-92} 
V.~Balevicius, N.~Weiden, and A.~Weiss,
Z.\ Naturforsch.\ {\bf A~47}, 583 (1992).

\bibitem{D-L-C-89}
P.~Damay, F.~Leclercq, and P.~Chieux, Phys.\ Rev.\ {\bf B~40},
4696 (1989).

\bibitem{B-Y-87} D.~P.~Belanger and H.~Yoshizawa,
Phys.\ Rev.\ {\bf B~35}, 4823 (1987).

\bibitem{B-N-K-J-L-B-83}
D.~P.~Belanger, P.~Nordblad, A.~R.~King, V.~Jaccarino, 
L.~Lundgren, and O.~Beckman,
J.\ Magn.\ Magn.\ Mater.\ {\bf 31-34}, 1095 (1983)

\bibitem{M-M-B-95}
M.~Marinelli, F.~Mercuri, and D.~P.~Belanger,
J.\ Magn.\ Magn.\ Mater.\ {\bf 140-144}, 1547 (1995).

\bibitem{M-D-N-94}
J.~Mattsson, C.~Djurberg, and P.~Nordblab,
J.\ Magn.\ Magn.\ Mater.\ {\bf 136}, L23 (1994).

\bibitem{S-A-A-S-C-E-93}
M.~A.~Salgueiro, B.~G.~Almeida, M.~M.~Amado, J.~B.~Sousa, B.~Chevalier,
and J.~\'Etourneau, J.\ Magn.\ Magn.\ Mater.\ {\bf 125}, 103 (1993).

\bibitem{S-P-K-T-93}
A.~M.~Strydom, P.~de V.\ du Plessis, D.~Kaczorowski,
and E.~Tro\'c, Physica {\bf B 186-188}, 785 (1993).

\bibitem{A-B-93}
R.~Aschauer and D.~Beysens, J.\ Chem.\ Phys.\ {\bf 98}, 8194 (1993).

\bibitem{A-B-93b}
R.~Aschauer and D.~Beysens, Phys.\ Rev.\  {\bf E~47}, 1850 (1993).

\bibitem{S-B-W-94} 
J.~Schimtz, L.~Belkoura, and D.~Woermann,
Ann.\ Phys.\ (Leipzig) {\bf 3}, 1 (1994).

\bibitem{Z-B-W-94}
A.~Zielesny, L.~Belkoura, and D.~Woermann,
Ber.\ Bunsenges.\ Phys.\ Chem.\ {\bf 98}, 579 (1994).

\bibitem{H-K-K-K-91}
K.~Hamano, N.~Kuwahara, I.~Mitsushima, K.~Kubota, and T.~Kamura,
J.\ Chem.\ Phys.\ {\bf 94}, 2172 (1991).

\bibitem{S-W-92}
C.~Sinn and D.~Woermann,
Ber.\ Bunsenges.\ Phys.\ Chem.\ {\bf 96}, 913 (1992).


\bibitem{Griffiths-66}
R.~B.~Griffiths,
Phys.\ Rev.\ {\bf 152}, 240  (1966); in
{\em Phase Transitions and Critical Phenomena},
vol.\ 1, C.~Domb and M.~S.~Green eds.
(Academic Press, New York, 1972).

\bibitem{P-V-br-99}
A.~Pelissetto and E.~Vicari, Nucl.\ Phys.\ {\bf B~540}, 639 (1999).

\bibitem{Parisi-80}
G.~Parisi, Carg\`{e}se Lectures (1973),
J.\ Stat.\ Phys.\ {\bf 23}, 49 (1980).

\bibitem{Nickel-91} 
B.~G.~Nickel, 
Physica {\bf A~117}, 189 (1991).

\bibitem{C-P-R-V-96}
M.~Campostrini, A.~Pelissetto,
P.~Rossi, and E.~Vicari, Nucl.\ Phys.\ {\bf B~459}, 207 (1996).

\bibitem{Tsypin-94} 
M.~M.~Tsypin, Phys.\ Rev.\ Lett.\
{\bf 73}, 2015 (1994); 
``The universal effective potential for three-dimensional massive scalar
field theory from the Monte Carlo study of the Ising model'',
preprint {\tt hep-lat/9401034}.

\bibitem{B-B-92}
C.~M.~Bender and S.~Boettcher,
Phys.\ Rev.\ {\bf D~48}, 4919 (1992);
Phys.\ Rev.\ {\bf D~51}, 1875 (1995).

\bibitem{Kim-99} 
J.-K.~Kim, 
``The critical renormalized coupling constants in the symmetric
phase of the Ising models'',
preprint {\tt cond-mat/9905138}.



\bibitem{S-O-U-K-98} 
A.~I.~Sokolov, E.~V.~Orlov, V.~A.~Ul'kov, and S.~S.~Kashtanov,
``Universal critical coupling constants for the three-dimensional
$n$-vector model from field theory'',
 preprint {\tt hep-th/9810082},
Phys.\ Rev.\ {\bf E}, in press.

\bibitem{B-K-96} G.~A.~Baker, Jr.\ and N.~Kawashima,
J.\ Phys.\ {\bf A~29}, 7183 (1996).

\bibitem{K-L-96} 
J.-K.~Kim and D.~P.~Landau,
Nucl.\ Phys.\ {\bf B} (Proc.\ Suppl.) {\bf 53}, 706 (1997).


\bibitem{Z-L-F-96} S.-Y.~Zinn, S.-N.~Lai, and M.~E.~Fisher,
Phys.\ Rev.\ {\bf E~54}, 1176 (1996).



\bibitem{Reisz-95} T.~Reisz, Phys.\ Lett.\ {\bf B~360}, 77 (1995).

\bibitem{R-L-J-98}
J.~Rudnick, W.~Lay, and D.~Jasnow,
Phys.\ Rev.\ {\bf E~58}, 2902 (1998).

\bibitem{P-V-ef-98}
A.~Pelissetto and E.~Vicari, Nucl.\ Phys.\ {\bf B~522}, 605 (1998).

\bibitem{L-F-96} 
S.-N.~Lai and M.~E.~Fisher,
Molec.\ Phys.\ {\bf 88}, 1373 (1996).

\bibitem{F-B-67} 
M.~E.~Fisher and R.~J.~Burford,
Phys.\ Rev.\ {\bf 156}, 583 (1967).

\bibitem{T-F-75} 
H.~B.~Tarko and M.~E.~Fisher,
Phys.\ Rev.\ Lett.\ {\bf 31}, 926 (1973);
Phys.\ Rev.\ {\bf B~11}, 1217 (1975).

\bibitem{Bray_76}
A. J. Bray, Phys. Rev. {\bf B 14}, 1248 (1976).

\bibitem{Ferrel-Scalapino_75}
R. A. Ferrel and D. J. Scalapino,
Phys. Rev. Lett. {\bf 34}, 200 (1975).

\bibitem{W-M-T-B-76}
T.~T.~Wu, B.~M.~McCoy, C.~A.~Tracy, and E.~Barouch,
Phys.\ Rev.\ {\bf B~13}, 316 (1976).

\bibitem{C-P-R-V-96-2} 
M.~Campostrini, A.~Pelissetto,
P.~Rossi, and E.~Vicari, Phys.\ Rev.\ {\bf B~54}, 7301 (1996).

\bibitem{Fisher-Langer_68}
M. E. Fisher and J. S. Langer, 
Phys. Rev. Lett. {\bf 20}, 665 (1968).

\bibitem{C-D-K-74} 
M.~Combescot, M.~Droz, and J. M.~Kosterlitz,
Phys.\ Rev.\ {\bf B~11}, 4661 (1974).

\bibitem{A-T-95} 
H.~Arisue and K.~Tabata,
Nucl.\ Phys.\ {\bf B~435}, 555 (1995).

\bibitem{C-H-P-99} 
M.~Caselle, M.~Hasenbusch, and P.~Provero,
``Non-perturbative states in the 3D $\phi^4$ theory'',
preprint {\tt hep-lat/9903011}.

\bibitem{singatcoexcurve}
M.~E.~Fisher, Physics {\bf 3}, 255 (1967);
A.~F.~Andreev, Sov.\ Phys.\ JETP {\bf 18}, 1415 (1964);
M.~E.~Fisher and B.~U.~Felderhof, Ann.\ Phys.\ 
(NY) {\bf 58} 176, 217 (1970);
S.~N.~Isakov, Comm.\ Math.\ Phys.\ {\bf 95}, 427
(1984).

\bibitem{B-W-W-72} 
E.~Br\'ezin, D.~J.~Wallace, and K.~G.~Wilson, 
Phys.\ Rev.\ Lett.\ {\bf 29}, 591 (1972);
Phys.\ Rev.\ {\bf B~7}, 232 (1973).

\bibitem{F-Z-98} 
M.~E.~Fisher and S.-Y.~Zinn, J.\ Phys.\ {\bf A~31}, L629 (1998).

\bibitem{Vohwinkel-93} 
C.~Vohwinkel,
Phys.\ Lett.\ {\bf B~301}, 208 (1993).

\bibitem{F-Z-U-98}
M.~E.~Fisher, S.-Y.~Zinn, and P.~J.~Upton,
``Trigonometric models for scaling behavior near criticality'',
preprint (1998).

\bibitem{L-F-89} 
A.~J.~Liu and M.~E.~Fisher,
Physica {\bf A~156}, 35 (1989).

\bibitem{B-H-K-75}
M.~Barmatz, P.~C.~Hohenberg, and A.~Kornblit, 
Phys.\ Rev.\ {\bf B~12}, 1947 (1975).

\bibitem{N-A-85} 
J.~F.~Nicoll and P.~C.~Albright,
Phys.\ Rev.\ {\bf B~31}, 4576 (1985).

\bibitem{Bervillier-86} 
C.~Bervillier,
Phys.\ Rev.\ {\bf B~34}, 8141 (1986).

\bibitem{L-M-S-D-98}
S.~A.~Larin, M.~M\"onnigman, M.~Str\"osser, and V.~Dohm, 
Phys.\ Rev.\ {\bf B~58}, 3394 (1998).

\bibitem{B-B-M-N-87} 
C.~Bagnuls, C.~Bervillier,
D.~I.~Meiron, and B.~G.~Nickel,
Phys.\ Rev.\ {\bf B~35}, 3585 (1987).

\bibitem{C-H-97} 
M.~Caselle and M.~Hasenbusch,
J.\ Phys.\ {\bf A~30}, 4963 (1997).

\bibitem{E-S-99}
J.~Engels and T.~Scheideler,
Nucl.\ Phys.\ {\bf B~539}, 557 (1999).

\bibitem{B-L-Z-74} E.~Br\'ezin, J.~C.~Le Guillou, and J.~Zinn-Justin,
Phys.\ Lett.\ {\bf A~47}, 285 (1974).

\bibitem{A-H-76} 
A.~Aharony and P.~C.~Hohenberg,
Phys.\ Rev.\ {\bf B~13}, 3081 (1976).

\bibitem{G-K-M-96} 
G.~M\"unster and J.~Heitger,
Nucl.\ Phys.\ {\bf B~424}, 582 (1994);
C.~Gutsfeld, J.~K\"uster, and G.~M\"unster,
Nucl.\ Phys.\ {\bf B~479}, 654 (1996).

\bibitem{Z-F-96} 
S.-Y.~Zinn and M.~E.~Fisher, 
Physica {\bf A~226}, 168 (1996).

\bibitem{Fisher-pv} 
M.~E.~Fisher, private communications.

\bibitem{R-Z-W-94}
C.~Ruge, P.~Zhu, and F.~Wagner,
Physica {\bf A~209}, 431 (1994).

\bibitem{B-C-99} 
P.~Butera and M.~Comi, 
``Critical specific heats of the $N$-vector spin models on the sc and
the bcc lattices'',
preprint {\tt hep-lat/9903010}.

\bibitem{B-G-80} 
C.~Bervillier and C.~Godr\`eche,
Phys.\ Rev.\ {\bf B~21}, 5427 (1980).

\bibitem{B-B-85} 
C.~Bagnuls and C.~Bervillier,
Phys.\ Rev.\ {\bf B~32}, 7209 (1985).

\bibitem{Tsypin-97} 
M.~M.~Tsypin, 
Phys.\ Rev.\ {\bf B~55}, 8911 (1997).

\bibitem{A-C-C-H-97} 
V.~Agostini, G.~Carlino, M.~Caselle, and M.~Hasenbusch, 
Nucl.\ Phys.\ {\bf B~484}, 331 (1997).

\bibitem{S-D-89}
R.~Schloms and V.~Dohm,
Nucl.\ Phys.\ {\bf B~328}, 639 (1989);
H.~J.~Krause, R.~Schloms, and V.~Dohm,
Z.\ Phys.\ {\bf B~79}, 287 (1990).

\bibitem{B-T-W-96} 
J.~Berges, N.~Tetradis, and C.~Wetterich, 
Phys.\ Rev.\ Lett.\ {\bf 77}, 873 (1996).

\bibitem{B-C-96} 
P.~Butera and M.~Comi, 
Phys.\ Rev.\ {\bf B~54}, 15828 (1996).  

\bibitem{Delfino-98} 
G.~Delfino, Phys.\ Lett.\ {\bf B~419}, 291 (1998).  

\bibitem{B-F-85} 
M.~Barma and M.~E.~Fisher, Phys.\ Rev.\ {\bf B~31}, 5954 (1985).  
\bibitem{Wortis} 
M.~Wortis, ``Linked cluster expansion'', in {\em Phase Transitions and
Critical Phenomena}, vol.~3, C.~Domb and M.~S.~Green eds.\
(Academic Press, London, 1974).

\bibitem{L-W-88}
M.~L\"uscher and P.~Weisz,
Nucl.\ Phys.\ {\bf B~300}, 325 (1988).

\bibitem{Reisz-95a}
T.~Reisz, Nucl.\ Phys.\ {\bf B~450}, 569 (1995).

\bibitem{Knuth}
D.~E.~Knuth, {\em The art of computer programming}, vol.\ 3
(Addison-Wesley publishing co., Reading, 1973).

\end{references}
\end{document}